\documentclass[%
twocolumn,
superscriptaddress,
 amsmath,amssymb,
 aps,prb,
]{revtex4-2}

\usepackage{graphicx}% Include figure files
\usepackage{dcolumn}% Align table columns on decimal point
\usepackage{bm}% bold math
\usepackage{color}
\usepackage{ulem}
\usepackage{xspace}
\usepackage{multirow}
\usepackage[unicode=true,colorlinks=true,linkcolor=blue,citecolor=blue,urlcolor=blue]{hyperref}
\usepackage{physics}
\usepackage{braket}
\usepackage{here}
\usepackage[utf8]{inputenc}

\usepackage{comment}
\usepackage{newtxtext}
\usepackage{newtxmath}

\definecolor{green}{rgb}{0,0.6,0.1}

\begin{document}

\preprint{APS/123-QED}

\title{Ab-initio structural optimization at finite temperatures based on anharmonic phonon theory: Application to the structural phase transitions of BaTiO$_3$}% Force line breaks with \\
\author{Ryota Masuki}
\email{masuki-ryota774@g.ecc.u-tokyo.ac.jp}
\affiliation{
Department of Applied Physics, The University of Tokyo,7-3-1 Hongo, Bunkyo-ku, Tokyo 113-8656, Japan
}

\author{Takuya Nomoto}
\email{nomoto@ap.t.u-tokyo.ac.jp}
\affiliation{
Research Center for Advanced Science and Technology, The University of Tokyo,
4-6-1 Komaba Meguro-ku, Tokyo 153-8904, Japan
}

\author{Ryotaro Arita}
\email{arita@riken.jp}
\affiliation{
Research Center for Advanced Science and Technology, The University of Tokyo,
4-6-1 Komaba Meguro-ku, Tokyo 153-8904, Japan
}
\affiliation{ 
RIKEN Center for Emergent Matter Science, 2-1 Hirosawa, Wako, Saitama 351-0198, Japan 
}
\author{Terumasa Tadano}
\email{TADANO.Terumasa@nims.go.jp }
\affiliation{ 
CMSM, National Institute for Materials Science (NIMS), 1-2-1 Sengen, Tsukuba, Ibaraki 305-0047, Japan
}


\date{\today}% It is always \today, today,

\begin{abstract}
We formulate a first-principle scheme for structural optimization at finite temperature ($T$) based on the self-consistent phonon (SCP) theory, which accurately takes into account the effect of strong phonon anharmonicity. The $T$-dependence of the shape of the unit cell and internal atomic configuration is determined by minimizing the variational free energy in the SCP theory. At each optimization step, the interatomic force constants in the new structure are calculated without running additional electronic structure calculations, which makes the method dramatically efficient. We demonstrate that the thermal expansion of silicon and the three-step structural phase transitions in BaTiO$_3$ and its pressure-temperature ($p$-$T$) phase diagram are successfully reproduced. The present formalism will open the way to the non-empirical prediction of physical properties at finite $T$ of materials having a complex structural phase diagram.
\end{abstract}

%\begin{description}
%\end{description}
\maketitle

%\tableofcontents

\section{Introduction}
Temperature dependence of crystal structure is one of the most fundamental problems in solid state physics. As the crystal structure changes, a wide variety of materials exhibit fascinating properties: Ferroelectric phase transitions in perovskite oxides~\cite{PhysRev.76.1221, doi:10.1080/14786444908561371, Cohen1992, Rabe2007}, charge density wave in transition metal dichalcogenides~\cite{Xu_2021, PhysRevB.77.165135, PhysRevB.86.155125, PhysRevLett.125.106101}, 
and Martensitic transformation in the shape-memory alloy~\cite{SHAW19951243, PhysRevB.94.214110, KO2018134} are a few representative examples. 
In addition to being an important topic for its fundamental interest and expected application, such as switches, memories, and piezoelectric devices~\cite{park_lee_mikolajick_schroeder_hwang_2018, doi:10.1063/1.4990046, Li_Huilin_el_al_Ferroelectric_Polymer}, various exotic phenomena occur in the vicinity of structural phase transition.
For example, the thermoelectric effect is enlarged by the transport anomaly due to the strong scattering of carriers~\cite{https://doi.org/10.1002/adma.201302660, doi:10.1126/sciadv.1601378, PhysRevB.100.195130}. Recent theoretical works claim that damped soft modes near structural phase transitions greatly enhance superconductivity~\cite{PhysRevB.105.L020506, PhysRevB.103.094519}. High-pressure high-$T_c$ hydride superconductors~\cite{Drozdov2015, Drozdov2019, PhysRevLett.122.027001, PhysRevLett.128.167001, FLORESLIVAS20201} and the halide perovskite photovoltaics~\cite{doi:10.1021/acs.chemrev.8b00539, Lanigan-Atkins2021} are also close to structural phase transitions.
Nonempirical determination of $T$-dependent crystal structure is crucial for quantitatively understanding these phenomena and searching for related materials.

While the state-of-the-art structural optimization based on density functional theory (DFT) accurately calculates the crystal structure at zero temperature
~\cite{
%VASP
PhysRevB.54.11169,
%QE
Giannozzi_2009, Giannozzi_2017,
%EPW
PONCE2016116, NOFFSINGER20102140,
%Wannier90
Pizzi_2020, MOSTOFI20142309},
it has been a significant challenge to predict crystal structures at finite temperatures because one needs to consider the strong anharmonicity in lattice vibrations.
Anharmonicity refers to the third and higher-order terms of the Taylor expansion of the Born-Oppenheimer potential energy surface, which corresponds to phonon-phonon interaction from a field-theoretical point of view. The finite-temperature phases of strongly anharmonic crystals often have unstable phonon modes, in which case the perturbative theory based on the harmonic approximation completely breaks down due to the imaginary frequency~\cite{PhysRevB.84.180301, PhysRevLett.100.095901, PhysRevB.97.014306, PhysRevB.92.054301}.
In addition, the nuclear quantum effect also has a significant impact on the structures of crystals and molecules that contains light atoms or those in the vicinity of the structural phase transition~\cite{PhysRevLett.117.115702, doi:10.1021/ja102004b, Errea2020, PhysRevLett.89.115503}.

{\it Ab initio} molecular dynamics (AIMD) exactly treats the anharmonic effect, but it neglects the nuclear quantum effect.
Additionally, AIMD calculations often suffer from the finite-size effect of the supercell and the stochastic errors of the sampling, which makes it infeasible to obtain accurate results with reasonable computational costs.
Recently, machine-learning potentials which can efficiently calculate atomic forces and energies with first-principles quality have been developed to overcome these difficulties~\cite{PhysRevLett.98.146401, https://doi.org/10.1002/qua.24890, PhysRevB.100.014105, Gigli2022}. Nonetheless, the computational cost to generate the potentials is still high because it requires a large set of training data, and assessing the quality of the generated potentials requires special expertise.

The self-consistent phonon (SCP) theory takes into account the nuclear quantum effect and the anharmonic effect in a self-consistent way~\cite{doi:10.1080/14786435808243224, PhysRevB.1.572, PhysRevLett.17.89, PhysRev.165.951}, which have been shown to accurately reproduce the vibrational properties of strongly anharmonic materials even with imaginary harmonic frequencies~\cite{PhysRevLett.100.095901, PhysRevB.92.054301, PhysRevLett.120.105901, PhysRevB.97.014306, PhysRevLett.122.075901}. 
While several different methods and implementations have been presented~\cite{PhysRevLett.100.095901, SOUVATZIS2009888, VANROEKEGHEM2021107945, PhysRevB.97.014306, Monacelli_2021, Tadano_2014, PhysRevB.92.054301, doi:10.7566/JPSJ.87.041015, PhysRevMaterials.3.033601}, we employ the momentum-space SCP theory~\cite{PhysRevB.92.054301, doi:10.7566/JPSJ.87.041015}, which is suitable for calculation on a fine temperature grid due to the efficiency of SCP calculations at each temperature. Additionally, the momentum-space formulation takes advantage of the Brillouin zone interpolation of being able to efficiently obtain results that are free from the finite size effect of the DFT supercells.

In this paper, we formulate a theory of structural optimization at finite temperature based on the SCP theory in momentum space. The internal coordinates and the shape of the unit cell are optimized to minimize the SCP free energy with the aid of its gradient with respect to the structural degrees of freedom.
At each optimization step, the interatomic force constants (IFCs) in the updated structure are calculated from the IFCs in the reference structure without running additional DFT calculations, which makes our method highly efficient. We call the process of updating the IFCs as IFC renormalization.
The momentum-space SCP theory combined with the IFC renormalization technique remarkably extends the applicability of our method to complex materials with many degrees of freedom.

We then apply the present method to the thermal expansion of silicon and the three-step structural phase transition of BaTiO$_3$, which shows good agreement with experimental results.
BaTiO$_3$ is a perovskite oxide which is known as one of the most representative ferroelectrics~\cite{doi:10.1063/1.4990046, jona1993ferroelectric, lines2001principles}. Under ambient pressure, BaTiO$_3$ displays the cubic phase at high temperatures. When cooled, it exhibits transitions to the tetragonal phase at around 390 K, then to the orthorhombic phase at around 270 K, and finally to the rhombohedral phase at around 180 K~\cite{doi:10.1063/1.4990046}. We show that the developed method successfully reproduces the three structural phase transitions and the temperature dependence of the lattice constants as well as the spontaneous polarization of BaTiO$_3$. In addition, we perform calculations on BaTiO$_3$ under static pressure and demonstrate that our method well reproduces the experimental results of its $p$-$T$ phase diagram.

\section{Theory}

In this section, we explain the formulation of structural optimization at finite temperature based on momentum-space SCP theory. Starting from the Taylor expansion of the potential energy surface, we explain the process of updating interatomic force constants (IFCs) at each step in structural optimization, which we call the IFC renormalization. The process is essential to the efficiency of our method. In the end, the theory of structural optimization at finite temperature is presented based on the preparations.
\label{sec_theory}

\subsection{Taylor expansion of the potential energy surface}
In crystalline solids, in which the atoms are bound near their equilibrium positions, the Born-Oppenheimer potential energy surface can be Taylor-expanded in terms of the atomic displacement operator $\hat{u}_{\bm{R}\alpha \mu}$. $\bm{R}$ denotes the position of the primitive cell to which the atom belongs. $\alpha$ is the atom index in the primitive cell and $\mu = x,y,z$. If we write the potential as $\hat{U}$, its Taylor expansion is
\begin{align}
\hat{U} = \sum_{n=0}^{\infty} \hat{U}_n,
\end{align}
\begin{align}
&\hat{U}_n \nonumber \\
&=\frac{1}{n!} \sum_{\{\bm{R}\alpha \mu\}} \Phi_{\mu_1 \cdots \mu_n}(\bm{R}_1\alpha_1, \cdots, \bm{R}_n \alpha_n) \hat{u}_{\bm{R}_1 \alpha_1 \mu_1} \cdots \hat{u}_{\bm{R}_n \alpha_n \mu_n}
  \nonumber \\
&=\frac{1}{n!} \frac{1}{N^{n/2-1}} \sum_{\{\bm{k}\lambda\}} \delta_{\bm{k}_1 + \cdots + \bm{k}_n} \widetilde{\Phi} (\bm{k}_1 \lambda_1, \cdots, \bm{k}_n \lambda_n ) \hat{q}_{\bm{k_1} \lambda_1} \cdots \hat{q}_{\bm{k_n }\lambda_n}.
  \label{eq_Un}
\end{align}
The expansion coefficients
\begin{equation}
\Phi_{\mu_1 \cdots \mu_n}(\bm{R}_1\alpha_1, \cdots, \bm{R}_n \alpha_n)
= \frac{\partial^n U}{\partial {u}_{\bm{R}_1 \alpha_1 \mu_1} \cdots \partial u_{\bm{R}_n \alpha_n \mu_n}}\Bigg|_{u=0},
\end{equation}
are called the interatomic force constants (IFCs). The second-order, third-order and fourth-order terms are called the harmonic, cubic, and quartic terms, respectively.
The normal coordinate operators
\begin{equation}
\hat{q}_{\bm{k}\lambda} = 
\frac{1}{\sqrt{N}}\sum_{\bm{R}\alpha \mu} 
e^{-i\bm{k}\cdot \bm{R}} 
\epsilon^*_{\bm{k}\lambda,\alpha\mu}\sqrt{M_\alpha} \hat{u}_{\bm{R}\alpha\mu},
\end{equation}
are introduced in the momentum space, where $N$ is the number of primitive cells.
Accordingly, the momentum-space representation of the interatomic force constants is defined as
\begin{align}
&
\widetilde{\Phi}(\bm{k}_1 \lambda_1, \cdots, \bm{k}_n \lambda_n ) 
\nonumber\\
&=\frac{1}{N}\sum_{\{\bm{R}\alpha\mu\}} \Phi_{\mu_1 \cdots \mu_n}(\bm{R}_1\alpha_1, \cdots, \bm{R}_n \alpha_n) 
\nonumber\\& \times
\frac{\epsilon_{\bm{k}_1 \lambda_1,\alpha_1 \mu_1}}{\sqrt{M_{\alpha_1}}} e^{i\bm{k}_1\cdot \bm{R}_1} \cdots \frac{\epsilon_{\bm{k}_n \lambda_n,\alpha_n \mu_n}}{\sqrt{M_{\alpha_n}}} e^{i\bm{k}_n \cdot \bm{R}_n}
  \nonumber \\
&=\sum_{\{\alpha \mu\}} \frac{\epsilon_{\bm{k}_1 \lambda_1,\alpha_1 \mu_1}}{\sqrt{M_{\alpha_1}}}  \cdots \frac{\epsilon_{\bm{k}_n \lambda_n,\alpha_n \mu_n}}{\sqrt{M_{\alpha_n}}} 
\nonumber\\& \times 
  \sum_{\bm{R}_1\cdots \bm{R}_{n-1}} 
  \Phi_{\mu_1 \cdots \mu_n}(\bm{R}_1\alpha_1, \cdots, \bm{R}_{n-1} \alpha_{n-1}, \bm{0}\alpha_n)
  e^{i(\bm{k}_1\cdot \bm{R}_1+\cdots+\bm{k}_{n-1} \cdot \bm{R}_{n-1})}.\,
  \nonumber
\end{align}
where $\epsilon_{\bm{k}\lambda, \alpha \mu}$ is the polarization vector of phonon mode $\bm{k}\lambda$, which diagonalizes the harmonic dynamical matrix in the reference structure.

\subsection{Changes in the crystal structure and the IFC renormalization}
\label{subsec_IFCrenormalization}
The SCP theory incorporates the vibrational free energy based on the Born-Oppenheimer approximation, which we use to consider the finite-temperature effect in crystal structures. 
Therefore, the IFCs, which determine the potential energy surface, play a crucial role in SCP calculations, as we later explain in more detail.
In this section, we explain that it is possible to calculate the IFCs in updated crystal structures from the IFCs in the reference crystal structure~\cite{wallace1972thermodynamics}.

We first consider the change of the internal coordinates of the atoms. We write the $\mu$ component of the atomic shift of the atom $\alpha$ as
\begin{align}
  q^{(0)}_{\alpha \mu} = \sqrt{M_\alpha} u^{(0)}_{\alpha \mu}.
\end{align}
Here, we assume that the atomic shift is commensurate to the primitive cell, i.e., $u^{(0)}_{\alpha \mu}$ is
independent of the choice of primitive cell $\bm{R}$. 
When the crystal structure is updated in the structural optimization process, we re-define the atomic displacement operators $\hat{u}_{\bm{R}\alpha \mu}$ or $\hat{q}_{\bm{k}\lambda}$ as the displacements from the new crystal structure. This re-definition of displacement operators corresponds to substituting
\begin{align}
  \hat{q}_{\bm{k}\lambda}
  \leftarrow
  \begin{cases}
    \hat{q}_{\bm{k}\lambda} & (\bm{k}\neq 0)\\
    \hat{q}_{\bm{k}\lambda} + \sqrt{N} q_{\lambda}^{(0)} & (\bm{k} = 0)
  \end{cases}
\end{align}
to Eq. (\ref{eq_Un}), where
\begin{align}
  q^{(0)}_{\lambda} = \sum_{\alpha \mu} \epsilon_{\bm{0}\lambda,\alpha\mu} q_{\alpha \mu}^{(0)}.
\end{align}
Thus, Eq. (\ref{eq_Un}) in the reference structure is rewritten as 
\begin{align}
\hat{U}_n^{(q^{(0)}=0)}
  &=
  \frac{1}{n!} \frac{1}{N^{n/2-1}} \sum_{\{\bm{k}\lambda\}} \delta_{\bm{k}_1 + \cdots + \bm{k}_n} \widetilde{\Phi}^{(q^{(0)}=0)} (\bm{k}_1 \lambda_1, \cdots, \bm{k}_n \lambda_n ) 
  \nonumber \\&
  \times(\hat{q}_{\bm{k_1} \lambda_1} + \delta_{\bm{k_1}} \sqrt{N}q^{(0)}_{\lambda_1}) \cdots (\hat{q}_{\bm{k_n }\lambda_n} + \delta_{\bm{k_n}} \sqrt{N}q^{(0)}_{\lambda_n}).
  \label{EqUnq00}
\end{align}
Note that the normal-coordinate operators $\hat{q}_{\bm{k}\lambda}$ are defined in accordance with the atomic displacements in the updated structure. Properly rearranging the terms, we get
\begin{align}
    \hat{U} = \sum_{n = 0}^{\infty} \hat{U}_n^{(q^{(0)})},
    \label{EqSumUnq0}
\end{align}
\begin{align}
  \hat{U}_n^{(q^{(0)})}
  &=
  \frac{1}{n!} \frac{1}{N^{n/2-1}} \sum_{\{\bm{k}\lambda\}} \delta_{\bm{k}_1 + \cdots + \bm{k}_n} 
  \nonumber\\&\times
  \widetilde{\Phi}^{( q^{(0)} )}(\bm{k}_1 \lambda_1, \cdots, \bm{k}_n \lambda_n) \hat{q}_{\bm{k_1} \lambda_1} \cdots \hat{q}_{\bm{k_n }\lambda_n},
  \label{EqUnWithq0}
\end{align}
\begin{align}
&
  \widetilde{\Phi}^{( q^{(0)})} (\bm{k}_1 \lambda_1, \cdots, \bm{k}_n \lambda_n)
  \nonumber \\=&
\sum_{m} \frac{1}{m!} \sum_{\{\rho\}}\widetilde{\Phi}^{( q^{(0)} =0)} (\bm{k}_1 \lambda_1, \cdots, \bm{k}_n \lambda_n, \bm{0}\rho_1, \cdots \bm{0}\rho_m) q^{(0)}_{\rho_1} \cdots q^{(0)}_{\rho_m}.
\label{EqRenormalizePhiByDisplace}
\end{align}
Thus, the IFCs in the structure with atomic shifts, $\widetilde{\Phi}^{( q^{(0)})}$, can be obtained using the IFCs in the reference structure without atomic shifts ($\widetilde{\Phi}^{( q^{(0)}=0)}$) and the values of $\{q_{\lambda}^{(0)}\}$.

Next, we consider the case where the shape of the unit cell is distorted from the reference structure. If the atom at $\bm{x}$ is moved to $\bm{X}$ by the deformation, the strain can be described by the displacement gradient tensor
\begin{align}
    u_{\mu \nu} = \frac{\partial X_{\mu}}{\partial x_{\nu}} - \delta_{\mu\nu}.
\end{align}
Under this strain, the atom $\alpha$ in the primitive cell $\bm{R}$ is shifted by
\begin{align}
    u^{(0)}_{\bm{R}\alpha \mu} 
    &= 
    \sum_{\nu} u_{\mu \nu} (R_{\nu} + d_{\alpha\nu} )
    \nonumber
    \\&=
    \sum_{\nu} u_{\mu \nu} R_{\alpha \nu},
    \label{Eq_uRau_uR}
\end{align}
in real space. $\bm{d}_{\alpha}$ is the atomic position of atom $\alpha$ in the primitive cell, and we define $\bm{R}_{\alpha} = \bm{R} + \bm{d}_{\alpha}$ for notational simplicity. Equation (\ref{Eq_uRau_uR}) expresses the strain by the shifts of constituting atoms.
Following the same procedure as that for the change of internal coordinates, we can derive the formulas of the IFC renormalization for deformations of the primitive cell.
\begin{align}
    \hat{U}_n^{(u_{\mu\nu})} 
&=\frac{1}{n!} \sum_{\{\bm{R}\alpha \mu\}} \Phi^{(u_{\mu\nu})}_{\mu_1 \cdots \mu_n}(\bm{R}_1\alpha_1, \cdots, \bm{R}_n \alpha_n) \hat{u}_{\bm{R}_1 \alpha_1 \mu_1} \cdots \hat{u}_{\bm{R}_n \alpha_n \mu_n},
\end{align}
\begin{align}
&
    \Phi^{(u_{\mu\nu})}_{\mu_1 \cdots \mu_n}(\bm{R}_1\alpha_1, \cdots, \bm{R}_n \alpha_n)
    \nonumber\\
    &= 
    \sum_{m=0}^{\infty} \frac{1}{m!} \sum_{\{\bm{R}' \alpha' \mu' \nu'\}}
    \nonumber\\&\times
    \Phi^{(u_{\mu\nu}=0)}_{\mu_1 \cdots \mu_n \mu'_1 \cdots \mu'_m}(\bm{R}_1\alpha_1, \cdots, \bm{R}_n \alpha_n, \bm{R}'_1\alpha'_1, \cdots, \bm{R}'_m \alpha'_m)
    \nonumber
    \\&
    \times
    u_{\mu'_1 \nu'_1} R'_{1 \alpha'_1 \nu'_1}
    \cdots
    u_{\mu'_m \nu'_m} R'_{m \alpha'_m \nu'_m}.
    \label{EqRenormalizePhiByStrain}
\end{align}
Combining Eqs. (\ref{EqRenormalizePhiByDisplace}) and (\ref{EqRenormalizePhiByStrain}), it is possible to calculate the IFCs for arbitrary crystal structures ($\Phi^{(q^{(0)}, u_{\mu\nu})}$ or $\widetilde{\Phi}^{(q^{(0)}, u_{\mu\nu})}$) as long as the structural change is reasonably small. These equations are the key to the efficiency of our methods, which enables us to run SCP calculations without running additional DFT calculations at each structure optimization step.

We found that the second-order IFC renormalization by cell deformation [Eq.~(\ref{EqRenormalizePhiByStrain})] appears to be less accurate than that by atomic shift [Eq.~(\ref{EqRenormalizePhiByDisplace})]. The former attempts to estimate the IFCs in the distorted unit cell which is different from the original one, but the atomic forces in the distorted cell are not included in the training data set used for extracting the original IFCs.
By contrast, the latter changes the center of the Taylor expansion without changing the functional form of the potential energy surface. 
Hence, we provide an alternative option that directly calculates the coupling between the strain and the harmonic IFCs
\begin{align}
    \frac{\partial \Phi_{\mu_1 \mu_2}(\bm{R}_1\alpha_1, \bm{R}_2 \alpha_2)}{\partial u_{\mu\nu}},
\end{align}
by finite displacement method in terms of the tensor $u_{\mu\nu}$, which we use in the calculation of BaTiO$_3$. This option ensures the accurate renormalization on the harmonic IFCs, which is vital to the quantitative description of structural phase transitions because the harmonic IFCs describe the stability/instability of the system. The additional computational cost is marginal compared to that for calculating the anharmonic IFCs.
The details of the method to calculate the IFCs are described in Section~\ref{subsec_SimMethods_IFC_calculation}.
Hereafter in this paper, $\Phi$ and $\widetilde{\Phi}$ without notes on $q^{(0)}$ and $u_{\mu\nu}$ in superscripts signifies $\Phi^{(q^{(0)}, u_{\mu\nu})}$ and $\widetilde{\Phi}^{(q^{(0)}, u_{\mu\nu})}$ respectively, unless stated otherwise.

\subsection{Structural optimization based on the SCP theory}
\label{subsec_strctopt_theory}
In SCP theory, the SCP frequency $\Omega_{\bm{k}\lambda'}$ and the SCP polarization vectors are obtained by solving the SCP equation
\begin{align}
&
  \Omega_{\bm{k}\lambda_1 \lambda_2}^2 
  \nonumber\\
  &=
  \sum_{n=1}^\infty \frac{1}{(n-1)! N^{n-1}} \sum_{\{\bm{k}\lambda'\}} \Bigl( \frac{\hbar}{2} \Bigr)^{n-1}
  \nonumber\\&
  \frac{
  \widetilde{\Phi}(-\bm{k}\lambda_1,\bm{k}\lambda_2, \bm{k}_1\lambda'_1, -\bm{k}_1\lambda'_1,\cdots, -\bm{k}_{n-1}\lambda'_{n-1})
  }{\Omega_{\bm{k}_1 \lambda'_1} \cdots \Omega_{\bm{k}_{n-1} \lambda'_{n-1}} }
  \nonumber
  \\&{\times}
  \Bigl( n_B(\hbar\Omega_{\bm{k}_1 \lambda'_1}) + \frac{1}{2} \Bigr) \cdots \Bigl( n_B(\hbar\Omega_{\bm{k}_{n-1} \lambda'_{n-1}}) + \frac{1}{2} \Bigr). 
  \label{EqSCPEqOriginalMode1}
\end{align}
Here, $n_B(\hbar\Omega)=[e^{\beta\hbar\Omega}-1]^{-1}$ is the Bose-Einstein distribution function, and $\beta=1/k_{B}T$ is the inverse temperature with $k_{B}$ being the Boltzmann constant.
Note that indices of phonon modes with primes such as $\lambda'$ denotes the updated modes, which diagonalizes the SCP dynamical matrix $\Omega_{\bm{k}\lambda_1 \lambda_2}$ at temperature $T$. The phonon modes without primes like $\lambda$ are fixed throughout the calculation. The SCP frequencies $\Omega_{\bm{k}\lambda'}$ are obtained by diagonalizing the SCP dynamical matrix as $\sum_{\lambda_1} \Omega_{\bm{k}\lambda \lambda_1}C_{\bm{k}\lambda_1 \lambda'} = \Omega_{\bm{k}\lambda'}C_{\bm{k}\lambda \lambda'}$ using the $C$ matrix, which is defined in Appendix~\ref{Sec_appendix_SCP}. See Appendix~\ref{Sec_appendix_SCP} for the detailed derivation of SCP equation.
Because the derivation relies on the variational principle, the solution of the SCP equation [Eq. (\ref{EqSCPEqOriginalMode1})] satisfies
\begin{align}
  \frac{\partial \mathcal{F}}{\partial \Omega_{\bm{k}\lambda_1 \lambda_2}} = 0,
\end{align}
for the variational free energy $\mathcal{F} 
  =     -k_B T \log \Tr e^{-\beta \hat{\mathcal{H}}_0}
    + \braket{\hat{H} - \hat{\mathcal{H}}_0}_{\hat{\mathcal{H}}_0}$.
Therefore, the gradient of the SCP free energy with respect to the atomic shift is calculated as 
\begin{align}
&
  \frac{1}{N}\frac{\partial \mathcal{F}(\widetilde{\Phi}^{(q^{(0)}, u_{\mu \nu})}, \Omega_{\bm{k}\lambda_1\lambda_2}(q^{(0)}, u_{\mu\nu}))}{\partial q^{(0)}_{\lambda}}
  \nonumber\\
  &=
  \frac{1}{N}
  \frac{\partial \mathcal{F}(
  \widetilde{\Phi}^{(q^{(0)}, u_{\mu \nu})},
  \Omega_{\bm{k}\lambda_1\lambda_2})}{\partial q^{(0)}_{\lambda}} 
  + 
  \frac{1}{N}
  \sum_{\bm{k}\lambda_1\lambda_2} 
  \frac{\partial \mathcal{F}}{\partial \Omega_{\bm{k}\lambda_1 \lambda_2}}
  \frac{\partial \Omega_{\bm{k}\lambda_1 \lambda_2} }{\partial q^{(0)}_\lambda}
  \nonumber
  \\&=
  \frac{1}{N}
  \frac{\partial \mathcal{F}(
  \widetilde{\Phi}^{(q^{(0)}, u_{\mu \nu})},
  \Omega_{\bm{k}\lambda_1\lambda_2})}{\partial q^{(0)}_{\lambda}} 
  \nonumber
  \\&=
  \sum_{n=0}^\infty \frac{1}{n! N^n} \sum_{\{\bm{k}\lambda\}} \Bigl( \frac{\hbar}{2} \Bigr)^{n}
  \frac{\widetilde{\Phi}(\bm{k}_1\lambda'_1,-\bm{k}_1\lambda'_1, \cdots, \bm{k}_n\lambda'_n,-\bm{k}_n\lambda'_n,\bm{0}\lambda)}{\Omega_{\bm{k}_1\lambda'_1} \cdots \Omega_{\bm{k}_n\lambda'_n}}
  \nonumber
  \\&\times
  \Bigl( n_B(\hbar \Omega_{\bm{k}_1\lambda'_1}) + \frac{1}{2} \Bigr) \cdots \Bigl( n_B(\hbar \Omega_{\bm{k}_n\lambda'_n}) + \frac{1}{2} \Bigr),
  \label{EqDelF1Delq0}
\end{align}
in the normal coordinate representation, where we used Eqs. (\ref{EqRenormalizePhiByDisplace}) and (\ref{EqSCPFreeEnergyFreePolarization}). $\Omega_{\bm{k}\lambda_1 \lambda_2}( q^{(0)}_{\lambda}, u_{\mu\nu})$ is the solution of the SCP equation for the crystal structure given by $q^{(0)}_{\lambda}$, $u_{\mu\nu}$.
As for the cell deformation, the gradient of the SCP free energy with respect to the displacement gradient tensor $u_{\mu \nu}$ can be derived in the same way. Here, we add the $pV$ term to the free energy to take into account the effect of pressure.
\begin{align}
&
  \frac{1}{N}\frac{\partial \mathcal{F}(
  \widetilde{\Phi}^{(q^{(0)}, u_{\mu \nu})},
  \Omega_{\bm{k}\lambda_1\lambda_2}(q^{(0)}, u_{\mu\nu}))}{\partial u_{\mu\nu}}
  \nonumber \\
  &=
  \frac{1}{N}\frac{\partial \mathcal{F}(
  \widetilde{\Phi}^{(q^{(0)}, u_{\mu \nu})},
  \Omega_{\bm{k}\lambda_1\lambda_2})}{\partial u_{\mu\nu}} 
  \nonumber
  \\&=
  p \frac{\partial v_{\text{cell}}}{\partial u_{\mu\nu}}
  \nonumber
  \\&+
  \sum_{n=0}^\infty \frac{1}{n! N^n} \sum_{\{\bm{k}\lambda\}} \Bigl( \frac{\hbar}{2} \Bigr)^{n}
  \frac{\partial \widetilde{\Phi}(\bm{k}_1\lambda'_1,-\bm{k}_1\lambda'_1, \cdots, \bm{k}_n\lambda'_n,-\bm{k}_n\lambda'_n)}{\partial u_{\mu\nu}}
  \nonumber
  \\&\times
  \frac{( n_B(\hbar \Omega_{\bm{k}_1\lambda'_1}) + 1/2)}{\Omega_{\bm{k}_1\lambda'_1}} \cdots \frac{( n_B(\hbar \Omega_{\bm{k}_n\lambda'_n}) + 1/2 )}{\Omega_{\bm{k}_n\lambda'_n}},
  \label{EqDelF1Delumunu}
\end{align}
where $p$ is the static pressure and $v_{\text{cell}}$ is the volume of the unit cell.
Using the above formulas, we can relax the crystal structure to minimize the SCP free energy in the following way.
Concerning the internal coordinates, the difference $\delta q^{(0)}_{\lambda}$ from the optimum values at a given temperature can be estimated by solving
\begin{align}
\frac{1}{N}
    \frac{\partial \mathcal{F}}{\partial q^{(0)}_{\lambda}} = \sum_{\lambda_1} \Omega_{\bm{0}\lambda \lambda_1} \delta q^{(0)}_{\lambda_1},
\end{align}
if we approximate the Hessian of the SCP free energy by the SCP dynamical matrix $\Omega_{\bm{0}\lambda_1 \lambda_2}$. The internal coordinate is updated as
\begin{align}
    q^{(0)}_{\lambda} \leftarrow q^{(0)}_{\lambda} - \beta_{\text{mix,ion}} \delta q^{(0)}_{\lambda}.
    \label{Eq_ion_update}
\end{align}
$\beta_{\text{mix,ion}}$ is introduced to make the calculation scheme more robust and is usually chosen to be $0<\beta_{\text{mix,ion}} < 1$. 
In updating the shape of the unit cell, we estimate the difference  $\delta u_{\mu \nu}$ from the optimum value by 
\begin{align}
    \frac{1}{N}
    \frac{\partial \mathcal{F}}{\partial u_{\mu\nu}} = \sum_{\mu_1 \nu_1} C_{\mu\nu, \mu_1 \nu_1} \delta u_{\mu_1 \nu_1},
    \label{EqLinearEqDeltaU}
\end{align}
where we approximate the Hessian of the SCP free energy by the second-order elastic constants $C_{\mu\nu, \mu_1 \nu_1}$, which is defined in Appendix~\ref{Sec_Appendix_Implementation}.
We restrict the solution $\delta u_{\mu_1 \nu_1}$ of Eq. (\ref{EqLinearEqDeltaU}) to be symmetric in order to fix the rotational degree of freedom and get an unique solution.
The displacement gradient tensor is updated by 
\begin{align}
    u_{\mu\nu} \leftarrow u_{\mu\nu} - \beta_{\text{mix,cell}} \delta u_{\mu\nu},
    \label{Eq_cell_update}
\end{align}
where $0 < \beta_{\text{mix,cell}}<1$ is a coefficient like $\beta_{\text{mix,ion}}$, which is introduced to improve robustness of the calculation.

\begin{figure}[h]
\vspace{0cm}
\begin{center}
\includegraphics[width=0.3\textwidth]{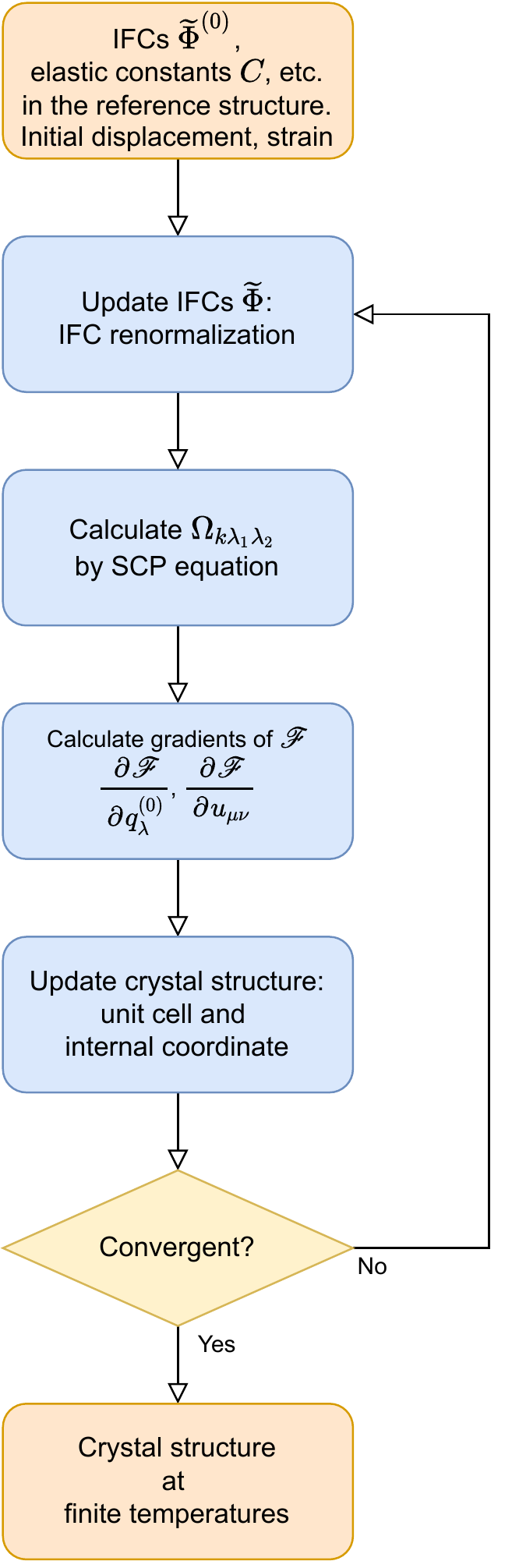}
\caption{
Schematic figure of the calculation flow of the structural optimization at finite temperature based on SCP theory. 
}
\label{Fig_Calculation_scheme}
\end{center}
\end{figure}

Putting together the above considerations, the calculation scheme of the structural optimization based on SCP theory is as follows, which is visualized with a flowchart in Fig.~\ref{Fig_Calculation_scheme}.

\begin{enumerate}
    \item Calculate IFCs of the current structure by the IFC renormalization.
    \item Solve SCP equation (Eq. (\ref{EqSCPEqOriginalMode}) )
    \item Calculate the gradient of the SCP free energy by Eqs. (\ref{EqDelF1Delq0}) and (\ref{EqDelF1Delumunu}).
    \item Update crystal structure by Eqs. (\ref{Eq_ion_update}) and (\ref{Eq_cell_update}).
    \item Check convergence. The calculation ends if the converged structure has been obtained.
    \item If convergence has not been achieved, return to 1.
\end{enumerate}
We discuss the details of the implementation in Appendix~\ref{Sec_Appendix_Implementation}.

\section{Simulation Details}
We apply the developed method to the thermal expansion of silicon and the three-step structural phase transition of BaTiO$_3$. In this section, we describe the detailed setting in the calculation of these materials.

\subsection{Calculation of the interatomic force constants (IFCs) and the SCP calculation}
\label{subsec_SimMethods_IFC_calculation}
The interatomic force constants (IFCs) and the elastic constants are calculated before performing the structural optimization at finite temperatures. The Taylor expansion of the potential energy surface is truncated at the fourth-order ($\hat{U} \simeq U_0 + \hat{U}_2 + \hat{U}_3 + \hat{U}_4$). The IFCs are calculated by fitting the relationship between atomic displacements and atomic forces, which are calculated by using external DFT packages. The fitting is performed by the compressive sensing method, which enables the efficient extraction of IFCs from a small number of displacement-force data~\cite{PhysRevB.92.054301, PhysRevLett.113.185501, PhysRevB.100.184308}. We impose on the IFCs the acoustic sum rule, the permutation symmetry, and the space group symmetry not only in the supercell model but also in the infinite real space. The details on the sum rules and symmetries of the IFCs are described in Appendix~\ref{Appendix_sumrule_symmetry_IFCs}.
The second-order and third-order elastic constants are calculated by fitting the strain-energy relation making use of the crystal symmetry~\cite{PhysRevB.75.094105, LIAO2021107777}.

For silicon, 2$\times$2$\times$2 conventional cubic supercell, which contains 64 atoms, is used in the phonon calculation. We use the lattice constant 5.4362 \AA~for calculating the IFCs and the elastic constants, which is obtained by the structural optimization based on DFT so that the stress tensor in the reference structure vanishes.
We generate 100 configurations by the random sampling at 300 K from harmonic phonon dispersion to obtain the displacement-force data and extract the anharmonic IFCs. The cutoff radius, which is introduced in Appendix~\ref{Appendix_conv_check_cutoff_Nmax}, for the third and fourth-order IFCs are 20.0 Bohr and 10.0 Bohr, respectively. The cutoff radius for the fourth-order IFCs can be relatively small because the anharmonicity is so weak in silicon that the higher-order IFCs are highly localized.
The fitting error of the displacement-force data is 0.52 \%. We use $8\times8\times8$ $k$-mesh for the SCP calculation.

A 2$\times$2$\times$2 supercell, which contains 40 atoms, is employed for the phonon calculation of BaTiO$_3$.
The high-symmetry cubic cell with the lattice constant 3.9855 \AA~is employed as the reference structure, which is used to calculate IFCs and the elastic constants, whose lattice constant is obtained by the DFT-based structural optimization. We choose the cubic phase as the reference structure to take full advantage of the crystal symmetry.
For BaTiO$_3$, we generated 300 configurations by adding random displacements to AIMD snapshots. The AIMD calculation is run with a smaller basis cutoff and a smaller number of $k$-points to efficiently sample the potential energy surface, whose details are shown in the next section. The displacement-force data are generated by DFT calculations with higher accuracy. We set the cutoff radius for the third and fourth-order IFCs as 15.0 Bohr, which is comparable to one side of the supercell. The fourth-order IFCs are restricted up to three-body terms, in which at least two of the four atoms regarding the IFC are the same. 
The fitting error of the forces is 3.51 \%, which shows that the calculated IFCs well reproduce the complex landscape of the potential energy surface of BaTiO$_3$.
We use $8\times8\times8$ $k$-mesh and do not add the nonanalytic correction to the dynamical matrix in the SCP calculation. The convergence of the calculation results with respect to the setting in the IFC calculation and the SCP calculation are thoroughly checked in Appendices~\ref{Appendix_conv_check_cutoff_Nmax}$\sim$\ref{Appendix_diffrence_of_Born_effcharge}.

\subsection{DFT calculation}
We employ the \textit{Vienna ab initio simulation package} (VASP)~\cite{PhysRevB.54.11169} for the electronic structure calculations. The PBEsol exchange-correlation functional~\cite{PhysRevLett.100.136406} and the PAW pseudopotentials~\cite{PhysRevB.50.17953, PhysRevB.59.1758} are used. The convergence criteria of the SCF loop is set $10^{-8}$ eV and accurate precision mode, which suppresses egg-box effects and errors, are used to calculate the forces accurately. We use a $4\times4\times4$ Monkhorst-Pack $k$-mesh for both materials. The basis cutoff is set 500 eV for silicon and 600 eV for BaTiO$_3$. A $2\times2\times2$ Monkhorst-Pack $k$-mesh and basis cutoff of 400 eV are used for the AIMD calculation of BaTiO$_3$, which is for sampling the potential energy surface to generate the random configurations.

\section{Result and Discussion}
We perform the first-principles calculation on the thermal expansion of silicon and the successive structural phase transitions of BaTiO$_3$.

\subsection{Silicon}
\label{Subsec_ResDis_silicon}
\begin{figure}[h]
\vspace{0cm}
\begin{center}
\includegraphics[width=0.48\textwidth]{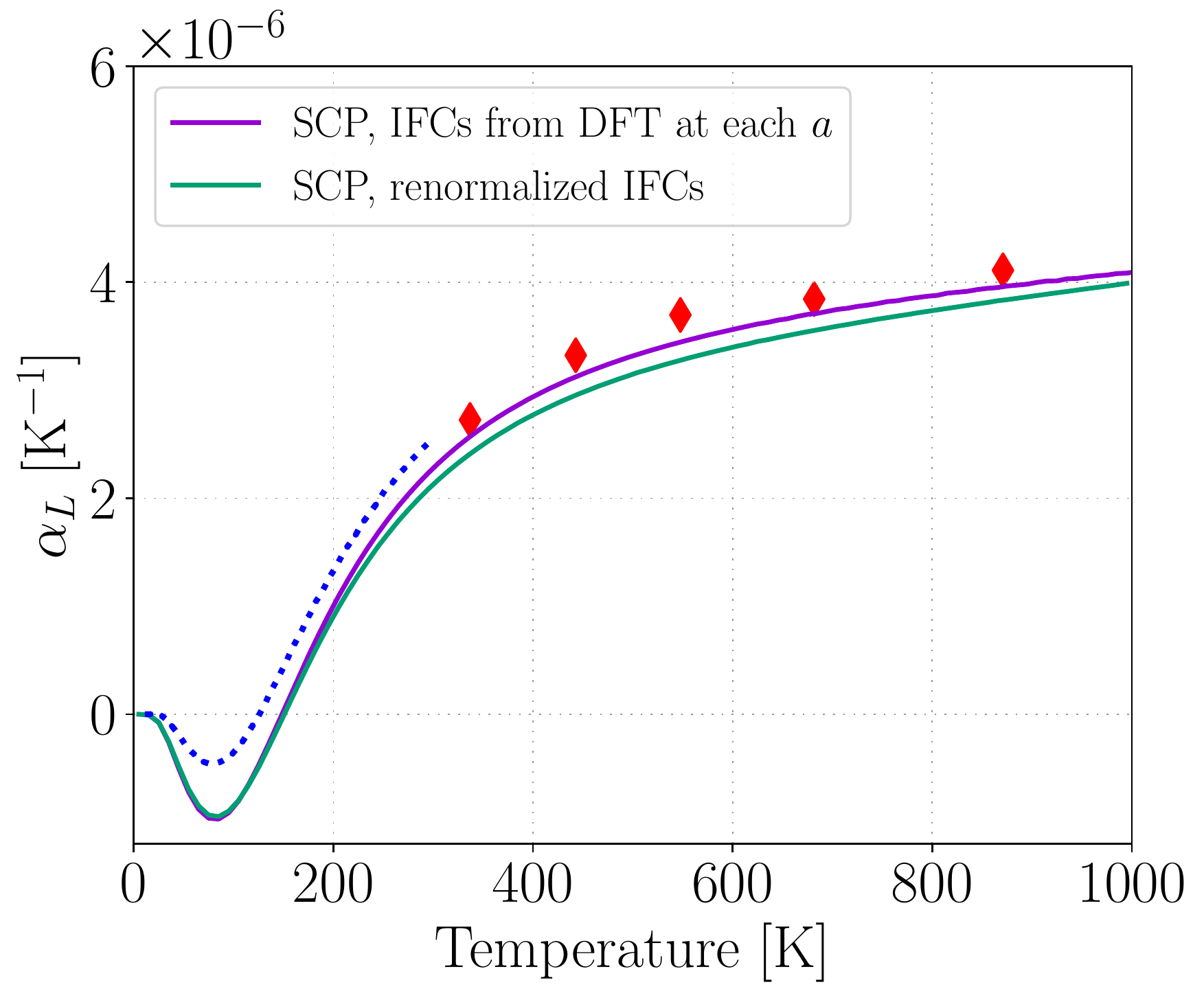}
\caption{
Calculation results of the temperature dependence of the linear thermal expansion coefficient $\alpha_L = \frac{1}{a}\frac{\partial a}{\partial T}$, where $a$
 is the lattice constant.
In the calculation of the purple line (SCP, IFCs from DFT at each $a$), the calculation of the IFCs and the free energy is performed at 14 different lattice constants and the optimum lattice constant at each temperature is obtained by fitting the free energy by the Birch-Murnaghan equation of state.
For the green line (SCP, renormalized IFCs), the lattice constant is directly calculated by the structural optimization at finite temperature including the nuclear quantum effect, which is developed in this research. The experimental data are taken from Ref.~\cite{PhysRevB.92.174113} (blue dotted line) and Ref.~\cite{doi:10.1063/1.333965} (red diamond).
}
\label{Fig_SiThermalExpansion}
\end{center}
\end{figure}

The thermal expansion of silicon is calculated by the structural optimization at finite temperatures developed in this research. Since the internal coordinate in the cubic cell is determined by the symmetry, only the lattice constant $a$ change with temperature. In Fig. \ref{Fig_SiThermalExpansion}, 
the linear thermal expansion coefficient $\alpha_L = \frac{1}{a}\frac{\partial a}{\partial T}$ calculated by two methods are compared, where $a$ is the lattice constant.
In the former method (SCP, IFCs from DFT at each $a$), the calculation of the IFCs and the free energy is performed at 14 different lattice constants, and the optimum lattice constant at each temperature is obtained by fitting the free energy by the Birch-Murnaghan equation of state~\cite{PhysRev.71.809, doi:10.1073/pnas.30.9.244}, as in the calculations in Refs.~\cite{PhysRevMaterials.3.033601, PhysRevB.105.064112}.
It is more accurate because the IFCs are calculated from DFT calculations at different lattice constants instead of estimating them by the IFC renormalization.
However, our method (SCP, renormalized IFCs) is much more efficient than the other because running the DFT calculations to generate the displacement-force data is the most expensive part of the calculation. The calculation results of the two methods in Fig.~\ref{Fig_SiThermalExpansion} are almost identical, which confirms that the IFC renormalization accurately calculates the change of IFCs in this case. The accuracy of the IFC renormalization is also tested in Appendix~\ref{Sec_Appendix_test_IFCremorm} by investigating the harmonic phonon dispersions in different lattice constants.

\subsection{BaTiO$_3$}
\label{Subsec_ResDis_BTO}

\subsubsection{Structural optimization at finite temperatures}
\label{subsec_str_opt_at_finiteT}
BaTiO$_3$ is a typical perovskite ferroelectrics, which shows a three-step structural phase transition at ambient pressure~\cite{doi:10.1063/1.4990046, doi:10.1080/14786444908561371, PhysRevB.52.6301, PhysRevLett.73.1861, Gigli2022, PhysRevLett.89.115503}. 
When the temperature is lowered, it exhibits structural phase transitions from the cubic phase to the tetragonal phase at around 390 K, then to the orthorhombic phase at around 270 K, and lastly to the rhombohedral phase at around 180 K~\cite{doi:10.1063/1.4990046}. 

\begin{figure}[h]
\vspace{0cm}
\begin{center}
\includegraphics[width=0.48\textwidth]{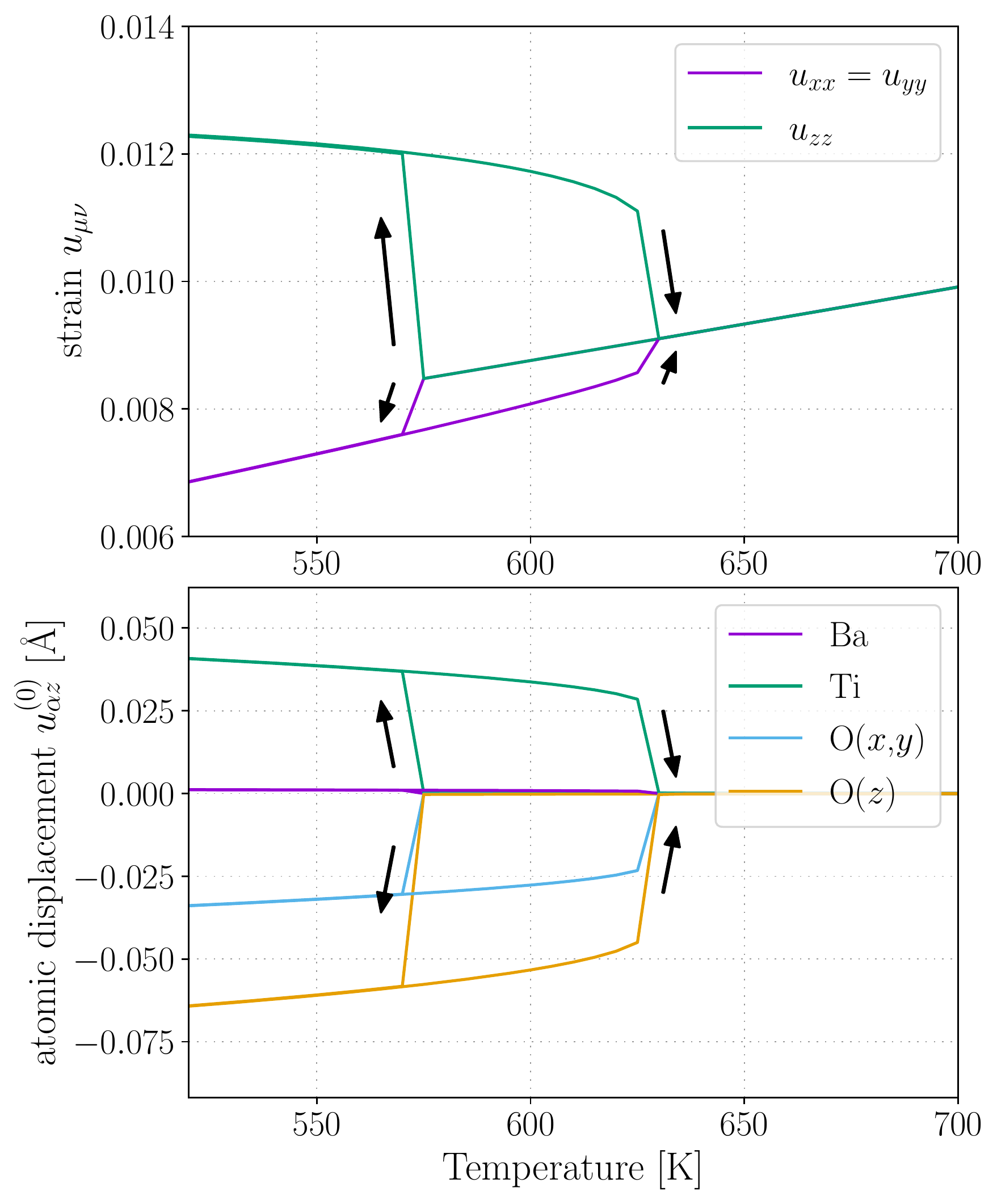}
\caption{
Temperature dependence of the crystal structure near the cubic-tetragonal phase transition of BaTiO$_3$. A hysteresis is observed between the cooling and heating calculations. In the cooling calculation, the initial structure is set by adding a small displacement to the high-symmetry cubic phase to induce symmetry breaking. In the heating calculation, we start from the lower temperatures and use the optimized structure at the previous temperature as the initial structure for the calculation at each temperature.
Both the unit cell and internal coordinates are relaxed considering the nuclear quantum effect.
}
\label{Fig_BTOCubicTetraNdata300Cutoff15Nbody233}
\end{center}
\end{figure}

We start with the calculation at ambient pressure, where we set the static pressure as 0 GPa.
We first calculate the cubic-tetragonal phase transition, which occurs at the highest temperature of the three structural phase transitions. The temperature dependence of the crystal structure (displacement gradient tensor $u_{\mu\nu}$ and the atomic displacements $u^{(0)}_{\alpha z}$) is shown in Fig.~\ref{Fig_BTOCubicTetraNdata300Cutoff15Nbody233}. 
In the cooling calculation, the initial structure is set by adding a small displacement to the high-symmetry cubic phase to induce symmetry breaking. In the heating calculation, we start from the lower temperatures and use the optimized structure at the previous temperature as the initial structure for the calculation at each temperature.
According to the figure, the atomic displacements are zero at high temperature while they become finite at low temperature, which shows the presence of the structural phase transition. The hysteresis between the cooling and heating calculations shows that the transition is first-order, which is consistent with the experiments although its temperature range is overestimated~\cite{doi:10.1080/14786444908561371, doi:10.1063/1.4990046}. 

\begin{figure}[h]
\vspace{0cm}
\begin{center}
\includegraphics[width=0.48\textwidth]{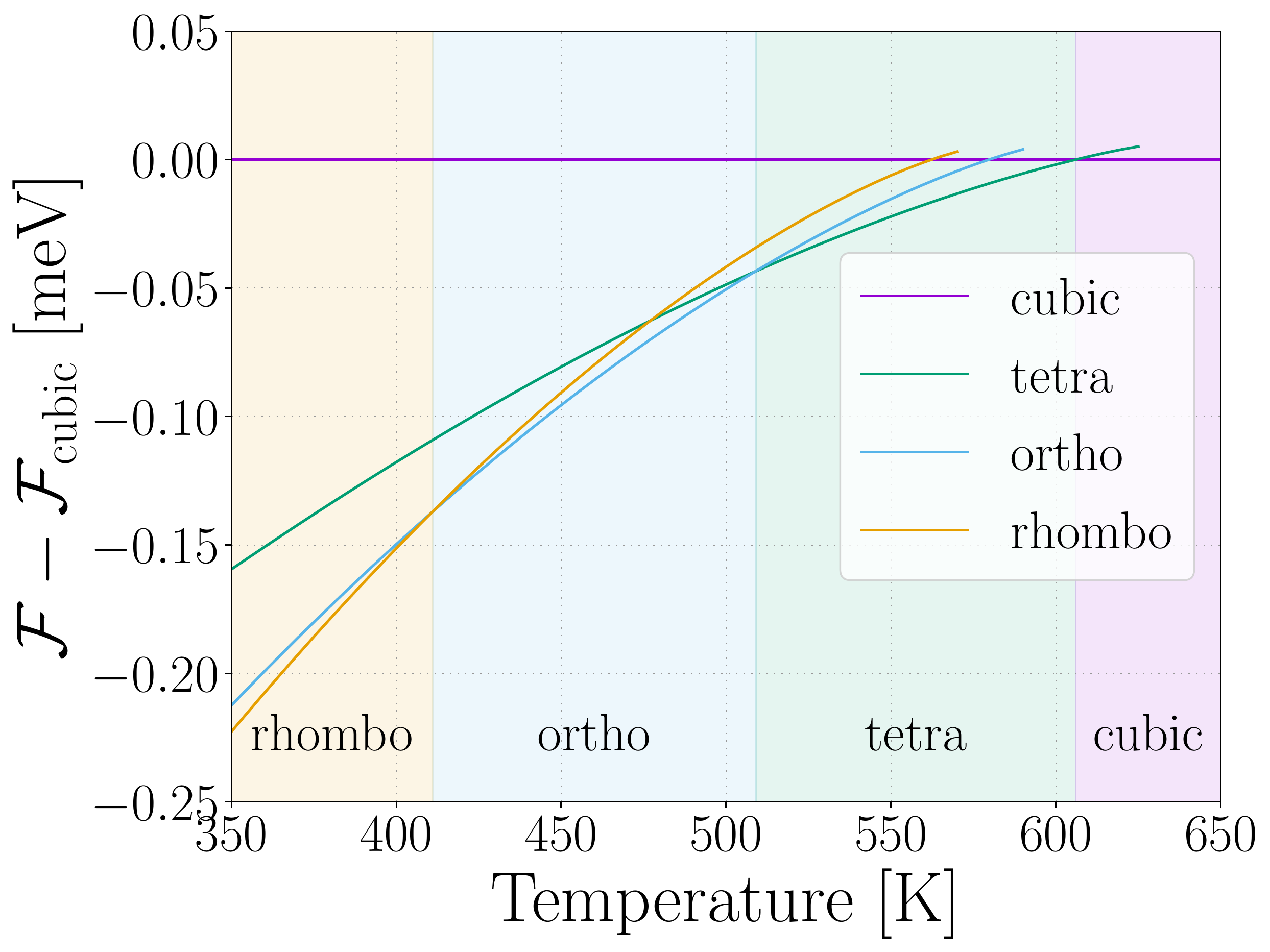}
\caption{
Temperature-dependence of SCP free energy of the four phases of BaTiO$_3$ calculated by the structural optimization at finite temperature. The difference from the SCP free energy of the cubic phase $\mathcal{F}_{\text{cubic}}$ is plotted to make the difference clearer.
The nuclear quantum effect is included in the calculation of the crystal structure and the free energy.}
\label{Fig_BTOFreeEnergyNdata300Cutoff15Nbody233}
\end{center}
\end{figure}

Figure~\ref{Fig_BTOFreeEnergyNdata300Cutoff15Nbody233} depicts the calculation result of the $T$-dependent SCP free energy of the four phases of BaTiO$_3$. As shown in Fig.~\ref{Fig_BTOFreeEnergyNdata300Cutoff15Nbody233}, the cubic phase is the most stable at high temperatures. As the temperature is lowered, the tetragonal phase, the orthorhombic phase, and then the rhombohedral phase become the most stable, whose order is consistent with the experiment. Therefore, we have successfully reproduced the three-step structural phase transition of BaTiO$_3$ with our method.
The transition temperatures ($T_c$) can be calculated from the crossing points of the free energy, which are comparable to the experimental results as summarized in Table~\ref{table_Tcs_BTO}. The origin of the differences between the experimental and calculation results are discussed later in this section in relation to the lattice constants.

\begin{table}[t]
  \caption{Transition temperatures of the three successive structural phase transitions of BaTiO$_3$. The comparison between the calculation results estimated from the crossing points of the SCP free energy and the experimental result~\cite{doi:10.1080/14786444908561371}. 
  The nuclear quantum effect is taken into account in the calculation.}
  \label{table_Tcs_BTO}
  \centering
  \begin{ruledtabular}
  
  \begin{tabular}{ccc}
      & our method [K] & experiment [K]\\
      \hline
      cubic-tetra & 606 & $\sim$390 \\
      tetra-ortho & 509 & $\sim$ 270 \\
      ortho-rhombo & 411 & $\sim$ 180
    \\ 
  \end{tabular}
    \end{ruledtabular}
\end{table}

\begin{figure}[h]
\vspace{0cm}
\begin{center}
\includegraphics[width=0.48\textwidth]{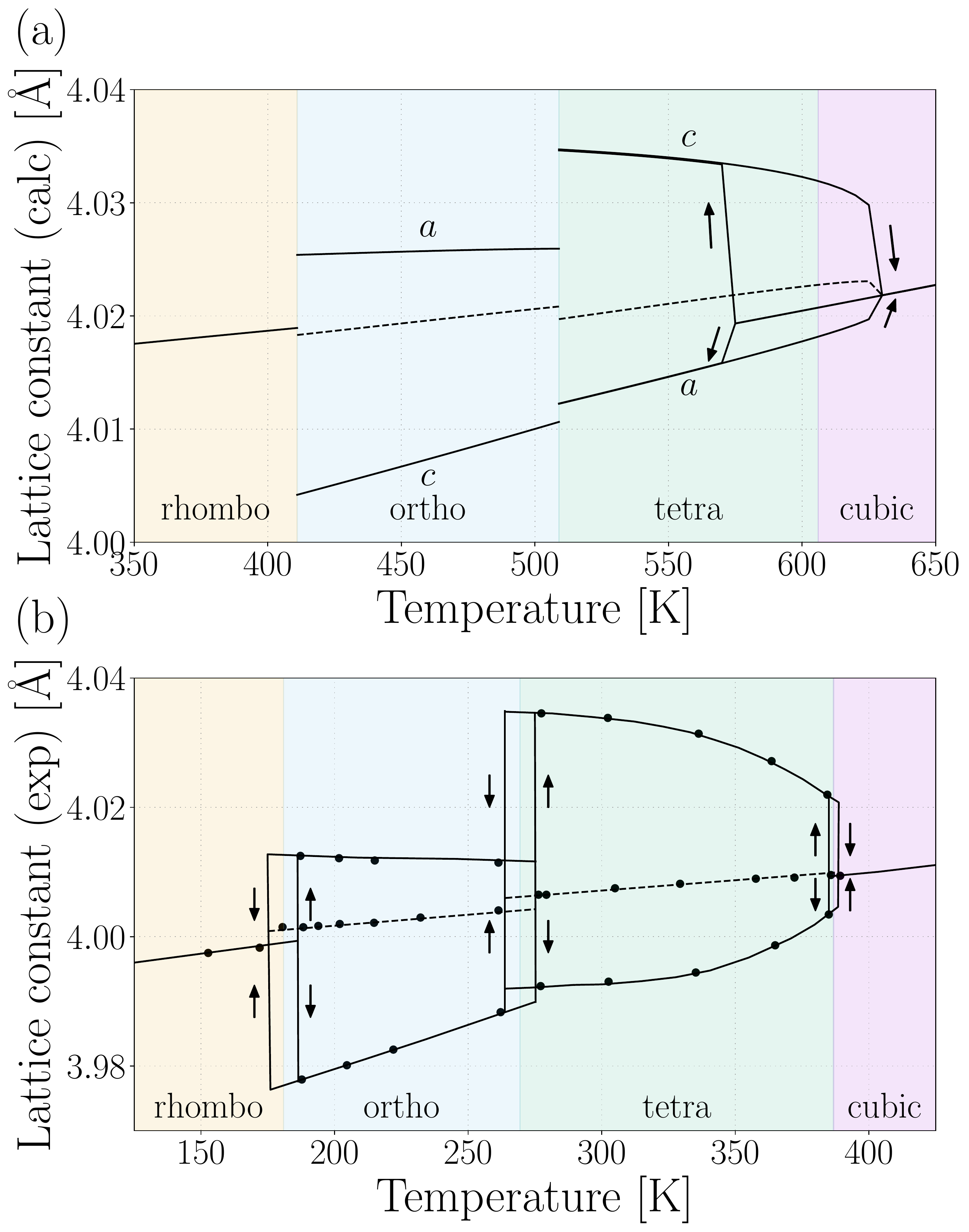}
\caption{
(a)
Calculation result of the temperature-dependent lattice constants of BaTiO$_3$. The results of the rhombohedral and the orthorhombic phases are plotted only in the temperature ranges where each phase is the most stable. 
The dotted lines represent the cubic root of the cell volume.
The hysteresis between the cooling and the heating calculations is plotted for the cubic-tetragonal phase transition. Both the unit cell and internal coordinates are relaxed including the nuclear quantum effect. (b) The experimental data taken from Ref.~\cite{doi:10.1080/14786444908561371}.
The boundaries of the background colors of each phase are taken to be the center of the hysteresis regions.
}
\label{BTOLatticeConstNdata300Cutoff15Nbody233}
\end{center}
\end{figure}

\begin{figure}[h]
\vspace{0cm}
\begin{center}
\includegraphics[width=0.48\textwidth]{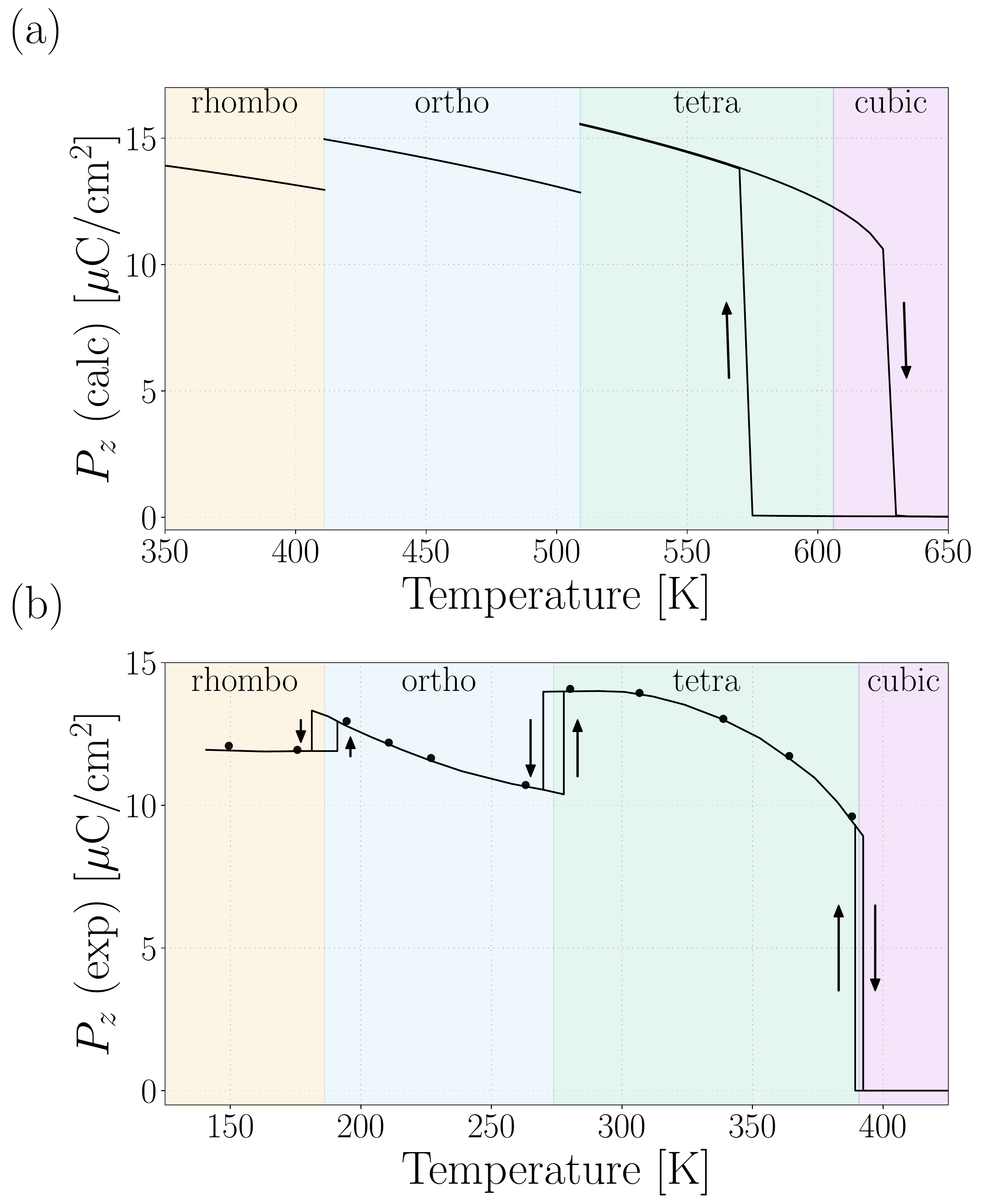}
\caption{
(a)
Calculation result of the temperature dependence of the spontaneous polarization $P_z$. The results of the rhombohedral and the orthorhombic phase are plotted only in the temperature ranges where each phase is the most stable. The hysteresis between the cooling and the heating calculations is plotted for the cubic-tetragonal phase transition. The spontaneous polarization is calculated from the atomic displacements and the Born effective charge. Both the unit cell and internal coordinates are relaxed including the nuclear quantum effect.
(b) The experimental data taken from Ref.~\cite{doi:10.1080/14786444908561371}. 
The boundaries of the background colors of each phase are taken to be the center of the hysteresis regions.
}
\label{BTOPolarizationNdata300Cutoff15Nbody233}
\end{center}
\end{figure}

The calculation results of the lattice constants and the spontaneous polarization are shown in Fig.~\ref{BTOLatticeConstNdata300Cutoff15Nbody233} and~\ref{BTOPolarizationNdata300Cutoff15Nbody233}, which are in agreement with the experimental results~\cite{ doi:10.1080/14786444908561371}
if the difference in transition temperatures are compensated. The spontaneous polarization is calculated by using the atomic displacements $u^{(0)}_{\alpha\nu}$ and the Born effective charge $Z^*_{\alpha, \mu\nu}$ as
\begin{align}
    P_{\mu} = \frac{e}{V_{\text{cell}}} \sum_{\alpha \nu} Z^*_{\alpha, \mu\nu} u^{(0)}_{\alpha\nu},
    \label{Eq_spontaneous_polarization_Borneffcharge}
\end{align}
where $V_{\text{cell}}$ is the volume of the primitive cell and $e$ is the elementary charge. The Born effective charge is calculated by  density functional perturbation theory~\cite{PhysRevB.33.7017, PhysRevB.73.045112} in the original high-symmetry cubic structure.
In Figs.~\ref{BTOLatticeConstNdata300Cutoff15Nbody233} and~\ref{BTOPolarizationNdata300Cutoff15Nbody233}, we show the hysteresis only for the cubic-tetragonal phase transition since we have not directly calculated the transitions between the tetragonal and the orthorhombic phase and between the orthorhombic and the rhombohedral phases. This is because we calculate the phase transitions between the cubic phase and each of the other three phases to take full advantage of the crystal symmetry. 

\begin{figure}[h]
\vspace{0cm}
\begin{center}
\includegraphics[width=0.48\textwidth]{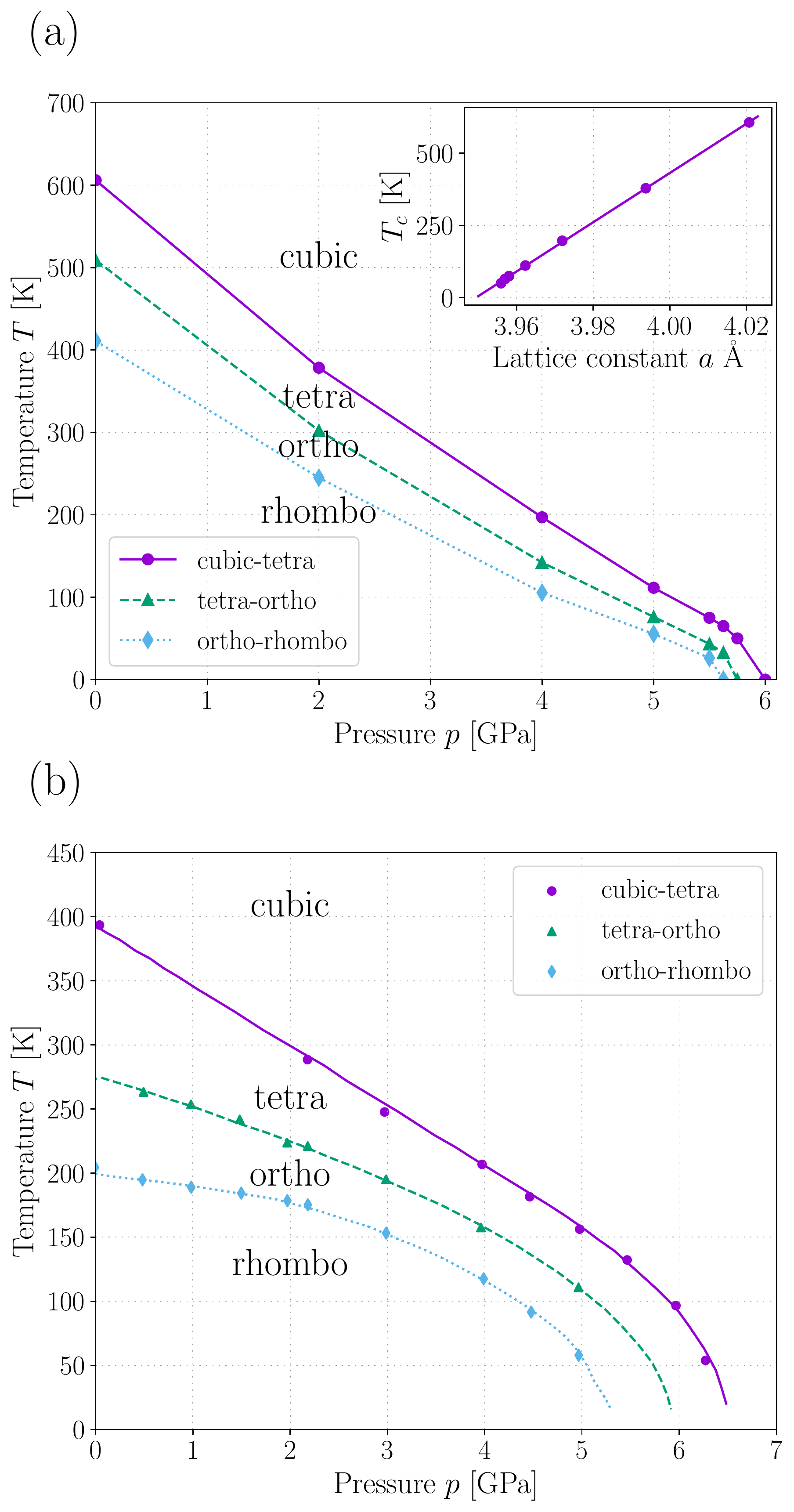}
\caption{
(a) The $p$-$T$ phase diagram of BaTiO$_3$ calculated by the structural optimization at finite temperature including the nuclear quantum effect. The inset shows the relation between the lattice constant $a$ and the transition temperature $T_c$ of the cubic-tetragonal phase transition. (b) The experimental result taken from Ref.~\cite{PhysRevLett.78.2397}.
}
\label{PT_phase_diagram}
\end{center}
\end{figure}

Let us discuss the origin of the overestimation of $T_c$ in Table~\ref{table_Tcs_BTO}.
The lattice constant is one of the primary factors that produce this deviation. 
In the inset in Fig.~\ref{PT_phase_diagram} (a), we show the relation between the calculated $T_c$ of the cubic-tetragonal phase transition in BaTiO$_3$ and the lattice constant $a$ at the corresponding temperature, which is derived from the finite-pressure calculations. The transition temperature at the experimental value of $a = 4.0093$ \AA \ is estimated as 511 K from the linear fitting in the figure. Thus, the error in the lattice constant explains around half of the error in the transition temperature.
The other possible sources of the errors in the transition temperatures can be the error in the DFT functional, the truncation of the Taylor expansion of the potential energy surface at the fourth order, and the approximation of the SCP theory.

The error in the lattice constant originates from the errors of the functional used in the DFT calculations. From the inset of Fig.~\ref{PT_phase_diagram} (a), we estimate that $\frac{dT_c}{da}\sim 8500$ K\;\AA$^{-1}$. Since the PBEsol functional predicts the lattice constants of materials with the accuracy of 0.3 \%~\cite{Zhang_2018}, the DFT error of the lattice constant can lead to an error in $T_c$ of around 102 K in this case. Comparing at the same temperature, the error of the lattice constant in our calculation is around 0.2 \%, which is within the DFT accuracy but significantly affects the calculated transition temperature.

We then investigate the pressure dependence of transition temperatures to draw the $p$-$T$ phase diagram of BaTiO$_3$. Similar calculations are run with different pressures, and the pressure ($p$)-dependent transition temperatures are estimated either from the crossing points of the $T$-dependent free energies or from the onset of finite atomic displacements when no hysteresis is observed between the cooling and the heating calculations. The resultant phase diagram is shown in Fig.~\ref{PT_phase_diagram}. The calculated phase diagram well reproduces the experimental result~\cite{PhysRevLett.78.2397}. At zero temperature, the rhombohedral phase is the most stable at low pressure. Applying the pressure, the orthorhombic phase, tetragonal phase, and then the cubic phase become the most stable, which is shown to be a consequence of the nuclear quantum effect~\cite{PhysRevLett.89.115503}.

Note that the temperature dependences of the crystal structures in Figs.~\ref{Fig_BTOCubicTetraNdata300Cutoff15Nbody233}$\sim$\ref{BTOPolarizationNdata300Cutoff15Nbody233} are calculated at every 5 K, and the gaps are connected by linear interpolation.
The plotted curves look smooth because we perform a deterministic calculation on a fine temperature grid. This is because we do not need to run time-consuming electronic structure calculations at each temperature or structural optimization step, which is enabled by the IFC renormalization as discussed in Section~\ref{subsec_IFCrenormalization}.
Therefore, the most costly part of the calculation is to run electronic structure calculations to get the force-displacement data for the 300 supercells. In comparison, the computational cost of each step of the structural optimization is negligible, which is one of the most significant advantages of the present method compared to other methods~\cite{SOUVATZIS2009888, VANROEKEGHEM2021107945, PhysRevB.97.014306, Monacelli_2021}. In addition, this property is also effective in the calculation of $p$-$T$ phase diagram in Fig.~\ref{PT_phase_diagram} because no additional electronic structure calculations are required for calculations at different pressures.

Some works proposed that the phase transitions of BaTiO$_3$ have an order-disorder nature~\cite{PhysRevB.13.207, PhysRevLett.120.167601, PhysRevB.94.134308, Gigli2022}, which is characterized by multiple-peak distribution of atomic positions above $T_c$. Our method based on SCP theory does not reproduce this property because the distribution is calculated within the effective harmonic Hamiltonian. However, softening of the optical phonons is observed in BaTiO$_3$~\cite{PhysRev.177.848, PhysRevB.26.5904}, which indicates that the treatment using the effective harmonic Hamiltonian is valid in this material. Since the main point of SCP theory is to calculate the best effective harmonic Hamiltonian by the variational principle, we consider that the SCP theory can be applied when the effective harmonic approach is reasonable, which is supported by the calculation results. Even when the transition is order-disorder or mixed type, SCP theory captures the general trend of the true distribution in a mean-field way because the displacement along the soft mode will be larger near the phase transition.

\subsubsection{Importance of relaxing the shape of the unit cell}
\label{Sec_Appendix_BTO_fixcell}
\begin{figure}[h]
\vspace{0cm}
\begin{center}
\includegraphics[width=0.48\textwidth]{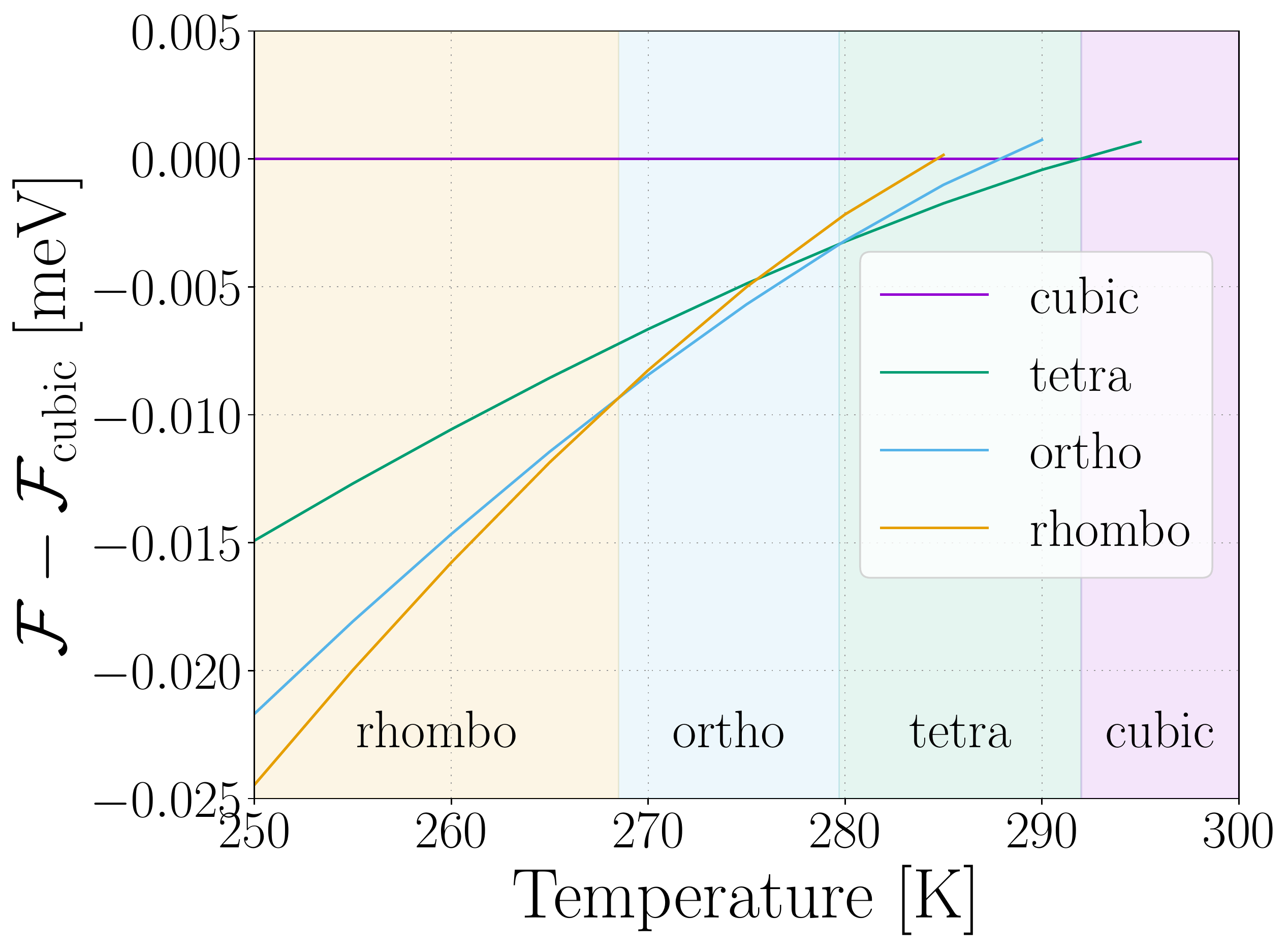}
\caption{
The temperature-dependence of SCP free energy of the four phases of BaTiO$_3$ calculated by the structural optimization at finite temperature. The unit cell is fixed to the original cubic cell and only the internal coordinates are relaxed. The difference from the SCP free energy of the cubic phase $\mathcal{F}_{\text{cubic}}$ is plotted to make the difference clearer.
The nuclear quantum effect is included in the calculation.}
\label{Fig_BTOFreeEnergyNdata300Cutoff15Nbody233_fix_cell}
\end{center}
\end{figure}

We perform the calculation of the structural phase transitions of BaTiO$_3$ with the unit cell fixed to the original cubic cell to investigate the importance of relaxing the unit cell. The calculation result of the free energy of the four phases is shown in Fig.~\ref{Fig_BTOFreeEnergyNdata300Cutoff15Nbody233_fix_cell}. In Fig~\ref{Fig_BTOFreeEnergyNdata300Cutoff15Nbody233_fix_cell}, the three transition temperatures are very close to each other, with gaps of around only 10 K. These gaps between transition temperatures are much smaller than in Fig.~\ref{Fig_BTOFreeEnergyNdata300Cutoff15Nbody233}, in which calculation the unit cell is relaxed.

We consider that there are two possible reasons for the underestimation of the gaps between transition temperatures when the unit cell is fixed.
Firstly, the stabilization effect by the cell deformation is underestimated in the tetragonal and orthorhombic phases. In both phases, the cell will get elongated to the polarization direction, which is in the $z$ direction for the tetragonal phase and in the $xy$ direction for the orthorhombic phase. This effect will be smaller in the cubic and the rhombohedral phase because the three lattice constants are the same as shown in Fig.~\ref{BTOLatticeConstNdata300Cutoff15Nbody233}. Therefore, if the unit cell is fixed, the crossing point of the free energy curves of the cubic and the tetragonal phase is shifted to a lower temperature while that of the orthorhombic and the rhombohedral phase is shifted to a higher temperature. 
The second possible explanation is the effect of thermal expansion. In general, the instability of the soft mode gets more significant if the unit cell expands, which can be seen from Figs.~\ref{Fig_BTOIFCRenormalize} and~\ref{Fig_BTOIFCRenormalizeStrainModeCoupling}. Hence, the transition temperatures will become higher in expanded unit cells. When the effect of thermal expansion is taken into account, the higher transition temperatures will go further higher than lower transition temperatures because high-symmetry phases become less favored in expanded cells. 
Therefore, it is vital to relax the shape of the unit cell to perform accurate calculations on the structural phase transition of materials, especially when three or more competing phases are involved.

\subsubsection{Significance of the nuclear quantum effect}
\label{Sec_Appendix_NQE}

We examine the importance of considering the nuclear quantum effect in calculating the structural phase transition. Here, we perform calculation without nuclear quantum effect, in which we replace $n_B(\hbar \omega) + \frac{1}{2}$ by $\frac{k_B T}{\hbar \omega}$ in the theory. The calculations are performed for four different phases of BaTiO$_3$, and the $T$-dependent crystal structure and free energy are calculated.
The three-step structural phase transition of BaTiO$_3$ has been reproduced by the calculation without nuclear quantum effect. However, as shown in Table~\ref{table_Tcs_BTO_NQE}, the transition temperatures are overestimated when we disregard the nuclear quantum effect. This overestimation occurs because the quantum fluctuation stabilizes the high-symmetry phases.
\begin{table}[t]
  \caption{
  Transition temperatures of the successive structural phase transitions of BaTiO$_3$. The comparison between the calculation results with and without nuclear quantum effect (NQE). The calculation which considers the nuclear quantum effect is explained in Section~\ref{Subsec_ResDis_BTO} in the main part.
  ~\cite{doi:10.1080/14786444908561371}.}
  \label{table_Tcs_BTO_NQE}
  \centering
  \begin{ruledtabular}
  
  \begin{tabular}{ccc}
      & with NQE [K] & without NQE [K]\\
      \hline
      cubic-tetra & 606 & 666 \\
      tetra-ortho & 509 & 572 \\
      ortho-rhombo & 411 & 475
    \\ 
  \end{tabular}
    \end{ruledtabular}
\end{table}

\begin{figure}[h]
\vspace{0cm}
\begin{center}
\includegraphics[width=0.48\textwidth]{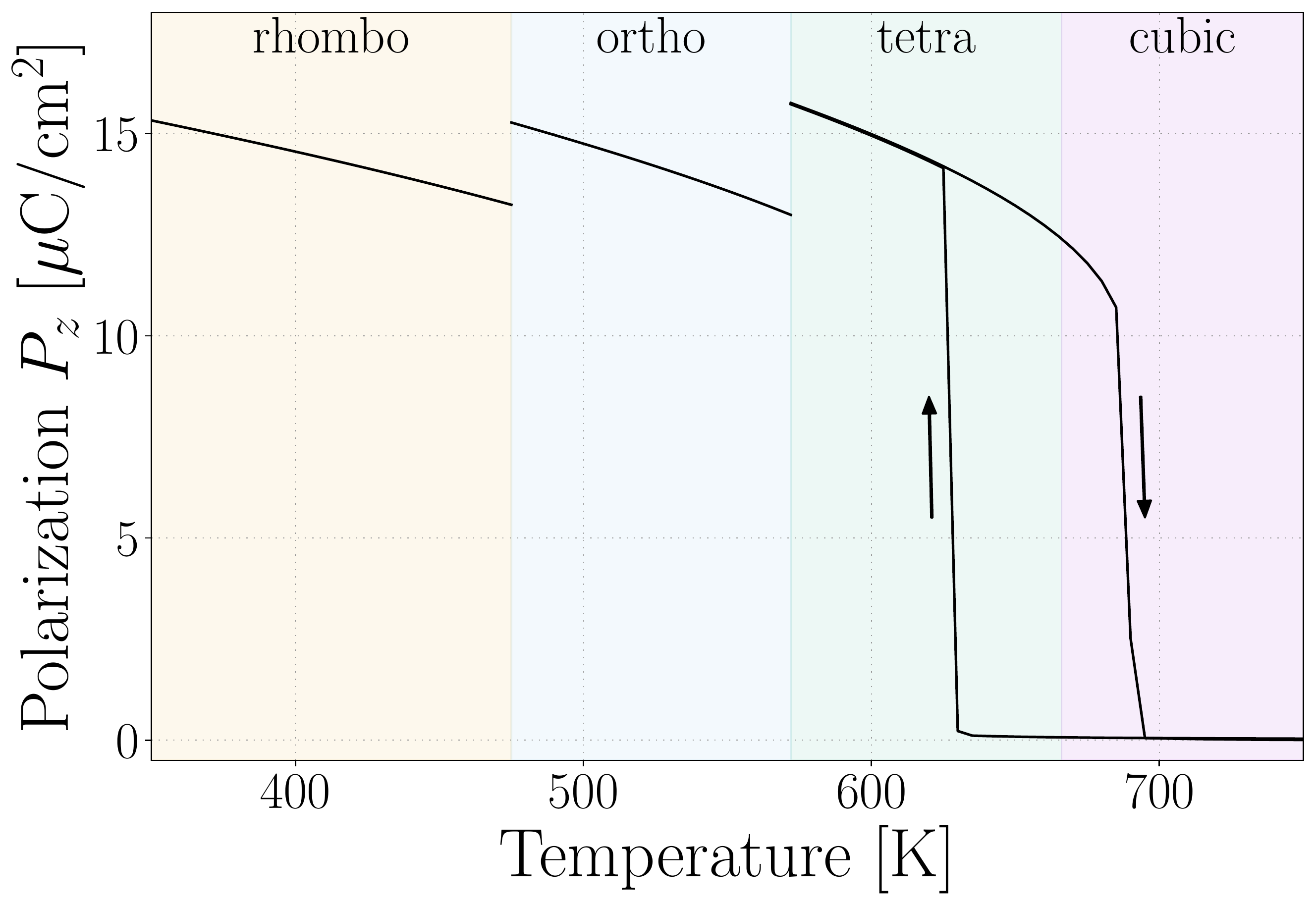}
\caption{
Temperature dependence of the spontaneous polarization $P_z$ calculated without nuclear quantum effect. The results of the rhombohedral and the orthorhombic phase are plotted only in the temperature ranges where each phase is the most stable. The hysteresis between the cooling and the heating calculations is plotted for the cubic-tetragonal phase transition. The spontaneous polarization is calculated from the atomic displacements and the Born effective charge as in Eq. (\ref{Eq_spontaneous_polarization_Borneffcharge}).}
\label{BTOPolarizationNdata300Cutoff15Nbody233_classical}
\end{center}
\end{figure}

\begin{figure}[h]
\vspace{0cm}
\begin{center}
\includegraphics[width=0.48\textwidth]{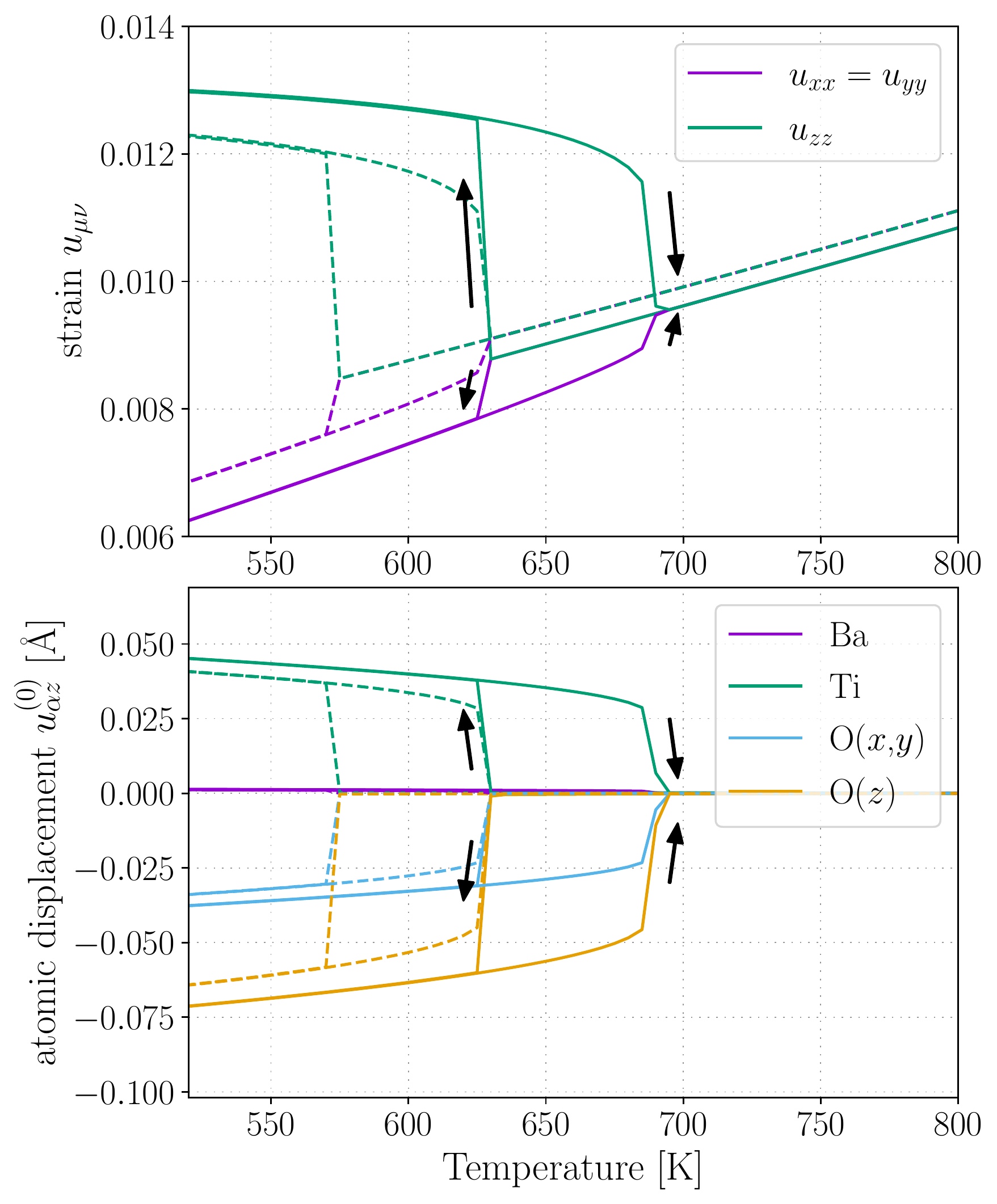}
\caption{
Temperature dependence of the crystal structure near the cubic-tetragonal phase transition of BaTiO$_3$.
The solid represents the calculation without nuclear quantum effect. The dashed line is the calculation results with nuclear quantum effect, which is plotted for comparison.
}
\label{BTOCubicTetraNdata300Cutoff15Nbody233_classical}
\end{center}
\end{figure}

In addition, the temperature dependence of the spontaneous polarization and the crystal structure near the cubic-tetragonal phase transition are shown in Figs.~\ref{BTOPolarizationNdata300Cutoff15Nbody233_classical} and~\ref{BTOCubicTetraNdata300Cutoff15Nbody233_classical}. We can see that the spontaneous polarization is overestimated compared to the experimental results or the calculation that includes the nuclear quantum effect, which are shown in Fig.~\ref{BTOLatticeConstNdata300Cutoff15Nbody233}. According to Fig.~\ref{BTOCubicTetraNdata300Cutoff15Nbody233_classical}, the deviation from the cubic structure gets larger in the classical calculation, which we identify as the reason for the overestimation of spontaneous polarization $P$.

From the above analysis, we can see that the nuclear quantum effect has a significant impact on the crystal structures near the structural phase transitions, and properly incorporating it is essential for the quantitative prediction of material properties at finite temperatures.

\section{Conclusion} 
We formulate the theory of structural optimization at finite temperature based on the SCP theory, which can accurately describe the lattice vibrations with strong anharmonic effect. The crystal structure is calculated by minimizing the SCP free energy with the help of its gradient with respect to the internal coordinates and the displacement gradient tensor $u_{\mu\nu}$. The interatomic force constants are updated without additional DFT calculations at each structural optimization step, which makes the calculation efficient.

We implement the theory to the ALAMODE package~\cite{Tadano_2014, PhysRevB.92.054301, PhysRevMaterials.3.033601} and perform calculations on silicon and BaTiO$_3$, which accurately reproduce the experimental results. As for BaTiO$_3$, we have successfully reproduced its three-step structural phase transition with quantitative agreement between theory and experiment for the $T$-dependent lattice constants and spontaneous polarization. Furthermore, the pressure-temperature ($p$-$T$) phase diagram of BaTiO$_3$ calculated by the developed method shows good agreement with experiment.

The present methodology can be generally applied to a wide range of materials. We expect that it lays the foundation for the non-empirical calculation of exotic finite temperature properties near the structural phase transitions.

\begin{acknowledgments}
We thank A. Togo and A. Seko for their fruitful comments. This work was supported by JSPS KAKENHI Grant Number 21K03424 and 19H05825, Grant-in-Aid for JSPS Fellows (22J20892), and JST-PRESTO (JPMJPR20L7).
\end{acknowledgments}

\appendix

\section{Self-consistent phonon (SCP) theory}
\label{Sec_appendix_SCP}
We derive the self-consistent phonon (SCP) equation in momentum-space representation from the variational principle of the free energy. The resultant SCP equation is equivalent to the previous derivations that use self-energy and Dyson's equation~\cite{PhysRevB.92.054301} or the ones in real-space representation~\cite{PhysRevB.89.064302}. The variational property plays an important role in calculating the gradient of the SCP free energy in Section~\ref{subsec_strctopt_theory}.

In preparation, we define the $C$ matrix, which describes the mode transformation, as
\begin{align}
  \hat{q}_{\bm{k}\lambda} = \sum_{\lambda'} C_{\bm{k}\lambda \lambda'} \hat{q}_{\bm{k}\lambda'},
\end{align}
\begin{align}
  C_{\bm{k}\lambda \lambda'} = \sum_{\alpha \mu} \epsilon^*_{\bm{k}\lambda, \alpha\mu} \epsilon_{\bm{k}\lambda', \alpha\mu},
\end{align}
where $\lambda'$ is the index of the updated polarization vector and the $\lambda$ of the original polarization vector. The updated mode $\lambda'$ diagonalizes the SCP dynamical matrix like Eq.~(\ref{Eq_SCP_dynamical_matrix}). The original mode $\lambda$ must be fixed throughout the structural optimization process, for which we take the mode that diagonalizes the harmonic dynamical matrix in the reference structure, $\widetilde{\Phi}^{(q^{(0)}=0, u_{\mu\nu}=0)}$. We distinguish the indices of the original mode and the updated mode by the absence and the presence of prime ($'$), as stated in Section~\ref{subsec_strctopt_theory}.

The variational principle that the trial free energy $\mathcal{F}$ defined below are always the same or larger than the true free energy $F$, is used in the derivation of the SCP equation.
\begin{align}
    \mathcal{F} 
  = 
  -k_B T \log \Tr e^{-\beta \hat{\mathcal{H}}_0} + \braket{\hat{H} - \hat{\mathcal{H}}_0}_{\hat{\mathcal{H}}_0}
  \geq F.
  \label{EqVariationalFreeEnergy}
\end{align}
Note that $\mathcal{F}$ and $F$ need to include $pV$ term when we consider systems under static pressure. However, we don't treat the $pV$ term explicitly in this Appendix because it does not play an important role here.
In SCP theory, we restrict the trial Hamiltonian $\hat{\mathcal{H}}_0$ to harmonic Hamiltonians
\begin{align}
  \hat{\mathcal{H}}_0 = \sum_{\bm{k}\lambda'} \hbar \Omega_{\bm{k}\lambda'} \hat{a}_{\bm{k}\lambda'}^\dag \hat{a}_{\bm{k}\lambda'}.
  \label{Eq_trial_harmonic_Hamiltonian}
\end{align}
The variational parameters $C_{\bm{k}\lambda \lambda'}$ (or $\epsilon_{\bm{k}\lambda', \alpha \mu}$) and $\Omega_{\bm{k}\lambda'}$ are optimized so as to minimize the variational free energy to get the best approximation of the original problem.
We first calculate the variational free energy by substituting Eq. (\ref{Eq_trial_harmonic_Hamiltonian}) to Eq. (\ref{EqVariationalFreeEnergy}).
\begin{align}
  \mathcal{F} 
  &= 
    -k_B T \log \Tr e^{-\beta \hat{\mathcal{H}}_0}
    + \braket{\hat{H} - \hat{\mathcal{H}}_0}_{\hat{\mathcal{H}}_0}
  \nonumber
  \\&=
  \sum_{\bm{k}\lambda'} \Bigl[
    \frac{1}{2} \hbar \Omega_{\bm{k}\lambda'} 
    + 
    k_B T \log(1 - e^{-\beta \hbar \Omega_{\bm{k}\lambda'}  })
  \Bigr]
  \nonumber
  \\&-
  \sum_{\bm{k}\lambda'}
  \frac{\hbar}{2}\Bigl(n_B(\hbar \Omega_{\bm{k}\lambda'}) + \frac{1}{2}\Bigr)
  \nonumber\\&\times
  \Bigl( \Omega_{\bm{k}\lambda'} - \sum_{\lambda_1 \lambda_2} \frac{C^{*}_{\bm{k}\lambda_1 \lambda'} \widetilde{\Phi}(-\bm{k}\lambda_1,\bm{k}\lambda_2) C_{\bm{k}\lambda_2 \lambda'}}{\Omega_{\bm{k}\lambda'}} \Bigr)
  \nonumber
  \\&+
  \sum_{n=2}^{\infty} \frac{1}{n!} \frac{1}{N^{n-1}} \sum_{\{\bm{k}\lambda'\}} \Bigl( \frac{\hbar}{2}\Bigr)^n
  \frac{\widetilde{\Phi}(\bm{k}_1\lambda'_1,-\bm{k}_1\lambda'_1,\cdots,\bm{k}_n\lambda'_n,-\bm{k}_n\lambda'_n)}{\Omega_{\bm{k}_1\lambda'_1}\cdots\Omega_{\bm{k}_n\lambda'_n}}
  \nonumber\\&\times
  \Bigl( n_B(\hbar \Omega_{\bm{k}_1 \lambda'_1}) + \frac{1}{2} \Bigr) \cdots \Bigl( n_B(\hbar \Omega_{\bm{k}_n \lambda'_n}) + \frac{1}{2} \Bigr).
  \label{EqSCPFreeEnergyFreePolarization}
\end{align}
The IFCs in the updated-mode representation are defined as 
\begin{align}
&
    \widetilde{\Phi}(\bm{k}_1 \lambda'_1, \cdots, \bm{k}_m \lambda'_m)
    \nonumber\\&
    =
    \sum_{\lambda_1 \cdots \lambda_m}
    \widetilde{\Phi}(\bm{k}_1 \lambda_1, \cdots, \bm{k}_m \lambda_m)
    C_{\bm{k}_1 \lambda_1 \lambda'_1} \cdots C_{\bm{k}_m \lambda_m \lambda'_m}
\end{align}
Now, let us consider minimizing the variational free energy of Eq. (\ref{EqSCPFreeEnergyFreePolarization}). 
We treat the SCP dynamical matrix
\begin{align}
  \Omega_{\bm{k}\lambda_1 \lambda_2} = \sum_{\lambda'} C_{\bm{k}\lambda_1 \lambda'} \Omega_{\bm{k}\lambda'} C^{*}_{\bm{k}\lambda_2 \lambda'} 
  \label{Eq_SCP_dynamical_matrix}
\end{align}
as the variational parameters because the $C$ matrix is difficult to handle due to the unitarity condition and the phase degrees of freedom. Abbreviating the momentum and the mode indices, the SCP dynamical matrix can be written as $\Omega = CDC^\dag$, where $D = \text{diag}(\Omega_{\bm{k}\lambda'})$. Here, we derive some formulas that are used in the following derivation of the SCP equation. 
We consider that the SCP dynamical matrix $\Omega = CDC^{\dag}$ is dependent on a parameter $s$.
From the unitarity of the $C$ matrix, we can show
\begin{align}
  \frac{\partial C^{\dag}}{\partial s} C + C^{\dag} \frac{\partial C}{\partial s} = 0.
\end{align}
Differentiating both hand sides of $\Omega = CDC^\dag$, we get
\begin{align}
  \frac{\partial D}{\partial s} 
  =
  C^{\dag} \frac{\partial \Omega}{\partial s} C - C^{\dag} \frac{\partial C}{\partial s} D - D \frac{\partial C^{\dag}}{\partial s} C.
  \label{EqDelDDelS}
\end{align}
The diagonal component in the update-mode representation is written as
\begin{align}
  \frac{\partial \Omega_{\bm{k}\lambda'}}{\partial s}
  =
  \Bigl(C^{\dag} \frac{\partial \Omega}{\partial s} C \Bigr)_{\lambda'\lambda'}.
\end{align}
Thus, substituting $s = \Omega_{\bm{k}\lambda_1 \lambda_2}$, we get
\begin{align}
  \frac{\partial \Omega_{\bm{k}\lambda'}}{\partial \Omega_{\bm{k}\lambda_1 \lambda_2}} = C^{*}_{\bm{k}\lambda_1 \lambda'} C_{\bm{k}\lambda_2 \lambda'}.
\end{align}
We also consider the off-diagonal components:
\begin{align}
  \Bigl( C^\dag \frac{\partial \Omega}{\partial s}C \Bigr)_{\lambda'_1 \lambda'_2} 
  &=
  \Bigl( C^\dag \frac{\partial C}{\partial s} \Bigr)_{\lambda'_1 \lambda'_2} \Omega_{\bm{k}\lambda'_2} +  \Omega_{\bm{k}\lambda'_1} \Bigl(\frac{\partial C^\dag}{\partial s} C\Bigr)_{\lambda'_1 \lambda'_2}
  \\&=
  (\Omega_{\bm{k}\lambda'_2} - \Omega_{\bm{k}\lambda'_1})\Bigl( C^\dag \frac{\partial C}{\partial s} \Bigr)_{\lambda'_1 \lambda'_2}.
\end{align}
We used that $( \frac{\partial D}{\partial s} )_{\lambda'_1 \lambda'_2} = 0$ if $\lambda'_1 \neq \lambda'_2$ because $D$ is taken to be diagonal. Substituting $s = \Omega_{\bm{k}\lambda_1 \lambda_2}$, we get
\begin{align}
  \frac{\partial C_{\bm{k}\lambda \lambda'}}{\partial \Omega_{\bm{k}\lambda_1 \lambda_2}} 
  &=
  \sum_{\lambda'_1} C_{\bm{k}\lambda \lambda'_1} \frac{C^*_{\bm{k}\lambda_1 \lambda'_1} C_{\bm{k}\lambda_2 \lambda'}}{\Omega_{\bm{k}\lambda'} -\Omega_{\bm{k}\lambda'_1} }.
  \label{Eq_delCdelOmega1}
\end{align}
In the same way, we get
\begin{align}
  \frac{\partial  C^{*}_{\bm{k}\lambda \lambda'}}{\partial \Omega_{\bm{k}\lambda_1 \lambda_2}}
  &=
  \sum_{\lambda'_1} \frac{C^*_{\bm{k}\lambda_1 \lambda'} C_{\bm{k}\lambda_2 \lambda'_1}}{\Omega_{\bm{k}\lambda'} -\Omega_{\bm{k}\lambda'_1} } C^*_{\bm{k}\lambda \lambda'_1}.
  \label{Eq_delCdelOmega2}
\end{align}
The sums in Eqs. (\ref{Eq_delCdelOmega1}) and (\ref{Eq_delCdelOmega2}) are taken over the modes that are not degenerate to $\Omega_{\bm{k}\lambda'}$ because the contribution from the degenerate modes can be made vanish by using the degrees of freedom in unitary transformation.

Using the formulas above, the gradient of the variational free energy with respect to the SCP dynamical matrix is derived as 
\begin{align}
  \frac{\partial \mathcal{F}}{\partial \Omega_{\bm{k}\lambda_1 \lambda_2}} 
  &=
  \sum_{\lambda'} \Bigl[
    \frac{\hbar}{2} \frac{\partial}{\partial \Omega_{\bm{k}\lambda'}} \Bigl( \frac{n_B + 1/2}{\Omega_{\bm{k}\lambda'}} \Bigr) C^{*}_{\bm{k}\lambda_1 \lambda'} C_{\bm{k}\lambda_2 \lambda'}
    \nonumber
    \\&
    \times\Bigl\{
      - \Omega_{\bm{k}\lambda'}^2 + \widetilde{\Phi}(-\bm{k}\lambda',\bm{k}\lambda') 
      \nonumber
      \\&+ 
      \sum_{n=2}^\infty \frac{1}{(n-1)! N^{n-1}} \sum_{\{\bm{k}\lambda'\}} \Bigl( \frac{\hbar}{2} \Bigr)^{n-1}
      \nonumber\\&\times
      \frac{\widetilde{\Phi}(-\bm{k}\lambda',\bm{k}\lambda', \bm{k}_1\lambda'_1, -\bm{k}_1\lambda'_1,\cdots, -\bm{k}_{n-1}\lambda'_{n-1})}{\Omega_{\bm{k}_1 \lambda'_1} \cdots \Omega_{\bm{k}_{n-1} \lambda'_{n-1}} }
\nonumber
    \\&
       \Bigl( n_B(\hbar\Omega_{\bm{k}_1 \lambda'_1}) + \frac{1}{2} \Bigr) \cdots \Bigl( n_B(\hbar\Omega_{\bm{k}_{n-1} \lambda'_{n-1}}) + \frac{1}{2} \Bigr) 
    \Bigr\}
  \Bigr]
  \label{EqSCPEqDiagonal}
  \\&+
  \sum_{\lambda' \lambda'_1}\Bigl[
    \frac{\hbar}{2} \frac{n_B(\hbar \Omega_{\bm{k}\lambda'}) + 1/2}{\Omega_{\bm{k}\lambda'}} \frac{C^*_{\bm{k}\lambda_1\lambda'} C_{\bm{k}\lambda_2\lambda'_1}}  {\Omega_{\bm{k}\lambda'} - \Omega_{\bm{k}\lambda'_1}}
    \nonumber
    \\&\times
    \Bigl\{
        \sum_{n=1}^\infty \frac{1}{(n-1)! N^{n-1}} \sum_{\{\bm{k}\lambda'\}} \Bigl( \frac{\hbar}{2} \Bigr)^{n-1}
        \nonumber\\&\times
        \frac{\widetilde{\Phi}(-\bm{k}\lambda'_1,\bm{k}\lambda', \bm{k}_2\lambda'_2, -\bm{k}_2\lambda'_2,\cdots, -\bm{k}_{n-1}\lambda'_{n-1})}{\Omega_{\bm{k}_2 \lambda'_2} \cdots \Omega_{\bm{k}_{n} \lambda'_{n}} }
      \nonumber
      \\&\times
          \Bigl( n_B(\hbar\Omega_{\bm{k}_2 \lambda'_2}) + \frac{1}{2} \Bigr) \cdots \Bigl( n_B(\hbar\Omega_{\bm{k}_n \lambda'_n}) + \frac{1}{2} \Bigr) 
    \Bigr\}
    \label{EqSCPEqOffDiagonal1}
    \\&
    +
    \frac{\hbar}{2} \frac{n_B(\hbar \Omega_{\bm{k}\lambda'}) + 1/2}{\Omega_{\bm{k}\lambda'}} \frac{C^*_{\bm{k}\lambda_1\lambda'_1} C_{\bm{k}\lambda_2\lambda'}}  {\Omega_{\bm{k}\lambda'} - \Omega_{\bm{k}\lambda'_1}}
    \nonumber
    \\&\times
    \Bigl\{
        \sum_{n=1}^\infty \frac{1}{(n-1)! N^{n-1}} \sum_{\{\bm{k}\lambda'\}} \Bigl( \frac{\hbar}{2} \Bigr)^{n-1}
        \nonumber\\&\times
        \frac{\widetilde{\Phi}(-\bm{k}\lambda',\bm{k}\lambda'_1, \bm{k}_2\lambda'_2, -\bm{k}_2\lambda'_2,\cdots, -\bm{k}_{n}\lambda'_{n})}{\Omega_{\bm{k}_2 \lambda'_2} \cdots \Omega_{\bm{k}_{n} \lambda'_{n}} }
      \nonumber
      \\&\times
          \Bigl( n_B(\hbar\Omega_{\bm{k}_2 \lambda'_2}) + \frac{1}{2} \Bigr) \cdots \Bigl( n_B(\hbar\Omega_{\bm{k}_n \lambda'_n}) + \frac{1}{2} \Bigr) 
    \Bigr\}
  \Bigr].
  \label{EqSCPEqOffDiagonal2}
\end{align}
The variational condition is that the terms in the curly brackets in Eqs. (\ref{EqSCPEqDiagonal})$\sim$(\ref{EqSCPEqOffDiagonal2}) all vanish. In the original-mode representation, these three conditions can be put together to a single SCP equation
\begin{align}
&
  \Omega_{\bm{k}\lambda_1 \lambda_2}^2 
  \nonumber\\
  &=
  \sum_{n=1}^\infty \frac{1}{(n-1)! N^{n-1}} \sum_{\{\bm{k}\lambda'\}} \Bigl( \frac{\hbar}{2} \Bigr)^{n-1}
  \nonumber\\&
  \frac{\widetilde{\Phi}(-\bm{k}\lambda_1,\bm{k}\lambda_2, \bm{k}_1\lambda'_1, -\bm{k}_1\lambda'_1,\cdots, -\bm{k}_{n-1}\lambda'_{n-1})}{\Omega_{\bm{k}_1 \lambda'_1} \cdots \Omega_{\bm{k}_{n-1} \lambda'_{n-1}} }
  \nonumber
  \\&{\times}
  \Bigl( n_B(\hbar\Omega_{\bm{k}_1 \lambda'_1}) + \frac{1}{2} \Bigr) \cdots \Bigl( n_B(\hbar\Omega_{\bm{k}_{n-1} \lambda'_{n-1}}) + \frac{1}{2} \Bigr) 
  \label{EqSCPEqOriginalMode}
\end{align}

\section{Calculation of the Hessian of the SCP free energy}

The Hessian of the SCP free energy is useful in discussing the stability of the crystal structure at given conditions. In this Appendix, we discuss the formulation to calculate the Hessian, whose implementation is a future task.
We derive the Hessian with respect to the internal coordinates. The Hessian for the strain and the mixed derivatives can be calculated in the same manner.
We start from the general formula for the derivative 
\begin{align}
&
  \frac{\partial^2 \mathcal{F}( 
  \widetilde{\Phi}^{(q^{(0)}, u_{\mu \nu})},
  \Omega_{\bm{k}\lambda_1 \lambda_2}( q^{(0)}_{\lambda}, u_{\mu\nu}))}{\partial q^{(0)}_{\lambda_1} \partial q^{(0)}_{\lambda_2}}
  \nonumber
  \\
  &=
  \Bigl( \frac{\partial}{\partial q^{(0)}_{\lambda_1}} + \sum_{\bm{k}\lambda_3 \lambda_4} \frac{\partial \Omega_{\bm{k}\lambda_3 \lambda_4}}{\partial q^{(0)}_{\lambda_1}} \frac{\partial}{\partial \Omega_{\bm{k}\lambda_3 \lambda_4}}\Bigr)
  \nonumber\\&\times
  \Bigl( \frac{\partial}{\partial q^{(0)}_{\lambda_2}} + \sum_{\bm{k}\lambda_3 \lambda_4} \frac{\partial \Omega_{\bm{k}\lambda_3 \lambda_4}}{\partial q^{(0)}_{\lambda_2}} \frac{\partial}{\partial \Omega_{\bm{k}\lambda_3 \lambda_4}}\Bigr)
  \mathcal{F}(q^{(0)}_{\lambda}, u_{\mu\nu}, \Omega_{\bm{k}\lambda_1 \lambda_2}),
\end{align}
where $\Omega_{\bm{k}\lambda_1 \lambda_2}( q^{(0)}_{\lambda}, u_{\mu\nu})$ is the solution of the SCP equation in the crystal structure given by $q^{(0)}_{\lambda}$, $u_{\mu\nu}$.
Since the solution of the SCP equation satisfies the variational condition, we can simplify the RHS by using
\begin{align}
  \Bigl( \frac{\partial}{\partial q^{(0)}_{\lambda}} + \sum_{\bm{k}\lambda_1 \lambda_2} \frac{\partial \Omega_{\bm{k}\lambda_1 \lambda_2}}{\partial q^{(0)}_{\lambda}} \frac{\partial}{\partial \Omega_{\bm{k}\lambda_1 \lambda_2}}\Bigr)
  \frac{\partial \mathcal{F}}{\partial \Omega_{\bm{k}\lambda_3 \lambda_4}} = 0,
\end{align}
as 
\begin{align}
&
  \frac{\partial^2 \mathcal{F}(  
  \widetilde{\Phi}^{(q^{(0)}, u_{\mu \nu})},
  \Omega( q^{(0)}_{\lambda}, u_{\mu\nu}))}{\partial q^{(0)}_{\lambda_1} \partial q^{(0)}_{\lambda_2}}
  \nonumber\\
  &=
  \frac{\partial^2 \mathcal{F}(
  \widetilde{\Phi}^{(q^{(0)}, u_{\mu \nu})},
  \Omega)}{\partial q^{(0)}_{\lambda_1} \partial q^{(0)}_{\lambda_2}}
  + 
  \sum_{\bm{k}\lambda_3 \lambda_4} \frac{\partial \Omega_{\bm{k}\lambda_3 \lambda_4}}{\partial q^{(0)}_{\lambda_1}} \frac{\partial^2 \mathcal{F}}{\partial q^{(0)}_{\lambda_2} \partial \Omega_{\bm{k}\lambda_3 \lambda_4}}.
  \label{EqF1q0Hessian}
\end{align}
Let us estimate each term in the RHS of Eq. (\ref{EqF1q0Hessian}).
The first term is rewritten by the SCP dynamical matrix
\begin{align}
&
  \frac{\partial^2 \mathcal{F}(
  \widetilde{\Phi}^{(q^{(0)}, u_{\mu \nu})},
  \Omega)}{\partial q^{(0)}_{\lambda_1} \partial q^{(0)}_{\lambda_2}}
 \nonumber\\ &=
  N\times \sum_{n = 0}^\infty \frac{1}{n!N^n}\sum_{\{\bm{k}\lambda'\}} 
  \Bigl( \frac{\hbar}{2} \Bigr)^n
  \nonumber \\&\times
  \frac{\widetilde{\Phi}(\bm{k}_1\lambda'_1,-\bm{k}_1\lambda'_1,\cdots,\bm{k}_n\lambda'_n,-\bm{k}_n\lambda'_n,\bm{0}{\lambda_1},\bm{0}{\lambda_2})}{\Omega_{\bm{k}_1\lambda'_1}\cdots \Omega_{\bm{k}_1\lambda'_1}}
  \nonumber
  \\&\times
  \Bigl( n_B(\hbar \Omega_{\bm{k}_1\lambda'_1}) + \frac{1}{2} \Bigr) \cdots \Bigl( n_B(\hbar \Omega_{\bm{k}_n\lambda'_n}) + \frac{1}{2} \Bigr)
  \nonumber
  \\&=
  N \Omega^2_{\bm{0}\lambda_1 \lambda_2}.
\end{align}
The second term is more complicated. $\frac{\partial^2 \mathcal{F}}{\partial q^{(0)}_{\lambda_2} \partial \Omega_{\bm{k}\lambda_3 \lambda_4}}$ is expanded as
\begin{align}
&
  \frac{\partial^2 \mathcal{F}}{\partial q^{(0)}_{\lambda} \partial \Omega_{\bm{k}\lambda_1 \lambda_2}}
  \nonumber\\&=
  N\times \sum_{n = 1}^\infty \frac{1}{(n-1)! N^n} \sum_{\{\bm{k}\lambda'\}} \Bigl( \frac{\hbar}{2} \Bigr)^n
  \nonumber
  \\&\times
  \frac{\widetilde{\Phi}(\bm{k} \lambda'_1, -\bm{k} \lambda'_1,\bm{k}_2 \lambda'_2, -\bm{k}_2 \lambda'_2, \cdots, \bm{k}_n \lambda'_n, -\bm{k}_n \lambda'_n, \bm{0}\lambda) }{\Omega_{\bm{k}_2\lambda'_2} \cdots \Omega_{\bm{k}_n\lambda'_n}}
  \nonumber
  \\&\times
  \Bigl( n_B(\hbar \Omega_{\bm{k}_2\lambda'_2}) + \frac{1}{2} \Bigr) \cdots \Bigl( n_B(\hbar \Omega_{\bm{k}_n\lambda'_n}) + \frac{1}{2} \Bigr)
  \nonumber
  \\&\times
  \frac{\partial \Omega_{\bm{k}\lambda'_1}}{\partial \Omega_{\bm{k}\lambda_1 \lambda_2}}
  \frac{\partial }{\partial \Omega_{\bm{k} \lambda'_1}} \Bigl( n_B(\hbar \Omega_{\bm{k} \lambda'_1}) + \frac{1}{2} \Bigr)
  \nonumber
  \\&+
  N\times \sum_{n = 1}^\infty \frac{1}{(n-1)! N^n} \sum_{\{\bm{k}\lambda'\}} \Bigl( \frac{\hbar}{2} \Bigr)^n
  \sum_{\lambda_3 \lambda_4}
  \nonumber \\&\times
  \frac{\widetilde{\Phi}(\bm{k} \lambda_3, -\bm{k} \lambda_4,\bm{k}_2 \lambda'_2, -\bm{k}_2 \lambda'_2, \cdots, \bm{k}_n \lambda'_n, -\bm{k}_n \lambda'_n, \bm{0}\lambda) }{\Omega_{\bm{k}_1\lambda'_1} \cdots \Omega_{\bm{k}_n\lambda'_n}}
  \nonumber
  \\&\times
  \Bigl( n_B(\hbar \Omega_{\bm{k}\lambda'_1}) + \frac{1}{2} \Bigr) \cdots \Bigl( n_B(\hbar \Omega_{\bm{k}_n\lambda'_n}) + \frac{1}{2} \Bigr)
  \nonumber
  \\&\times
  \frac{\partial }{\partial \Omega_{\bm{k}\lambda_1 \lambda_2}}
  (C_{\bm{k} \lambda_3 \lambda'_1} C^*_{\bm{k}\lambda_4 \lambda'_1}).
  \label{Eq_del2FdelqdelOmega}
\end{align}
The derivative $\frac{\partial \Omega_{\bm{k}\lambda_3 \lambda_4}}{\partial q^{(0)}_{\lambda_1}}$ cannot be analytically calculated. Thus, we derive a self-consistent equation of $\frac{\partial \Omega_{\bm{k}\lambda_3 \lambda_4}}{\partial q^{(0)}_{\lambda_1}}$ by differentiating the both sides of the SCP equation.
\begin{align}
&
  \frac{\partial \Omega^2_{\bm{k}\lambda_1 \lambda_2}}{\partial q^{(0)}_\lambda}
  \nonumber \\&=
  \sum_{n = 0}^\infty \frac{1}{n! N^n} \sum_{\{\bm{k}\lambda'\}} \Bigl( \frac{\hbar}{2} \Bigr)^n
  \nonumber\\&\times
  \frac{\widetilde{\Phi}(-\bm{k} \lambda_1, -\bm{k} \lambda_2,\bm{k}_1 \lambda'_1, -\bm{k}_1 \lambda'_1, \cdots, \bm{k}_n \lambda'_n, -\bm{k}_n \lambda'_n, \bm{0}\lambda) }{\Omega_{\bm{k}_1\lambda'_1} \cdots \Omega_{\bm{k}_n\lambda'_n}}
  \nonumber
  \\&\times
  \Bigl( n_B(\hbar \Omega_{\bm{k}_1\lambda'_1}) + \frac{1}{2} \Bigr) \cdots \Bigl( n_B(\hbar \Omega_{\bm{k}_n\lambda'_n}) + \frac{1}{2} \Bigr)
  \nonumber
  \\&+
  \sum_{n = 1}^\infty \frac{1}{(n-1)! N^n} \sum_{\{\bm{k}\lambda'\}} \Bigl( \frac{\hbar}{2} \Bigr)^n
  \nonumber\\&\times
  \frac{\widetilde{\Phi}(-\bm{k} \lambda_1, -\bm{k} \lambda_2,\bm{k}_1 \lambda'_1, -\bm{k}_1 \lambda'_1, \cdots, \bm{k}_n \lambda'_n, -\bm{k}_n \lambda'_n, \bm{0}\lambda) }{\Omega_{\bm{k}_2\lambda'_2} \cdots \Omega_{\bm{k}_n\lambda'_n}}
  \nonumber
  \\&\times
  \Bigl( n_B(\hbar \Omega_{\bm{k}_2\lambda'_2}) + \frac{1}{2} \Bigr) \cdots \Bigl( n_B(\hbar \Omega_{\bm{k}_n\lambda'_n}) + \frac{1}{2} \Bigr)
  \nonumber
  \\&\times
  \sum_{\lambda_3 \lambda_4} 
  \frac{\partial \Omega_{\bm{k}_1\lambda_3 \lambda_4}}{\partial q^{(0)}_\lambda}
  \frac{\partial \Omega_{\bm{k}_1\lambda'_1}}{\partial \Omega_{\bm{k}_1 \lambda_3 \lambda_4}}
  \frac{\partial}{\partial \Omega_{\bm{k}_1\lambda'_1} } \Bigl( \frac{n_B(\Omega_{\bm{k}_1\lambda'_1})+ 1/2}{\Omega_{\bm{k}_1\lambda'_1}} \Bigr)
  \nonumber
  \\&+
  \sum_{n = 1}^\infty \frac{1}{(n-1)! N^n} \sum_{\{\bm{k}\lambda'\}} \Bigl( \frac{\hbar}{2} \Bigr)^n
  \sum_{\lambda_3 \lambda_4}
  \nonumber\\&\times
  \frac{\widetilde{\Phi}(-\bm{k} \lambda_1, -\bm{k} \lambda_2,\bm{k}_1 \lambda_3, -\bm{k}_1 \lambda_4, \cdots, \bm{k}_n \lambda'_n, -\bm{k}_n \lambda'_n, \bm{0}\lambda) }{\Omega_{\bm{k}_1\lambda'_1} \cdots \Omega_{\bm{k}_n\lambda'_n}}
  \nonumber
  \\&\times
  \Bigl( n_B(\hbar \Omega_{\bm{k}_1\lambda'_1}) + \frac{1}{2} \Bigr) \cdots \Bigl( n_B(\hbar \Omega_{\bm{k}_n\lambda'_n}) + \frac{1}{2} \Bigr)
  \nonumber
  \\&\times
  \sum_{\lambda_5 \lambda_6} 
  \frac{\partial \Omega_{\bm{k}_1\lambda_5 \lambda_6}}{\partial q^{(0)}_\lambda}
  \frac{\partial }{\partial \Omega_{\bm{k}_1 \lambda_5 \lambda_6}}
  (C_{\bm{k}_1 \lambda_3 \lambda'_1} C^*_{\bm{k}_1 \lambda_4 \lambda'_1})
  \label{EqDelOmegaDelq0}
\end{align} 
We get $\frac{\partial \Omega^2_{\bm{k}\lambda_1 \lambda_2}}{\partial q^{(0)}_\lambda}$ or $\frac{\partial \Omega_{\bm{k}\lambda_1 \lambda_2}}{\partial q^{(0)}_\lambda}$ by solving Eq. (\ref{EqDelOmegaDelq0}) self-consistently. Substituting the solution to Eq. (\ref{Eq_del2FdelqdelOmega}) and then to Eq. (\ref{EqF1q0Hessian}), it is possible to calculate the Hessian of the SCP free energy and discuss the stability of crystal structures at finite temperature.

\section{Implementation details of the IFC renormalization.}
\label{Sec_Appendix_Implementation}
In this Appendix, we explain the details of the implementation of the IFC renormalization to the ALAMODE package. We truncate the Taylor expansion of the potential energy surface at the fourth-order, which is the approximation in the ALAMODE implementation of the SCP calculation. 

The IFC renormalization by strain is written down in real-space representation as shown in Eq.~(\ref{EqRenormalizePhiByStrain}). However, the Fourier transformation of the anharmonic IFCs is so computationally costly that it is not feasible to Fourier-transform the anharmonic IFCs at every step of the structural optimization. Therefore, the derivatives of the IFCs with respect to the strain and their Fourier transformation are calculated in the pre-calculation before the structural optimization process. For the first-order IFCs, the following quantities are calculated.
\begin{widetext}
\begin{align}
    \frac{\partial \widetilde{\Phi}(\bm{0}\lambda_1)}{\partial u_{\mu\nu}} \Bigr|_{q^{(0)}=0, u_{\mu\nu} = 0} = 
    \sum_{\alpha_1 \mu_1}
    \frac{\epsilon_{\bm{0}\lambda_1,\alpha_1\mu_1}}{\sqrt{M_{\alpha_1}}}
    \sum_{\bm{R}'_1 \alpha'_1}
    \Phi^{(0)}_{\mu_1 \mu}(\bm{0}\alpha_1, \bm{R}'_1\alpha'_1)
    R'_{1 \alpha'_1 \nu}
\end{align}
\begin{align}
    \frac{\partial^2 \widetilde{\Phi}(\bm{0}\lambda_1)}{\partial u_{\mu_1\nu_1} \partial u_{\mu_2 \nu_2}} \Bigr|_{q^{(0)}=0, u_{\mu\nu} = 0} = 
    \sum_{\alpha_1 \mu'_1}
    \frac{\epsilon_{\bm{0}\lambda_1,\alpha_1\mu'_1}}{\sqrt{M_{\alpha_1}}}
    \sum_{\{\bm{R}' \alpha'\}}
    \Phi^{(0)}_{\mu'_1 \mu_1 \mu_2}(\bm{0}\alpha_1, \bm{R}'_1\alpha'_1, \bm{R}'_2 \alpha'_2)
    R'_{1 \alpha'_1 \nu_1} R'_{2 \alpha'_2 \nu_2}
\end{align}
\begin{align}
    \frac{\partial^3 \widetilde{\Phi}(\bm{0}\lambda_1)}{\partial u_{\mu_1\nu_1} \partial u_{\mu_2 \nu_2} \partial u_{\mu_3 \nu_3}} \Bigr|_{q^{(0)}=0, u_{\mu\nu} = 0} = 
    \sum_{\alpha_1 \mu'_1}
    \frac{\epsilon_{\bm{0}\lambda_1,\alpha_1\mu'_1}}{\sqrt{M_{\alpha_1}}}
    \sum_{\{\bm{R}' \alpha'\}}
    \Phi^{(0)}_{\mu'_1 \mu_1 \mu_2 \mu_3}(\bm{0}\alpha_1, \bm{R}'_1\alpha'_1, \bm{R}'_2 \alpha'_2, \bm{R}'_3 \alpha'_3)
    R'_{1 \alpha'_1 \nu_1} R'_{2 \alpha'_2 \nu_2} R'_{3 \alpha'_3 \nu_3}
\end{align}
\end{widetext}
${\Phi}^{(0)} = \Phi^{(q^{(0)}=0, u_{\mu\nu}=0)}$ is the IFC in the reference structure without atomic displacements nor cell deformation. For the harmonic and cubic IFCs, we calculate
\begin{widetext}
\begin{align}
    \frac{\partial \widetilde{\Phi}(\bm{k}_1 \lambda_1, -\bm{k}_1 \lambda_2)}{\partial u_{\mu_1 \nu_1}}\Bigr|_{q^{(0)}=0, u_{\mu\nu} = 0}
    &=\sum_{\{\alpha \mu'\}} \frac{\epsilon_{\bm{k}_1 \lambda_1,\alpha_1 \mu'_1}}{\sqrt{M_{\alpha_1}}} \frac{\epsilon_{-\bm{k}_1 \lambda_2,\alpha_2 \mu'_2}}{\sqrt{M_{\alpha_2}}}
  \sum_{\bm{R}_1} 
  e^{i\bm{k}_1\cdot \bm{R}_1}
  \nonumber
  \\&\times
  \sum_{\bm{R}'_1\alpha'_1}
  \Phi^{(0)}_{\mu'_1 \mu'_2 \mu_1}(\bm{R}_1\alpha_1, \bm{0}\alpha_2,\bm{R}'_1\alpha'_1)
    R'_{1 \alpha'_1 \nu_1}
\end{align}
\begin{align}
    \frac{\partial^2 \widetilde{\Phi}(\bm{k}_1 \lambda_1, -\bm{k}_1 \lambda_2)}{\partial u_{\mu_1 \nu_1} \partial u_{\mu_2 \nu_2}}\Bigr|_{q^{(0)}=0, u_{\mu\nu} = 0}
    &=\sum_{\{\alpha \mu'\}} \frac{\epsilon_{\bm{k}_1 \lambda_1,\alpha_1 \mu'_1}}{\sqrt{M_{\alpha_1}}} \frac{\epsilon_{-\bm{k}_1 \lambda_2,\alpha_2 \mu'_2}}{\sqrt{M_{\alpha_2}}}
  \sum_{\bm{R}_1} 
  e^{i\bm{k}_1\cdot \bm{R}_1}
  \nonumber
  \\&\times
  \sum_{\{\bm{R}'\alpha'\}}
  \Phi^{(0)}_{\mu'_1 \mu'_2 \mu_1 \mu_2}(\bm{R}_1\alpha_1, \bm{0}\alpha_2, \bm{R}'_1\alpha'_1, \bm{R}'_2\alpha'_2)
    R'_{1 \alpha'_1 \nu_1} R'_{2 \alpha'_2 \nu_2}
\end{align}
\begin{align}
    \frac{\partial \widetilde{\Phi}(\bm{0}\lambda_1, \bm{k}_1 \lambda_2, -\bm{k}_1 \lambda_3)}{\partial u_{\mu_1 \nu_1}}\Bigr|_{q^{(0)}=0, u_{\mu\nu} = 0}
    &=\sum_{\{\alpha \mu'\}} 
    \frac{\epsilon_{\bm{0} \lambda_1,\alpha_1 \mu'_1}}{\sqrt{M_{\alpha_1}}} 
    \frac{\epsilon_{\bm{k}_1 \lambda_2,\alpha_2 \mu'_2}}{\sqrt{M_{\alpha_2}}} 
    \frac{\epsilon_{-\bm{k}_1 \lambda_3,\alpha_3 \mu'_3}}{\sqrt{M_{\alpha_3}}}
  \sum_{\bm{R}_1 \bm{R}_2} 
  e^{i\bm{k}_1\cdot \bm{R}_2}
  \nonumber
  \\&\times
  \sum_{\bm{R}'_1\alpha'_1}
  \Phi^{(0)}_{\mu'_1 \mu'_2 \mu'_3 \mu_1}(\bm{R}_1\alpha_1, \bm{R}_2\alpha_2, \bm{0}\alpha_3, \bm{R}'_1\alpha'_1)
    R'_{1 \alpha'_1 \nu_1}
    \label{Eq_delPhi3_delu}
\end{align}
\end{widetext}
For the cubic IFCs, it is enough to calculate the derivatives of the IFCs of the form $\widetilde{\Phi}(\bm{0}\lambda_1, \bm{k}\lambda_2, -\bm{k}\lambda_3)$. The quartic IFCs are not changed because the higher-order IFCs are truncated in the Taylor expansion.
It should be noted that the fractional coordinates of the $k$-points in the Brillouin zone are kept fixed in these derivatives. This makes the $k$-mesh adapt to the deformed Brillouin zone, which is convenient for calculating the free energy per unit cell. In addition, this convention simplifies the formulas of the IFC renormalization because it keeps the $\bm{k}\cdot \bm{R}$ terms invariant when the lattice vectors are changed.

Special care needs to be taken for the zeroth-order IFC (constant term of the potential energy surface) because the surface effect must be properly considered to derive similar formulas~\cite{wallace1972thermodynamics}, which is quite complicated. Therefore, we directly calculate the second-order and third-order elastic constants by fitting the strain-energy relation. Although the elastic constants are usually defined as quantities per unit volume, we define them as the quantities per unit cell as
\begin{align}
    C_{\mu_1 \nu_1, \mu_2 \nu_2} = \frac{1}{N} \frac{\partial^2 U_0}{\partial \eta_{\mu_1 \nu_1} \partial \eta_{\mu_2 \nu_2}},
\end{align}
\begin{align}
    C_{\mu_1 \nu_1, \mu_2 \nu_2, \mu_3 \nu_3} = \frac{1}{N} \frac{\partial^2 U_0}{\partial \eta_{\mu_1 \nu_1} \partial \eta_{\mu_2 \nu_2} \partial \eta_{\mu_3 \nu_3}}, 
\end{align}
for later convenience. The strain tensor $\eta_{\mu\nu}$ is defined as
\begin{align}
    \eta_{\mu\nu} 
    &= 
    \frac{1}{2}\Bigl(\sum_{\mu'} (\delta_{\mu \mu'}+u_{\mu \mu'}) (\delta_{\nu\mu'}+u_{\nu \mu'}) - \delta_{\mu \nu}\Bigr)
    \\&=
    \frac{1}{2} 
    \Bigl(u_{\mu\nu} + u_{\nu \mu} + \sum_{\mu'} u_{\mu \mu'}u_{\nu \mu'}
    \Bigr).
\end{align}
There is one-to-one correspondence between $u_{\mu\nu}$ and $\eta_{\mu\nu}$ as long as we restict $u_{\mu\nu}$ to be symmetric.
The number of strain modes to calculate can be decreased using the crystal symmetry~\cite{PhysRevB.75.094105, LIAO2021107777}. Note that although the internal coordinates are relaxed in calculations to reproduce experimentally observed elastic constants~\cite{PhysRevB.75.094105, LIAO2021107777}, we run the DFT calculations with fixed internal coordinates because the internal coordinates are treated separately as independent degrees of freedom.

Therefore, the IFCs in the reciprocal space representation for the strained cell can be calculated by 
\begin{widetext}

\begin{align}
    \frac{1}{N}U_0^{(q^{(0)}=0, u_{\mu\nu})} = \frac{1}{2}\sum_{\mu_1 \nu_1, \mu_2 \nu_2} C_{\mu_1 \nu_1, \mu_2 \nu_2} \eta_{\mu_1 \nu_1} \eta_{\mu_2 \nu_2}
    +
    \frac{1}{6}\sum_{\mu_1 \nu_1, \mu_2 \nu_2, \mu_3 \nu_3} C_{\mu_1 \nu_1, \mu_2 \nu_2, \mu_3 \nu_3} \eta_{\mu_1 \nu_1} \eta_{\mu_2 \nu_2} \eta_{\mu_3 \nu_3},
    \label{EqUpdateU0withStrain}
\end{align}
\begin{align}
    \widetilde{\Phi}^{(q^{(0)}=0, u_{\mu\nu})}(\bm{0}\lambda) 
    &= 
    \widetilde{\Phi}^{( q^{(0)}=0, u_{\mu\nu}=0)}(\bm{0}\lambda)
    +
    \sum_{\mu_1 \nu_1} \frac{\partial \widetilde{\Phi}(\bm{0}\lambda_1)}{\partial u_{\mu_1\nu_1}} u_{\mu_1 \nu_1}
    \nonumber
    \\&
    + \frac{1}{2}\sum_{\{\mu_1 \nu\}} \frac{\partial^2 \widetilde{\Phi}(\bm{0}\lambda_1)}{\partial u_{\mu_1\nu_1} \partial u_{\mu_2\nu_2}} u_{\mu_1 \nu_1} u_{\mu_2 \nu_2}
    + \frac{1}{6}\sum_{\{\mu_1 \nu\}} \frac{\partial^3 \widetilde{\Phi}(\bm{0}\lambda_1)}{\partial u_{\mu_1\nu_1} \partial u_{\mu_2\nu_2} \partial u_{\mu_3\nu_3}} u_{\mu_1 \nu_1} u_{\mu_2 \nu_2} u_{\mu_3 \nu_3},
\end{align}
\begin{align}
\widetilde{\Phi}^{(q^{(0)}=0, u_{\mu\nu})}(\bm{k}_1 \lambda_1, -\bm{k}_1 \lambda_2)
&=
\widetilde{\Phi}^{( q^{(0)}=0, u_{\mu\nu}=0)}(\bm{k}_1 \lambda_1, -\bm{k}_1 \lambda_2)
+ \sum_{\mu_1 \nu_1}
    \frac{\partial \widetilde{\Phi}(\bm{k}_1 \lambda_1, -\bm{k}_1 \lambda_2)}{\partial u_{\mu_1 \nu_1}} u_{\mu_1 \nu_1}
    \nonumber\\&
    + \frac{1}{2}
    \sum_{\{\mu \nu\}}
    \frac{\partial \widetilde{\Phi}(\bm{k}_1 \lambda_1, -\bm{k}_1 \lambda_2)}{\partial u_{\mu_1 \nu_1}\partial u_{\mu_2 \nu_2}} u_{\mu_1 \nu_1} u_{\mu_2 \nu_2},
    \label{EqUpdatePhi2withStrain}
\end{align}
\begin{align}
    \widetilde{\Phi}^{(q^{(0)}=0, u_{\mu\nu})}(\bm{0}\lambda_1, \bm{k}_1 \lambda_2, -\bm{k}_1 \lambda_3)
    =
    \widetilde{\Phi}^{( q^{(0)}=0, u_{\mu\nu}=0)}(\bm{0}\lambda_1, \bm{k}_1 \lambda_2, -\bm{k}_1 \lambda_3)
    +
    \sum_{\mu_1 \nu_1}
    \frac{\partial \widetilde{\Phi}(\bm{0}\lambda_1, \bm{k}_1 \lambda_2, -\bm{k}_1 \lambda_3)}{\partial u_{\mu_1 \nu_1}} u_{\mu_1 \nu_1},
\end{align}
\begin{align}
    \widetilde{\Phi}^{(q^{(0)}=0, u_{\mu\nu})}(\bm{k}_1\lambda_1, -\bm{k}_1 \lambda_2, \bm{k}_2 \lambda_3, -\bm{k}_2 \lambda_4)
    =
    \widetilde{\Phi}^{( q^{(0)}=0, u_{\mu\nu}=0)}(\bm{k}_1\lambda_1, -\bm{k}_1 \lambda_2, \bm{k}_2 \lambda_3, -\bm{k}_2 \lambda_4).
    \label{EqUpdatePhi4withStrain}
\end{align}
\end{widetext}

The first-order term is not included in Eq. (\ref{EqUpdateU0withStrain}) because the stress tensor is assumed to vanish in the reference structure, which assumption can be relaxed straightforwardly.
We also set $U_0^{(q^{(0)}=0,u_{\mu\nu}=0)}=0$ without loss of generality.
Substituting the $\widetilde{\Phi}^{(q^{(0)}=0, u_{\mu\nu})}$ obtained in the above formulas to the RHS of Eq. (\ref{EqRenormalizePhiByDisplace}), we get the IFCs in arbitrary structures $\widetilde{\Phi}^{(q^{(0)}, u_{\mu\nu})}$.

Similarly, the derivatives of the even-order IFCs with respect to the structural degrees of freedom are calculated as
\begin{widetext}
\begin{align}
    \frac{1}{N}\frac{\partial U_0^{(q^{(0)}, u_{\mu\nu})}}{\partial u_{\mu\nu}}
    &=
    \sum_{\mu'\nu'}
    \frac{\partial \eta_{\mu'\nu'}}{\partial u_{\mu\nu}}
    \Bigl(
    \sum_{\mu_1 \nu_1} C_{\mu_1 \nu_1, \mu' \nu'} \eta_{\mu_1 \nu_1} 
    +
    \frac{1}{3}\sum_{\mu_1 \nu_1, \mu_2 \nu_2} C_{\mu_1 \nu_1, \mu_2 \nu_2, \mu' \nu'} \eta_{\mu_1 \nu_1} \eta_{\mu_2 \nu_2}
    \Bigr)
    \nonumber
    \\&
    +
    \sum_{m=1}^{3} \frac{1}{m!} 
    \sum_{\{\lambda\}} \frac{\partial \widetilde{\Phi}^{(q^{(0)}=0, u_{\mu\nu})}(\bm{0}\lambda_1, \cdots, \bm{0}\lambda_m)}{\partial u_{\mu\nu}} q^{(0)}_{\lambda_1} \cdots  q^{(0)}_{\lambda_m},
\end{align}
\begin{align}
    \frac{\partial \widetilde{\Phi}^{(q^{(0)}, u_{\mu\nu})}(\bm{k}_1\lambda_1, -\bm{k}_1 \lambda_2)}{\partial u_{\mu\nu}}
    &=
    \frac{\partial \widetilde{\Phi}(\bm{k}_1\lambda_1, -\bm{k}_1 \lambda_2)}{\partial u_{\mu\nu}}
    +
    \sum_{\mu' \nu'} \frac{\partial^2 \widetilde{\Phi}(\bm{k}_1\lambda_1, -\bm{k}_1 \lambda_2)}{\partial u_{\mu\nu} \partial u_{\mu'\nu'}} u_{\mu'\nu'}
    \nonumber
    \\&+
    \sum_{\rho_1} \frac{\partial \widetilde{\Phi}(\bm{k}_1\lambda_1, -\bm{k}_1 \lambda_2, \bm{0}\rho_1)}{\partial u_{\mu\nu}} q^{(0)}_{\rho_1},
    \label{EqDelPhiDelUWithStrainDisplace}
\end{align}
\end{widetext}
which are used in the calculation of the gradient of the SCP free energy.

\section{Sum rules, symmetries, and mirror image conventions in IFC calculations}
\label{Appendix_sumrule_symmetry_IFCs}
The IFCs need to satisfy several constraints to obtain physically reasonable results in phonon calculations. The permutation symmetry
\begin{align}
&
    \Phi_{\cdots \mu_i \cdots \mu_j \cdots}(\cdots,\bm{R}_i \alpha_i, \cdots, \bm{R}_j \alpha_j, \cdots)
    \nonumber
    \\&
    = 
    \Phi_{\cdots \mu_j \cdots \mu_i \cdots}(\cdots,\bm{R}_j \alpha_j, \cdots, \bm{R}_i \alpha_i, \cdots),
\end{align}
the acoustic sum rule (ASR) 
\begin{align}
    \sum_{\bm{R}_n \alpha_n} \Phi_{\mu_1 \cdots \mu_{n-1} \mu_n}(\bm{R}_1\alpha_1, \cdots, \bm{R}_{n-1}\alpha_{n-1}, \bm{R}_n \alpha_n ) = 0,
\end{align}
and the space group symmetry are considered as they are especially important. The permutation symmetry is essential for the Hermiticity of the dynamical matrix. The acoustic sum rule must be satisfied to make the potential energy invariant under rigid translation of the whole system. 

In first-principles phonon calculations, the size of the supercell is usually taken to be comparable to the interaction range because using larger supercells drastically increases the computational cost. Thus, the IFCs in the finite supercell and those in the infinite real space needs to be carefully distinguished in implementing a phonon calculation code. This distinction also plays an important role in considering the constraints on IFCs because they should be imposed in infinite real space, not in the finite supercell model, which have been frequently overlooked.  In this research, we implement a program to the ALAMODE package~\cite{Tadano_2014, PhysRevB.92.054301} to obtain IFCs that satisfy the permutation symmetry, ASR, and the space group symmetry in the infinite real space.

\begin{figure}[h]
\vspace{0cm}
\begin{center}
\includegraphics[width=0.48\textwidth]{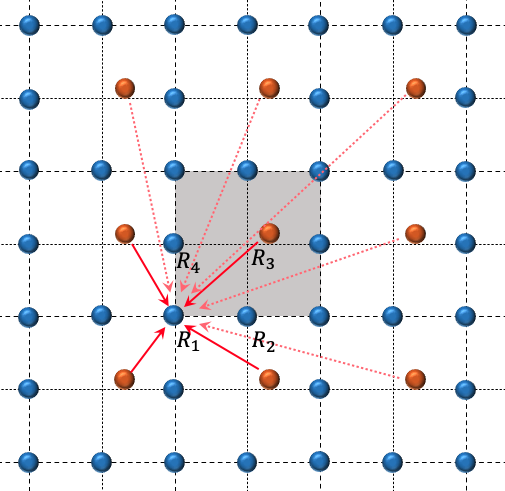}
\caption{
Schematic figure of a two dimensional monatomic $2\times2$ supercell. The shaded part is the supercell considered in the calculation, which includes four primitive cells denoted $\bm{R}_1, \cdots, \bm{R}_4$. We consider that an atom in the supercell, which is colored orange, is displaced, which corresponds to displacing all the atoms in the real space which are colored orange. 
}
\label{Fig_SupercellSchematic2}
\end{center}
\end{figure}

We begin with the definition and the relation of the IFCs in the supercell and those in the infinite real space.
Consider that an atom in a two-dimensional $2\times2$ monatomic supercell (shaded area) is displaced, as shown in Fig.~\ref{Fig_SupercellSchematic2}. In infinite real space, this situation corresponds to displacing all the mirror images of the atom, which are colored orange.
Here, we temporarily assume that the potential is harmonic to discuss the relation between IFCs in the supercell and those in the infinite real space. The force that acts on the atom at $\bm{R}_1$ in Fig.~\ref{Fig_SupercellSchematic2} (lower-left corner in the supercell) is
\begin{align}
    f_{\bm{0}\alpha\mu} 
    &= 
    - \sum_{\nu} \Phi^{\text{supercell}}_{\mu \nu}(\bm{0}\alpha, \bm{R}_3 \alpha) u_{\bm{R}_3\alpha\nu}
    \nonumber
    \\&=
    - \sum_J \sum_{\nu} \Phi_{\mu \nu}(\bm{0}\alpha, \bm{R}_3(J) \alpha) u_{\bm{R}_3\alpha\nu},
\end{align}
where $\Phi^{\text{supercell}}$ are the IFCs in the supercell, and $\Phi$ are the IFCs in the infinite real space. In this appendix, we specify atoms in the infinite space by the combination $(\bm{R}(J), \alpha)$. $\bm{R}$ is the position of the primitive cell in the considering supercell. 
The infinite space consists of periodically aligning supercells. $J$ specifies the supercell which includes the considering atom.
$\alpha$ is the atom number in the primitive cell. $J$ for the first atom in IFCs is abbreviated because it can be fixed to the original supercell due to the translational invariance. From the above discussion, the relation between the harmonic IFCs in the real space and those in the supercell is 
\begin{align}
    \Phi^{\text{supercell}}_{\mu_1 \mu_2}(\bm{0}\alpha_1, \bm{R}\alpha_2) = \sum_J \Phi_{\mu_1 \mu_2}(\bm{0}\alpha_1, \bm{R}(J)\alpha_2).
    \label{EqAppendixSCIFCRSIFCharm}
\end{align}
The similar formula can be derived for the anharmonic IFCs.
\begin{align}
&
    \Phi^{\text{supercell}}_{\mu_1 \cdots \mu_n}(\bm{0}\alpha_1, \bm{R}_2\alpha_2 \cdots \bm{R}_n \alpha_n) 
    \nonumber
    \\&= \sum_{J_2 \cdots J_n} \Phi_{\mu_1 \mu_2\cdots \mu_n}(\bm{0}\alpha_1, \bm{R}_2(J_2)\alpha_2, \cdots, \bm{R}_n(J_n)\alpha_n).
    \label{EqAppendixSCIFCRSIFC}
\end{align}
These formulas show that the IFCs calculated in the supercell need to be distributed to the mirror images in infinite space. 
It is crucial to choose a proper convention when distributing IFCs because the reciprocal-space IFCs at incommensurate $k$-points are dependent on how to treat the mirror images. For the harmonic IFCs, equal values are usually assigned to the nearest mirror images, but the definition of nearest mirror images is not straightforward for anharmonic IFCs.

We then consider the physical constraints on the IFCs in the supercell and in the real space.
We denote the distribution of IFCs to the mirror images as
\begin{align}
&
    \Phi_{\mu_1 \cdots \mu_n}(\bm{0}\alpha_1 , \cdots \bm{R}_n(J_n) \alpha_n)
    \nonumber
    \\&
    =
    c_{\bm{R}_2 \cdots \bm{R}_n, \alpha_1 \cdots \alpha_n, \mu_1 \cdots \mu_n}^{J_2 \cdots J_n} 
    \Phi^{\text{supercell}}_{\mu_1 \cdots \mu_n}(\bm{0}\alpha_1 , \cdots \bm{R}_n \alpha_n),
    \label{EqAppendixccoefDef}
\end{align}
where
\begin{align}
    \sum_{\{J\}}
    c_{\bm{R}_2 \cdots \bm{R}_n, \alpha_1 \cdots \alpha_n, \mu_1 \cdots \mu_n}^{J_2 \cdots J_n} =1.
\end{align}
We start with the discussion on the permutation symmetry. The permutation symmetry of the IFCs in the supercell is
\begin{align}
    \Phi^{\text{supercell}}_{\mu_1 \cdots \mu_n}(\bm{0}\alpha_1 , \cdots \bm{R}_n \alpha_n)
    =
    \Phi^{\text{supercell}}_{\mu'_1 \cdots \mu'_n}(\bm{0}\alpha'_1 , \cdots \bm{R}'_n \alpha'_n)
    \label{EqPermSymmSupercell}
\end{align}
where $(\bm{0}\alpha'_1 \mu'_1, \cdots \bm{R}'_n \alpha'_n \mu'_n)$ is a combination of the atoms and the $xyz$ components such that $(\bm{0}\alpha_1 \mu_1 , \cdots \bm{R}_n \alpha_n \mu_n)$ is permuted and translated to bring the first atom to the original primitive cell $\bm{0}$.
On the other hand, the permutation symmetry of the IFCs in the real space is 
\begin{align}
    \Phi_{\mu_1 \cdots \mu_n}(\bm{0}\alpha_1 , \cdots \bm{R}_n(J_n) \alpha_n)
    =
    \Phi_{\mu'_1 \cdots \mu'_n}(\bm{0}\alpha'_1 , \cdots \bm{R}'_n(J'_n) \alpha'_n),
    \label{EqPermSymmRealSpace}
\end{align}
where $(\bm{0}\alpha'_1 \mu'_1, \cdots \bm{R}'_n (J'_n)\alpha'_n \mu'_n)$ is a combination of the atoms and $xyz$ components such that $(\bm{0}\alpha_1 \mu_1 , \cdots \bm{R}_n(J_n) \alpha_n \mu_n)$ is permuted and translated to bring the first atom to the original primitive cell in the original supercell $\bm{R}(J) = \bm{0}(0)$.
Thus, in addition to Eq. (\ref{EqPermSymmRealSpace}), the $c$ coefficients need to satisfy 
\begin{align}
    c_{\bm{R}_2 \cdots \bm{R}_n, \alpha_1 \cdots \alpha_n, \mu_1 \cdots \mu_n}^{J_2 \cdots J_n} 
    =
    c_{\bm{R}'_2 \cdots \bm{R}'_n, \alpha'_1 \cdots \alpha'_n, \mu'_1 \cdots \mu'_n}^{J'_2 \cdots J'_n} 
    \label{Eq_coef_permutation_symmetry}
\end{align}
for the IFCs to satisfy permutation symmetry in the infinite space.

The acoustic sum rule in the supercell is 
\begin{widetext}
\begin{align}
&
    \sum_{\bm{R}_n \alpha_n} \Phi^{\text{supercell}}_{\mu_1 \cdots \mu_{n-1} \mu_n}(\bm{0}\alpha_1, \cdots, \bm{R}_{n-1}\alpha_{n-1}, \bm{R}_n \alpha_n )
        \nonumber
    \\&= 
    \sum_{\bm{R}_n \alpha_n  J_n}
    \Bigl(
    \sum_{J_2 \cdots J_{n-1}}
    \Phi_{\mu_1 \cdots \mu_{n-1} \mu_n}(\bm{0}\alpha_1, \cdots, \bm{R}_{n-1}(J_{n-1})\alpha_{n-1}, \bm{R}_n(J_n) \alpha_n )
    \Bigr)
    \nonumber
    \\&= 0,
    \label{EqASRSupercell}
\end{align}
\end{widetext}
for all $\mu_1, \cdots, \mu_n, \alpha_1, \cdots, \alpha_{n-1}, \bm{R}_2, \cdots, \bm{R}_{n-1}$,
which is a weaker condition than ASR in the real space.
\begin{align}
&
    \sum_{\bm{R}_n J_n \alpha_n} \Phi_{\mu_1 \cdots \mu_{n-1} \mu_n}(\bm{0}\alpha_1, \cdots, \bm{R}_{n-1}(J_{n-1})\alpha_{n-1}, \bm{R}_n (J_n)\alpha_n )
    \nonumber
    \\&=
    \sum_{\bm{R}_n J_n \alpha_n}
    c_{\bm{R}_2 \cdots \bm{R}_n, \alpha_1 \cdots \alpha_n, \mu_1 \cdots \mu_n}^{J_2 \cdots J_n} 
    \Phi^{\text{supercell}}_{\mu_1 \cdots \mu_n}(\bm{0}\alpha_1 , \cdots \bm{R}_n \alpha_n)
    \nonumber\\&
    = 0,
    \label{EqASRRealSpace}
\end{align}
for all $\mu_1, \cdots, \mu_n, \alpha_1, \cdots, \alpha_{n-1}, \bm{R}_2(J_2), \cdots, \bm{R}_{n-1}(J_{n-1})$.
In general, the coefficients $c_{\bm{R}_2 \cdots \bm{R}_n, \alpha_1 \cdots \alpha_n, \mu_1 \cdots \mu_n}^{J_2 \cdots J_n}$ need to be determined before calculating the IFCs in the supercell to impose the constraint of Eq. (\ref{EqASRRealSpace}).
For the IFCs to satisfy the space group symmetry in the real space, the coefficients $c_{\bm{R}_2 \cdots \bm{R}_n, \alpha_1 \cdots \alpha_n, \mu_1 \cdots \mu_n}^{J_2 \cdots J_n} $ needs to be compatible with the symmetry operations, like the above discussion on the permutation symmetry.

Two conventions had been implemented in the ALAMODE package to distribute IFCs to the mirror images, which we discuss in relation with constraints on IFCs in the infinite real space. Both conventions distribute IFCs to the mirror images after calculating the IFCs in the supercell, which made it impossible to satisfy the permutation symmetry and the acoustic sum rule at the same time. 
In one convention, the IFCs are distributed to the mirror images that the sum of the distance between all pairs
\begin{align}
    \sum_{i\neq j} d_{\bm{R}_i(J_i)\alpha_i, \bm{R}_j(J_j)\alpha_j}
\end{align}
is minimum. $d_{\bm{R}_i(J_i)\alpha_i, \bm{R}_j(J_j)\alpha_j}$ is the distance between the atoms $\bm{R}_i(J_i)\alpha_i$ and $\bm{R}_j(J_j)\alpha_j$. This way of determining $c_{\bm{R}_2 \cdots \bm{R}_n, \alpha_1 \cdots \alpha_n, \mu_1 \cdots \mu_n}^{J_2 \cdots J_n}$ is compatible with the permutation symmetry and the space group symmetry. The IFCs made in this convention satisfies the permutation symmetry and the space group symmetry in infinite space. However, ASR is broken in the infinite real space because Eq. (\ref{EqASRRealSpace}) is not imposed.
In the other convention, the IFCs are distributed to the mirror images in which the distances between the first atom and each of the other atoms ($d_{\bm{0}(0)\alpha_1, \bm{R}_i(J_i) \alpha_i}$ for $i \neq 1$) are the minimum. The distance between atoms but the first atom ($d_{\bm{R}_i(J_i) \alpha_i, \bm{R}_j(J_j) \alpha_j}$ for $i,j \neq 1$) are not considered. This convention breaks the permutation symmetry in the infinite space because the first atom is treated differently. However, it satisfies ASR in the real space because the choice of $J_2, \cdots, J_n$ are uncorrelated so that 
\begin{align}
    \sum_{\bm{R}_n J_n \alpha_n} \Phi_{\mu_1 \cdots \mu_{n-1} \mu_n}(\bm{0}\alpha_1, \cdots, \bm{R}_{n-1}(J_{n-1})\alpha_{n-1}, \bm{R}_n (J_n)\alpha_n )
\end{align} is indepdent of $J_2, \cdots, J_{n-1}$.

In this research, we implement a program to the ALAMODE package to calculate IFCs which satisfy the permutation symmetry, ASR, and the space group symmetry in the infinite space. We first calculate $c_{\bm{R}_2 \cdots \bm{R}_n, \alpha_1 \cdots \alpha_n, \mu_1 \cdots \mu_n}^{J_2 \cdots J_n}$ to distribute equal values to the mirror images that the sum of the distance between all pairs
is minimum.  Then the IFCs in the supercell is calculated to satisfy the permutation symmetry and the space group symmetry in the supercell and the ASR in the real space (Eq. (\ref{EqASRRealSpace}) ).

 \section{Test of the IFC renormalization by the strain}
\label{Sec_Appendix_test_IFCremorm}

The IFC renormalization associated with the change of internal coordinates does not affect the fitting accuracy of the potential energy surface. This is because it does not alter the functional form of the potential energy surface. However, the IFC renormalization by the strain estimates the set of IFCs in deformed unit cells from the IFCs in the reference structure. Therefore, thorough tests are necessary to confirm the accuracy of the method. 

\begin{figure}[h]
\vspace{0cm}
\begin{center}
\includegraphics[width=0.48\textwidth]{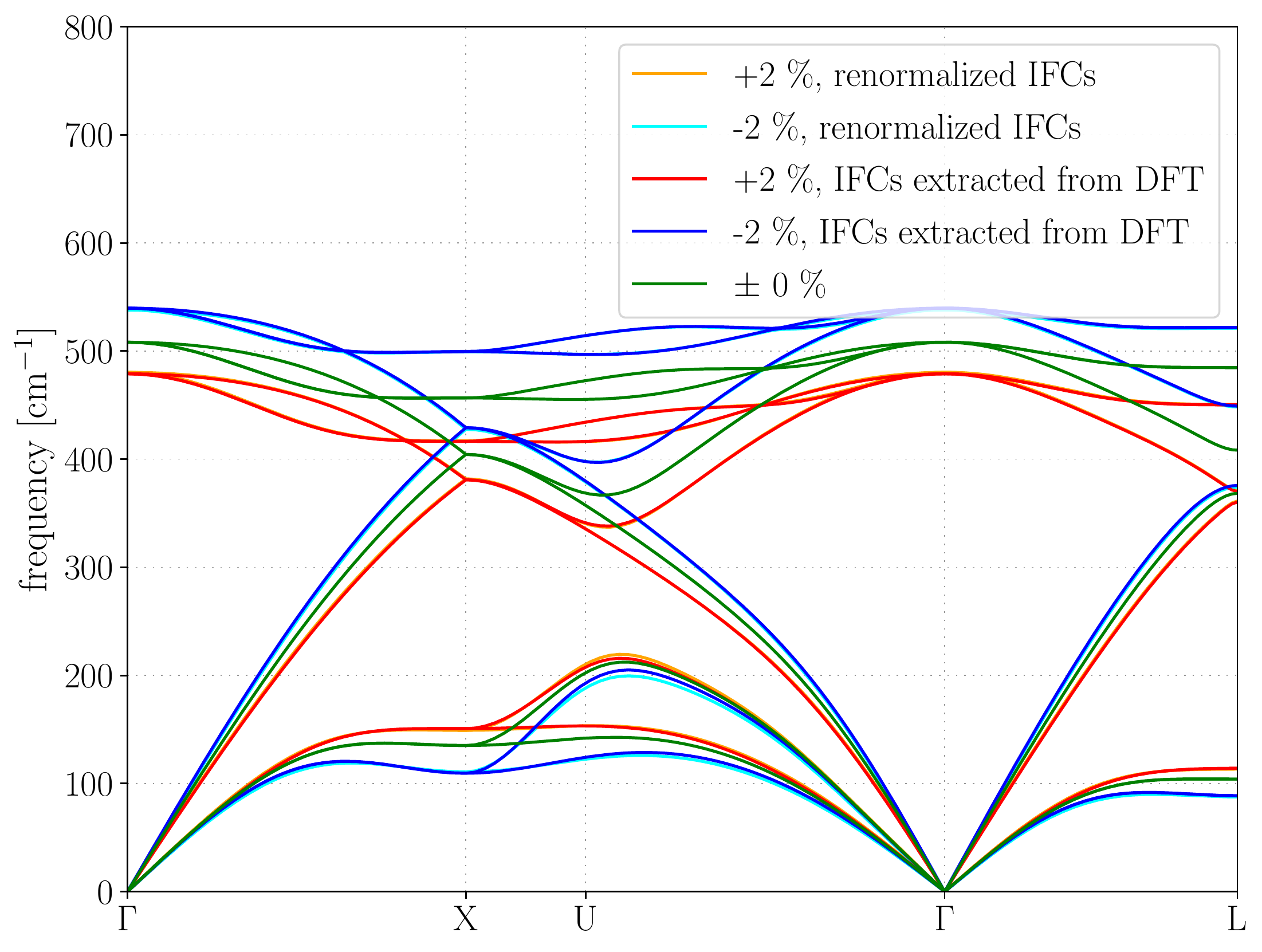}
\caption{
The harmonic phonon dispersions of silicon in different lattice constants. $\pm 2$\% means that lattice constant is expanded by $\pm 2$\% from the reference structure.
The calculation results of the IFC renormalization (renormalized IFCs) are compared with the results of DFT calculations at each lattice constant (IFCs extracted from DFT).
}
\label{Fig_SiIFCrenormalize}
\end{center}
\end{figure}

We first cosider the harmonic IFCs of silicon. The harmonic phonon dispersions of silicon are calculated for lattice constants expanded by $\pm 2$ \% from the reference structure. As shown in Fig. \ref{Fig_SiIFCrenormalize}, the phonon dispersions calculated by the IFC renormalization (renormalized IFCs) correctly reproduce the results of DFT calculations at each lattice constant (IFCs extracted from DFT). In conjunction with the success of the IFC renormalization in calculating the thermal expansion ( Fig.~\ref{Fig_SiThermalExpansion} in Section~\ref{Subsec_ResDis_silicon}), we can conclude that the IFC renormalization is accurate for weakly anharmonic materials, in which the low-order Taylor expansion accurately describes the potential energy surface. 

\begin{figure}[h]
\vspace{0cm}
\begin{center}
\includegraphics[width=0.48\textwidth]{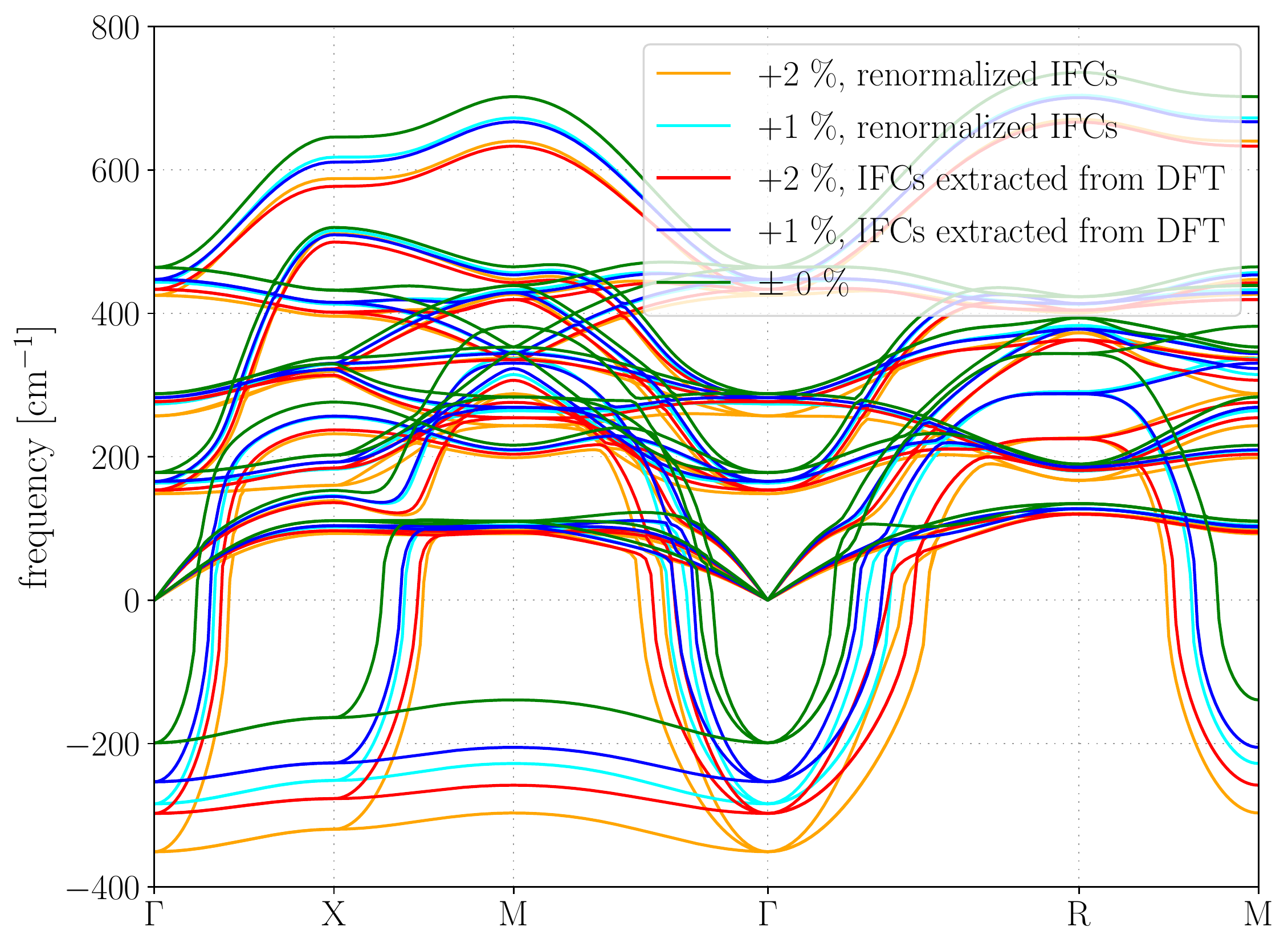}
\caption{
Harmonic phonon dispersions of cubic BaTiO$_3$ in lattice constants expanded by 1\% and 2\% from the reference structure.
The calculation results of the IFC renormalization (renormalized IFCs) are compared with the results of DFT calculations at each lattice constant (IFCs extracted from DFT).
}
\label{Fig_BTOIFCRenormalize}
\end{center}
\end{figure}

On the contrary, as shown in Fig.~\ref{Fig_BTOIFCRenormalize}, the IFC renormalization by Eq.~(\ref{Fig_BTOIFCRenormalize}) overestimates the frequency shift of the soft mode of BaTiO$_3$. This can be because the up-to-quartic IFCs does not perfectly reproduce every detail of the complicated potential energy surface. Thus, the calculated IFCs cannot be extended for deformed cells, which is not included in the training data for the IFC calculation. 

\begin{figure}[h]
\vspace{0cm}
\begin{center}
\includegraphics[width=0.48\textwidth]{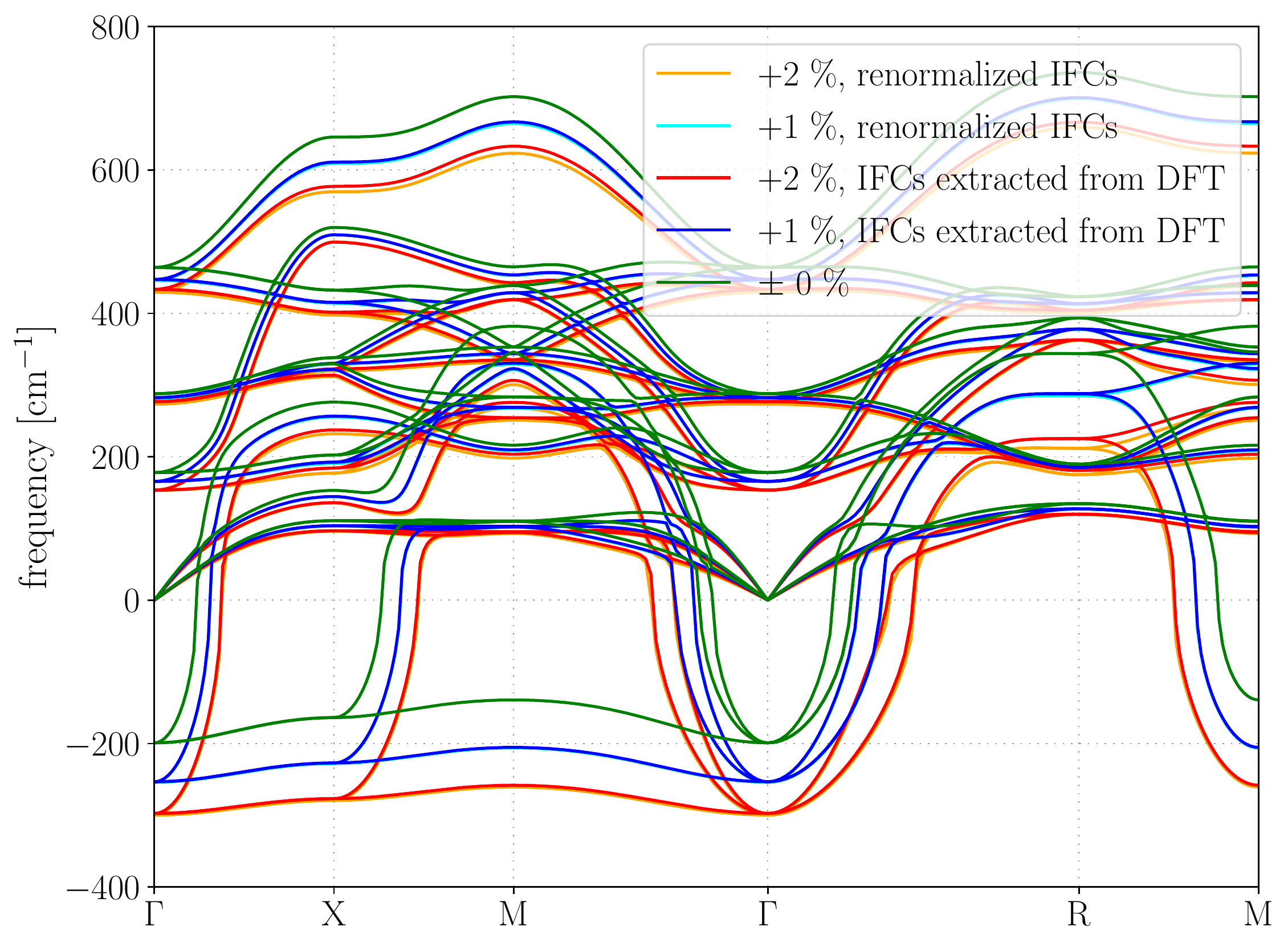}
\caption{
Harmonic phonon dispersions of cubic BaTiO$_3$ in lattice constants expanded by 1\% and 2\% from the reference structure.
The calculation results of the IFC renormalization (renormalized IFCs) are compared with the results of DFT calculations at each lattice constant (IFCs extracted from DFT). The strain-mode coupling constants are given as independent inputs in IFC renormalization.
}
\label{Fig_BTOIFCRenormalizeStrainModeCoupling}
\end{center}
\end{figure}

To avoid the error in the IFC renormalization, we directly calculate the strain-mode coupling $\dfrac{\partial \widetilde{\Phi}_{\mu_1 \mu_2}(\bm{R}_1 \alpha_1, \bm{R}_2 \alpha_2)}{\partial u_{\mu'_1 \nu'_1}}\Bigr|_{q^{(0)}=0, u_{\mu\nu} = 0}$ by fitting the relation between the strain and the harmonic IFCs. These coupling constants are given as inputs independent from the IFCs. The harmonic IFCs are expected to change smoothly with cell deformation because the number of independent degrees of freedom is small and nonanalytic regularization terms are not used in the calculation. Replacing the strain-mode coupling in the RHS of Eq. (\ref{EqUpdatePhi2withStrain}), we get Fig.~\ref{Fig_BTOIFCRenormalizeStrainModeCoupling}, in which the change of harmonic dispersion is accurately reproduced with the IFC renormalization.

\begin{figure}[h]
\vspace{0cm}
\begin{center}
\includegraphics[width=0.48\textwidth]{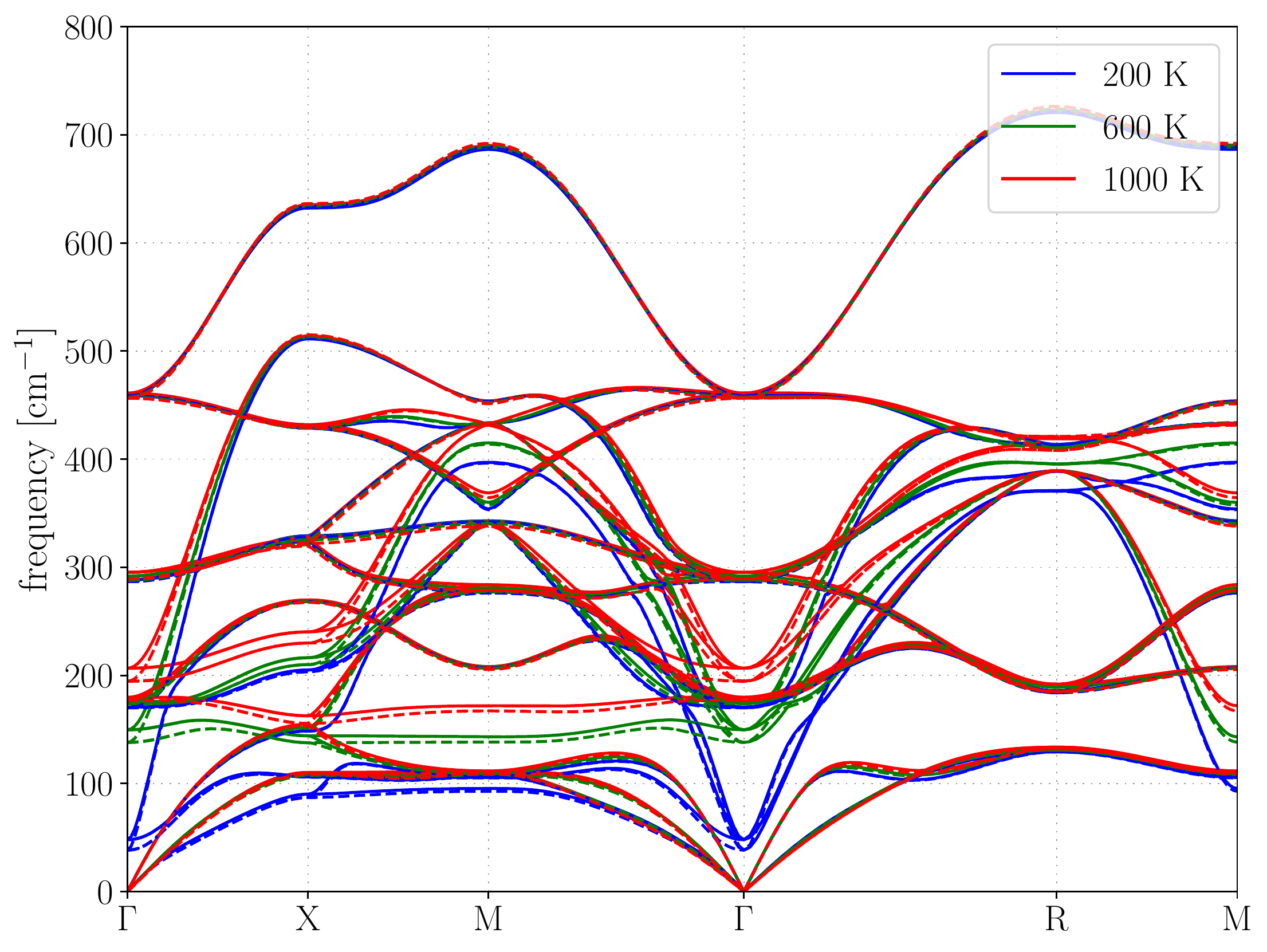}
\caption{
Temperature dependence of the SCP frequency $\Omega_{\bm{k}\lambda}$ of cubic BaTiO$_3$ with the lattice constant expanded by 1 \% from the reference structure. For the calculation of the solid lines, the harmonic IFCs for the expanded lattice constant and the quartic IFCs in the reference lattice constant are used. This corresponds to calculating quartic IFCs with IFC renormalization because the quartic IFCs are kept fixed in our calculation. For the calculation of dotted lines, the IFCs at the expanded lattice constant are used.}
\label{Fig_BTOIFCRenormalizeQuartic}
\end{center}
\end{figure}

Besides, we check the accuracy of the renormalized anharmonic IFCs in BaTiO$_3$. Because it is difficult to directly visualize the anharmonic IFCs, we investigate the temperature dependence of the SCP frequency $\Omega_{\bm{k}\lambda}$, which reflects the significance of the quartic IFCs. The quartic IFCs are fixed in our calculation of IFC renormalization because the Taylor expansion is truncated at fourth order. As shown in Fig.~\ref{Fig_BTOIFCRenormalizeQuartic}, using fixed quartic IFCs has little effect on the SCP frequencies, which indicates that the IFC renormalization works accurately for the anharmonic IFCs of BaTiO$_3$.

\section{Test of convergence with respect to the cutoff radius and $N_{\text{max}}$ in the IFC calculation}
\label{Appendix_conv_check_cutoff_Nmax}
The cutoff radius and the maximum number of distinct atoms $N_{\text{max}}$ involved in each IFC are adjustable parameters in the IFC calculation. The cutoff radius sets the maximum distance between the atoms. The IFCs are automatically set to zero for the combination of atoms that contains a pair whose distance is larger than the cutoff radius.
Setting a larger cutoff enables more accurate fitting, but it also leads to higher computational costs. 

Correspondingly, larger $N_{\text{max}}$ means more degrees of freedom in fitting the displacement-force data. It makes the fitting more accurate but calculating the IFCs become more costly.

\begin{figure}[h]
\vspace{0cm}
\begin{center}
\includegraphics[width=0.48\textwidth]{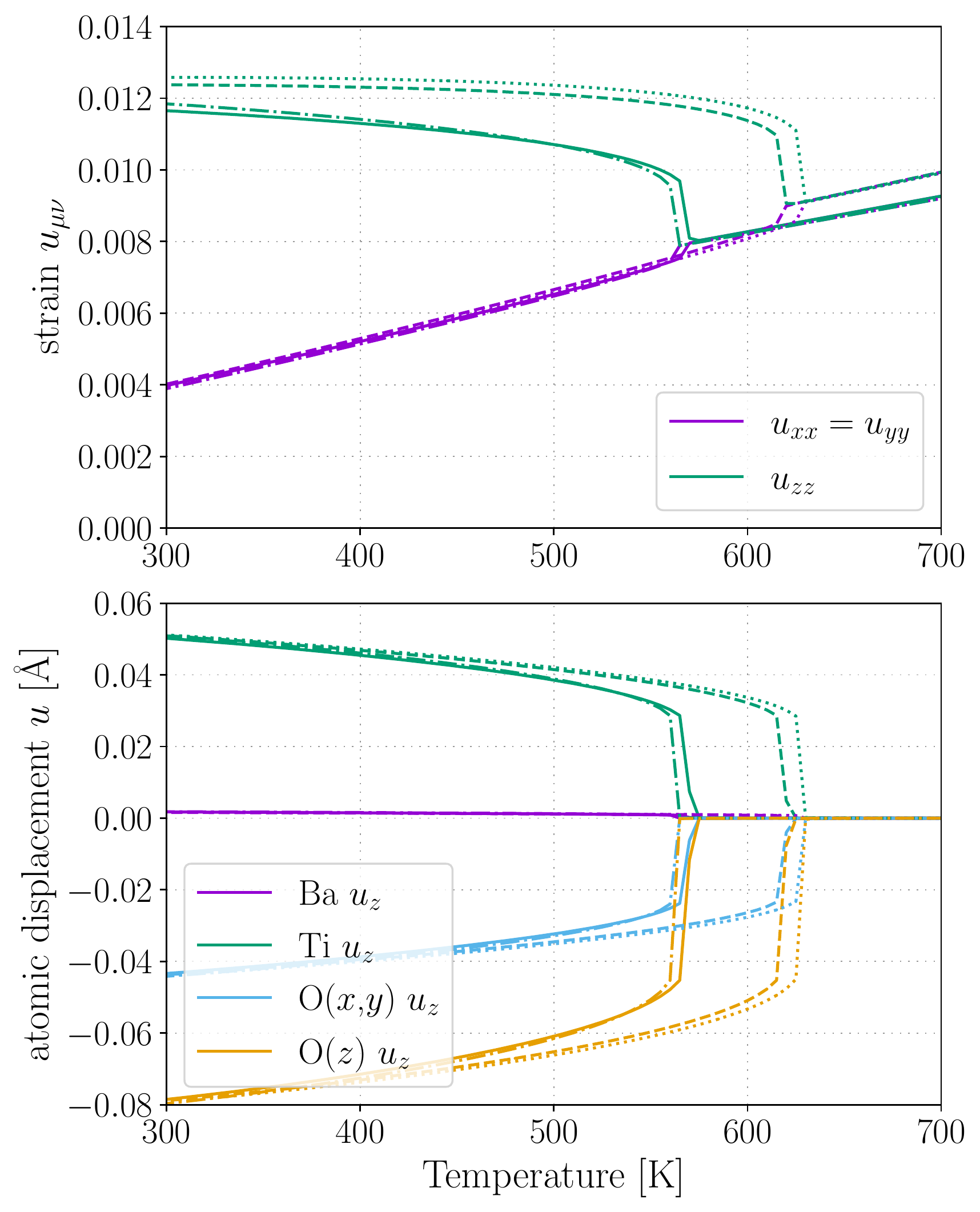}
\caption{
Comparison of the $T$-dependent crystal structure of BaTiO$_3$ in its cubic-tetragonal phase transition with different settings in the IFC calculation. The pairs (cutoff radius, $N_{\text{max}}$) are (cutoff$=$9 Bohr, $N_{\text{max}}$=3) for the solid line, (cutoff$=12$ Bohr, $N_{\text{max}}$=3) for the dashed line, (cutoff$=15$ Bohr, $N_{\text{max}}$=3) for the dotted line, (cutoff$=9$ Bohr, $N_{\text{max}}$=4) for the chain line.
The nuclear quantum effect is taken into account in the calculations.
}
\label{Fig_BTOCubicTetraNdata300CompareCutoff}
\end{center}
\end{figure}

\begin{figure}[h]
\vspace{0cm}
\begin{center}
\includegraphics[width=0.48\textwidth]{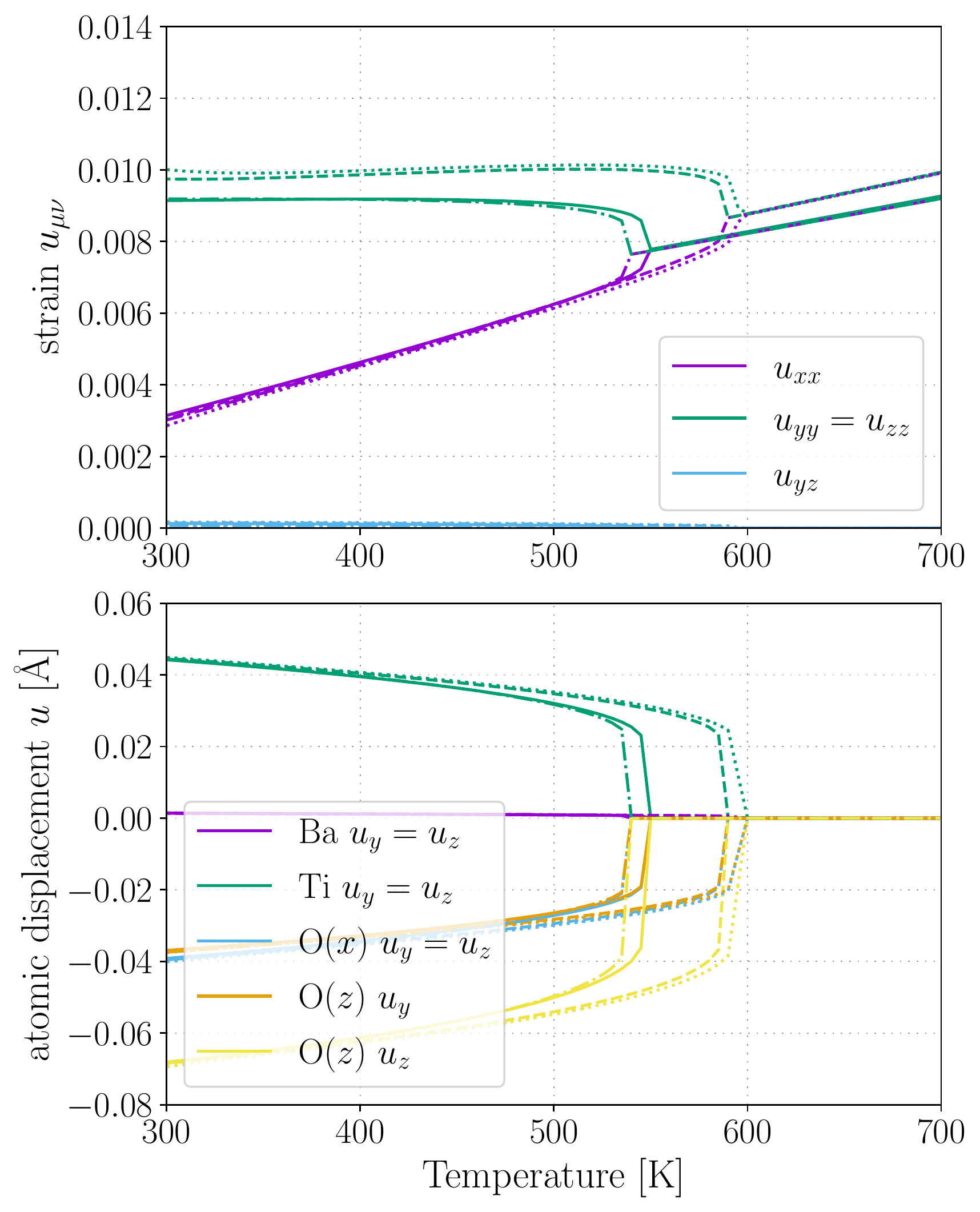}
\caption{
The comparison of the $T$-dependent crystal structure of BaTiO$_3$ in the virtual cubic-orthorhombic phase transition with different settings in the IFC calculation. The pairs (cutoff radius, $N_{\text{max}}$) are (cutoff$=$9 Bohr, $N_{\text{max}}$=3) for the solid line, (cutoff$=12$ Bohr, $N_{\text{max}}$=3) for the dashed line, (cutoff$=15$ Bohr, $N_{\text{max}}$=3) for the dotted line, (cutoff$=9$ Bohr, $N_{\text{max}}$=4) for the chain line.
The nuclear quantum effect is taken into account in the calculations.
}
\label{Fig_BTOCubicOrthoNdata300CompareCutoff}
\end{center}
\end{figure}

\begin{figure}[h]
\vspace{0cm}
\begin{center}
\includegraphics[width=0.48\textwidth]{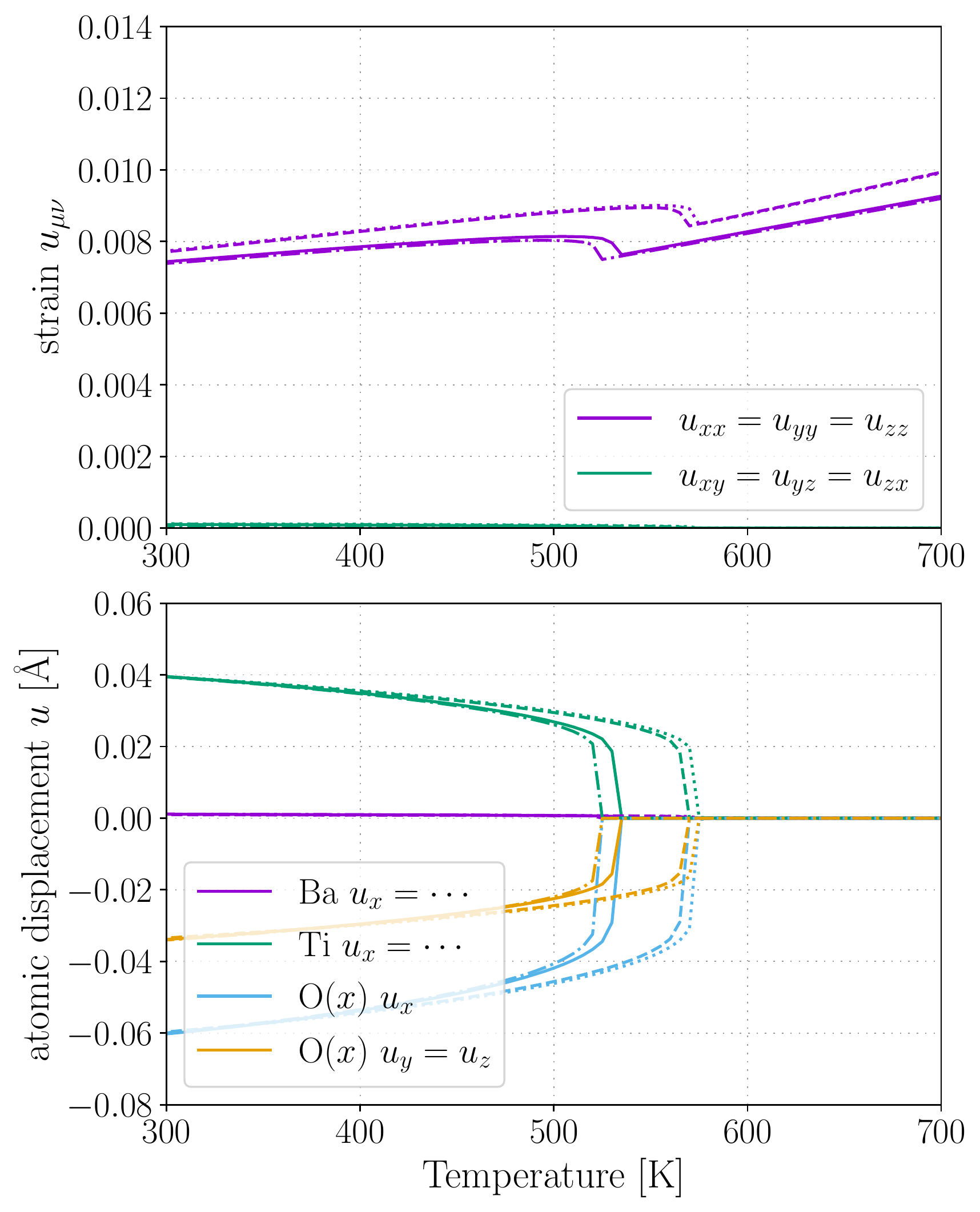}
\caption{
The comparison of the $T$-dependent crystal structure of BaTiO$_3$ in the virtual cubic-rhombohedral phase transition with different settings in the IFC calculation. The pairs (cutoff radius, $N_{\text{max}}$)  are (cutoff$=$9 Bohr, $N_{\text{max}}$=3) for the solid line, (cutoff$=12$ Bohr, $N_{\text{max}}$=3) for the dashed line, (cutoff$=15$ Bohr, $N_{\text{max}}$=3) for the dotted line, (cutoff$=9$ Bohr, $N_{\text{max}}$=4) for the chain line.
The nuclear quantum effect is taken into account in the calculations.
}
\label{Fig_BTOCubicRhomboNdata300CompareCutoff}
\end{center}
\end{figure}

We change the cutoff radius and $N_{\text{max}}$ of the quartic IFCs of BaTiO$_3$ to check the convergence of the calculation results, which results are shown in Figs.~\ref{Fig_BTOCubicTetraNdata300CompareCutoff}$\sim$\ref{Fig_BTOCubicRhomboNdata300CompareCutoff}. 
The cutoff radius for the cubic IFCs is fixed to 15.0 Bohr, which is comparable to the size of the supercell. No cutoff radius is set for harmonic IFCs. $N_{\text{max}} = 2,3$ for harmonic and cubic IFCs, respectively, is employed, which does not impose any restrictions on these IFCs. Later in this Appendix, the cutoff and $N_{\text{max}}$ refers to those for the quartic IFCs. We calculate the transitions between the cubic phase and each of the other three phases to take full advantage of crystal symmetry because the symmetry groups of rhombohedral and orthorhombic phases are not subgroups of that of orthorhombic and tetragonal phases respectively. The results in Figs.~\ref{Fig_BTOCubicTetraNdata300CompareCutoff}--\ref{Fig_BTOCubicRhomboNdata300CompareCutoff} are obtained by heating calculations, which is explained in Section~\ref{Subsec_ResDis_BTO}.
In Figs.~\ref{Fig_BTOCubicTetraNdata300CompareCutoff}--\ref{Fig_BTOCubicRhomboNdata300CompareCutoff}, the IFCs calculated by setting cutoff$=$9 Bohr, $N_{\text{max}}$=3 (solid line) and cutoff$=9$ Bohr, $N_{\text{max}}$=4 (chain line) produces almost the same result. This indicates that restricting the quartic IFCs to three-body ones ($N_{\text{max}} = 3$) does not affect the calculation result, which is fortunate because setting $N_{\text{max}} = 4$ with larger cutoff makes the calculation expensive.
Comparing the results of different cutoff radii, we can see that the calculation results are convergent when cutoff$=12$ Bohr. This also implies that the supercell is large enough because the calculation result converges with a cutoff which is smaller than the size of the supercell. Note that the length of a side of the $2\times2\times2$ supercell of BaTiO$_3$ is 7.9711~\AA$=$15.063 Bohr. We use the IFCs calculated with cutoff$=15$ Bohr, $N_{\text{max}}$=3 in the main part of this paper.

\section{Test of convergence with respect to the number of $k$-points in the SCP calculation}
\label{Appendix_conv_check_SCP_kpoints}

\begin{figure}[h]
\vspace{0cm}
\begin{center}
\includegraphics[width=0.48\textwidth]{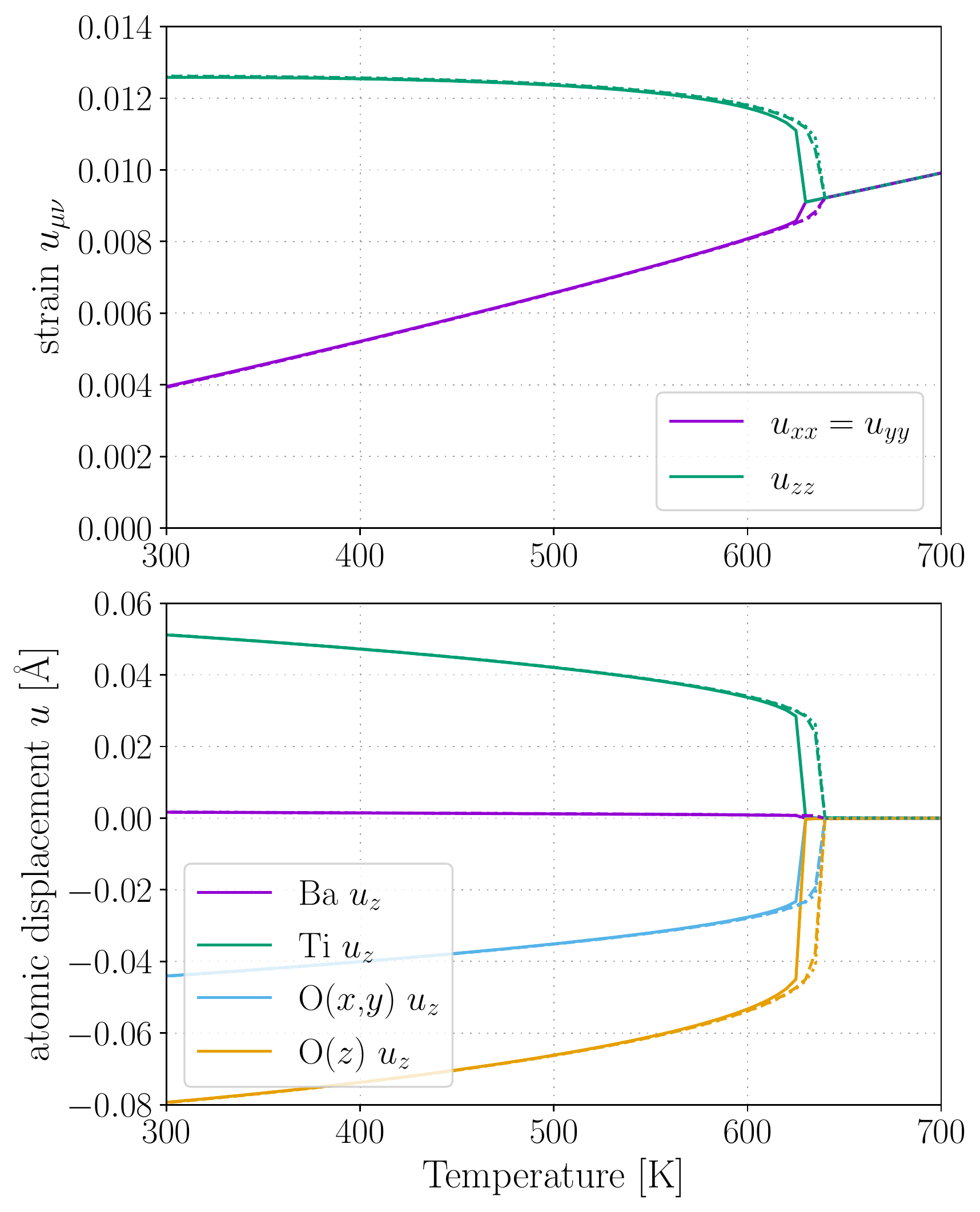}
\caption{
The comparison of the $T$-dependent crystal structure of BaTiO$_3$ in the cubic-tetragonal phase transition with different number of $k$-points in the SCP calculation. The $k$-point mesh is $8\times8\times8$ for the solid line, $12\times12\times12$ for the dashed line, $16\times16\times16$ for the dotted line. The dotted line ($16\times16\times16$) almost completely overlaps with the dashed line ($12\times12\times12$).
The nuclear quantum effect is taken into account in the calculations.
}
\label{Fig_BTOCubicTetraNdata300CompareKpoints}
\end{center}
\end{figure}

\begin{figure}[h]
\vspace{0cm}
\begin{center}
\includegraphics[width=0.48\textwidth]{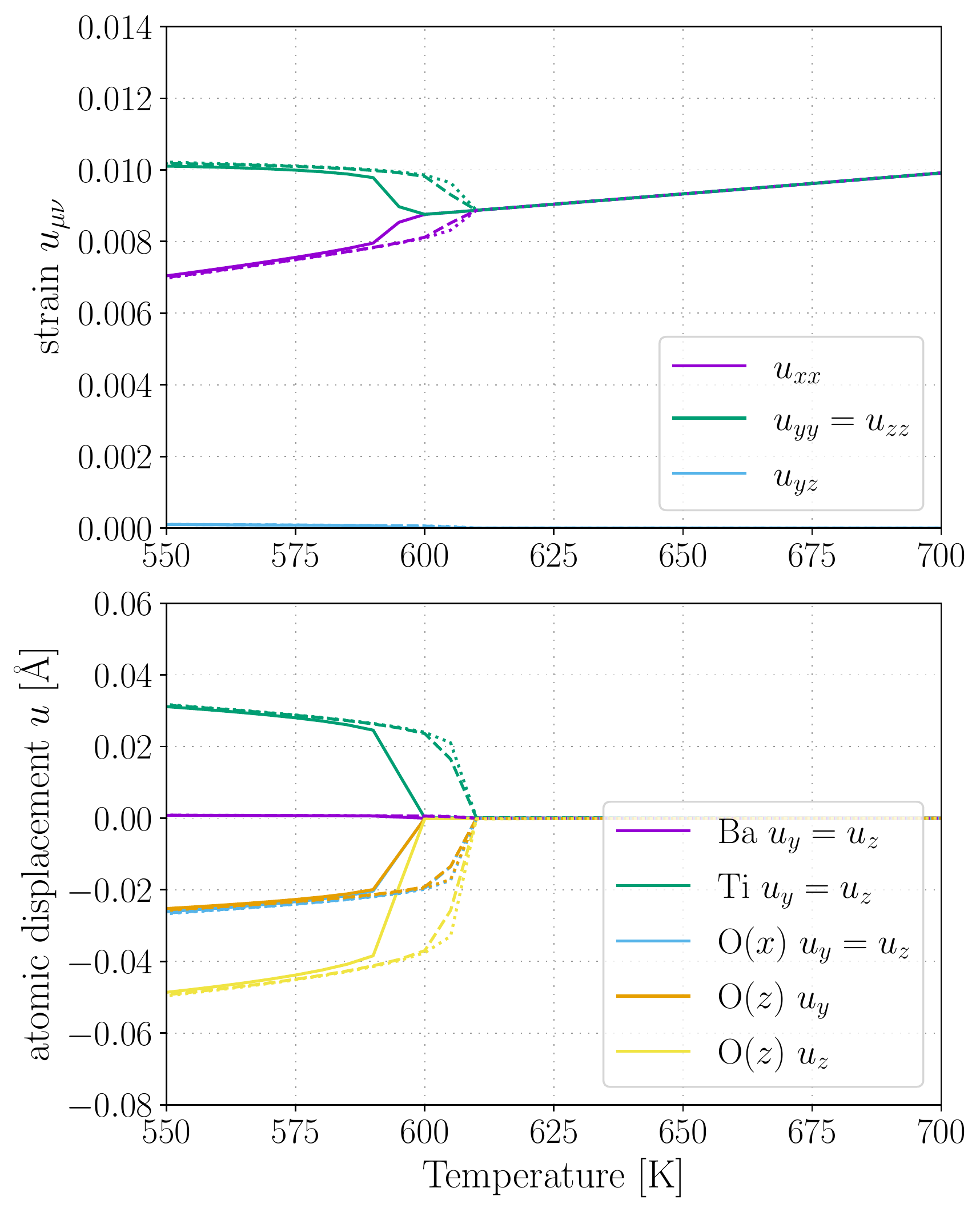}
\caption{
The comparison of the $T$-dependent crystal structure of BaTiO$_3$ in the virtual cubic-orthorhombic phase transition with different number of $k$-points in the SCP calculation. The $k$-point mesh is $8\times8\times8$ for the solid line, $12\times12\times12$ for the dashed line, $16\times16\times16$ for the dotted line. The dotted line ($16\times16\times16$) almost completely overlaps with the dashed line ($12\times12\times12$). 
The nuclear quantum effect is taken into account in the calculations.
}
\label{Fig_BTOCubicOrthoNdata300CompareKpoints}
\end{center}
\end{figure}

\begin{figure}[h]
\vspace{0cm}
\begin{center}
\includegraphics[width=0.48\textwidth]{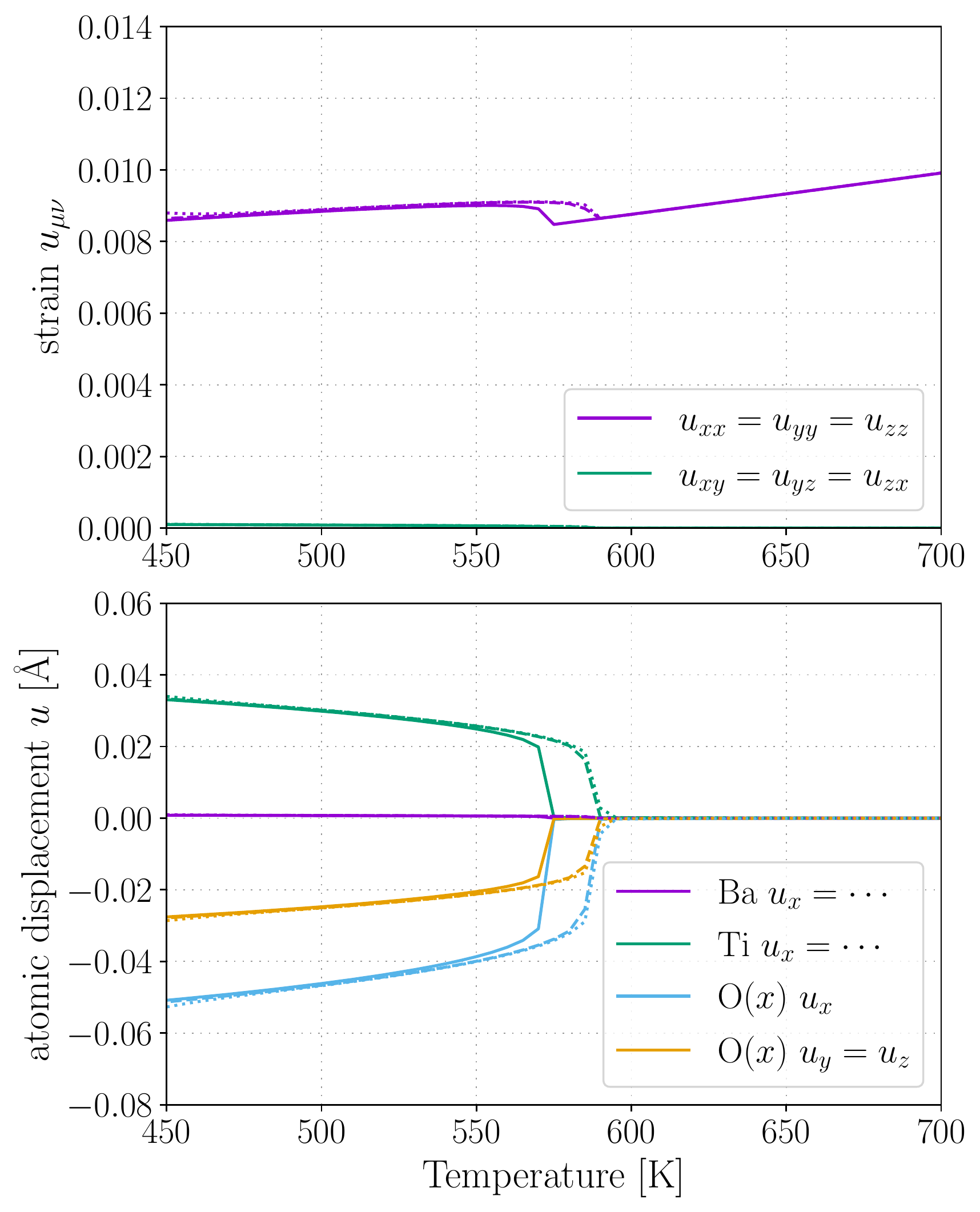}
\caption{
The comparison of the $T$-dependent crystal structure of BaTiO$_3$ in the virtual cubic-rhombohedral phase transition with different number of $k$-points in the SCP calculation. The $k$-point mesh is $8\times8\times8$ for the solid line, $12\times12\times12$ for the dashed line, $16\times16\times16$ for the dotted line. The dotted line ($16\times16\times16$) almost completely overlaps with the dashed line ($12\times12\times12$). 
The nuclear quantum effect is taken into account in the calculations.
}
\label{Fig_BTOCubicRhomboNdata300CompareKpoints}
\end{center}
\end{figure}

In this Appendix, we check the convergence of the calculation results with respect to the number of $k$-points in the SCP calculation. We calculate the temperature-dependent crystal structure of BaTiO$_3$ in the transitions between the cubic phase and each of the other three phases, as in Appendix~\ref{Appendix_conv_check_cutoff_Nmax}. We compare the result of heating calculations to check the convergence in low-symmetry phases in a wider temperature range.
As depicted in Figs.~\ref{Fig_BTOCubicTetraNdata300CompareKpoints}--\ref{Fig_BTOCubicRhomboNdata300CompareKpoints}, the calculation results is convergent with the $8\times8\times8$ $k$-mesh except in the very vicinity of the structural phase transition. We use $8\times8\times8$ $k$-mesh in the main text because the SCP calculation did not always show robust convergence when increasing the number of $k$-points. Improving the SCP solver to achieve more robust and efficient convergence is a topic of future research. 

\section{Test on the effect of the nonanalytic correction to the dynamical matrix}
\label{Appendix_nonanalytic}
\begin{figure}[h]
\vspace{0cm}
\begin{center}
\includegraphics[width=0.48\textwidth]{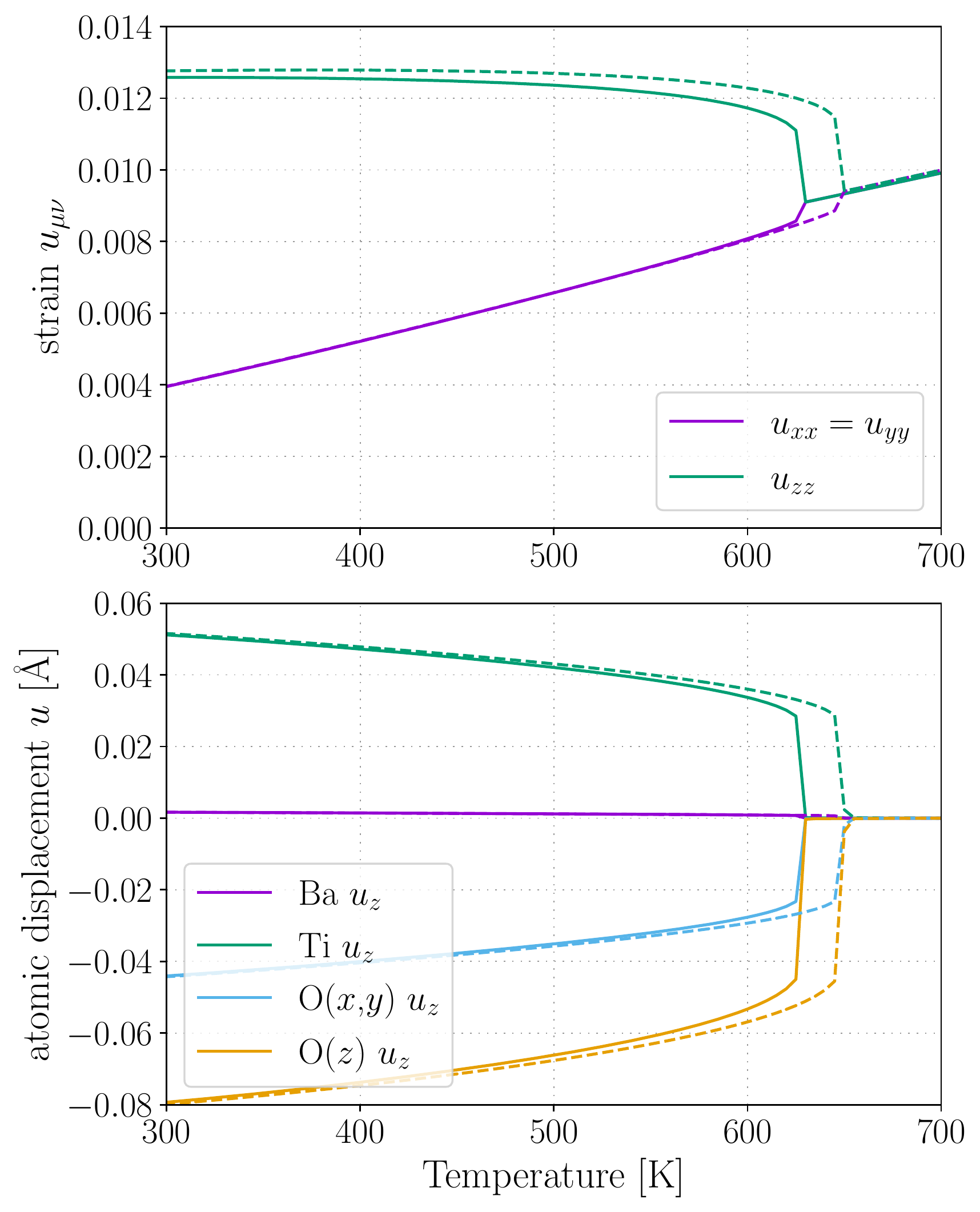}
\caption{
The comparison of the $T$-dependent crystal structure of BaTiO$_3$ in the cubic-tetragonal phase transition with (the solid line) and without (the dashed line) nonanalytic correction to the dynamical matrix in SCP calculations. The nonanalytic correction is considered in the mixed-space approach~\cite{Wang_2010}.
The nuclear quantum effect is taken into account in the calculations.
}
\label{Fig_BTOCubicTetraNdata300CompareNonanalytic}
\end{center}
\end{figure}

In the main text, we do not add the nonanalytic correction to the dynamical matrix in SCP calculations to enhance the convergence of the SCP calculations. Here, we check the effect of nonanalytic contribution by comparing the calculation results of the cubic-tetragonal transition, in which the calculation converged the most robustly. The nonanalytic contribution makes the calculation more unstable, presumably because the mixed-space approach~\cite{Wang_2010} causes unreasonable curves in the low-energy harmonic dispersions. As shown in Fig.~\ref{Fig_BTOCubicTetraNdata300CompareNonanalytic}, the nonanalytic contribution makes a small difference when the calculation converges, which validates the calculations without nonanalytic correction in the main text.

\section{Test of convergence with respect to the size of the size of the supercell in IFC calculation}
In order to check the convergence of the calculation results with respect to the size of the supercell in IFC calculation, we obtain the harmonic IFCs using the 4$\times$4$\times$4 supercell and use them in the structural optimization. We use the anharmonic IFCs calculated in the 2$\times$2$\times$2 supercell since the higher-order IFCs are more short-ranged in general. In addition, as discussed in Appendix~\ref{Appendix_conv_check_cutoff_Nmax}, the calculation results are fairly converged on the cutoff radius, which indirectly demonstrates the convergence on the supercell size for the anharmonic IFCs.

\begin{figure}[h]
\vspace{0cm}
\begin{center}
\includegraphics[width=0.48\textwidth]{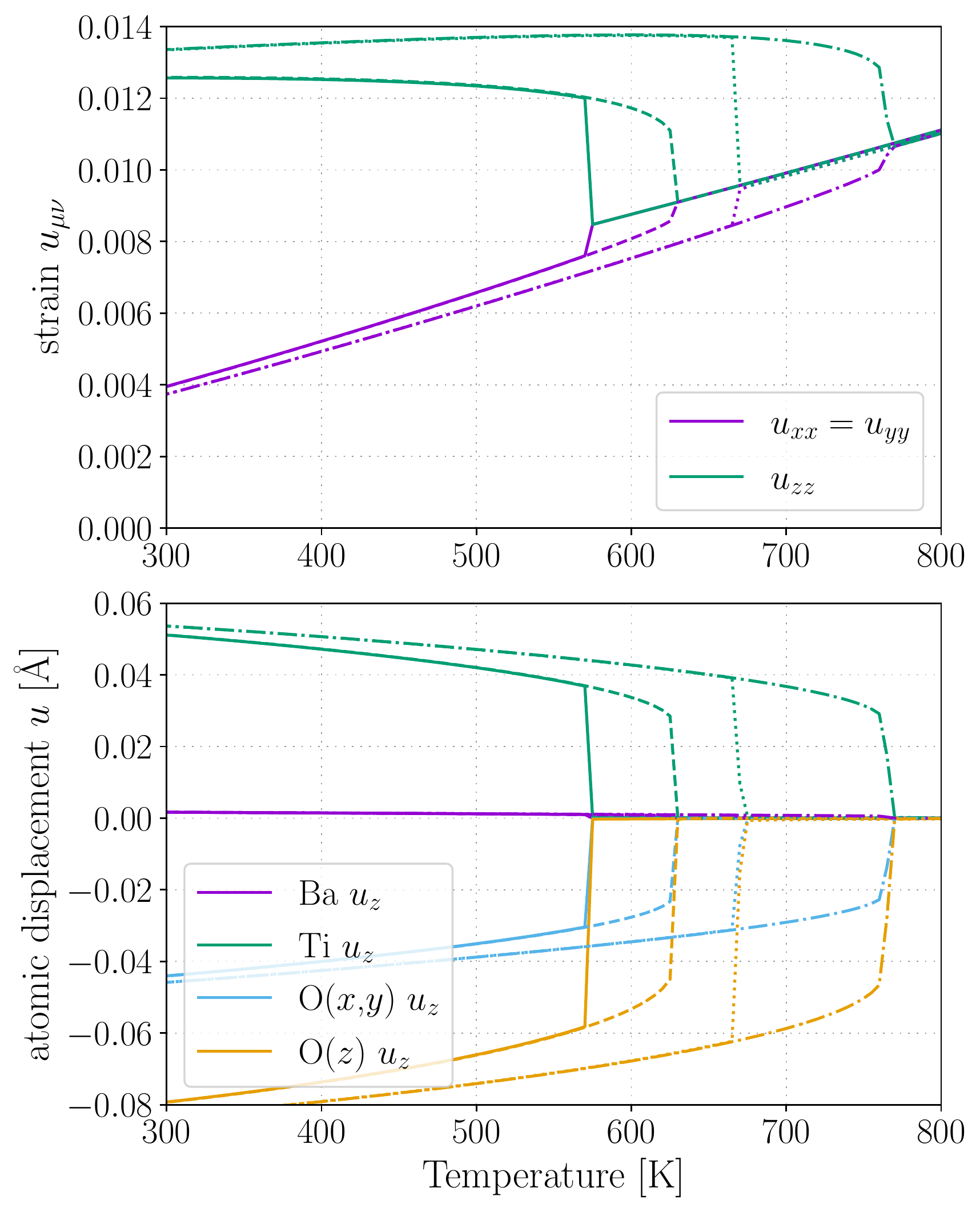}
\caption{
The comparison of the $T$-dependent crystal structure of BaTiO$_3$ in the cubic-tetragonal phase transition using IFCs calculated in different supercells. The solid line and the dashed lines represent the cooling and heating calculation with harmonic IFCs calculated in the 2$\times$2$\times$2 supercell, respectively. The dotted line and the chain line are the cooling and heating calculations with harmonic IFCs calculated in the 4$\times$4$\times$4 supercell.
The nuclear quantum effect is taken into account in the calculations.
}
\label{Fig_BTOCubicTetraNdata300CompareSupercell}
\end{center}
\end{figure}

\begin{figure}[h]
\vspace{0cm}
\begin{center}
\includegraphics[width=0.48\textwidth]{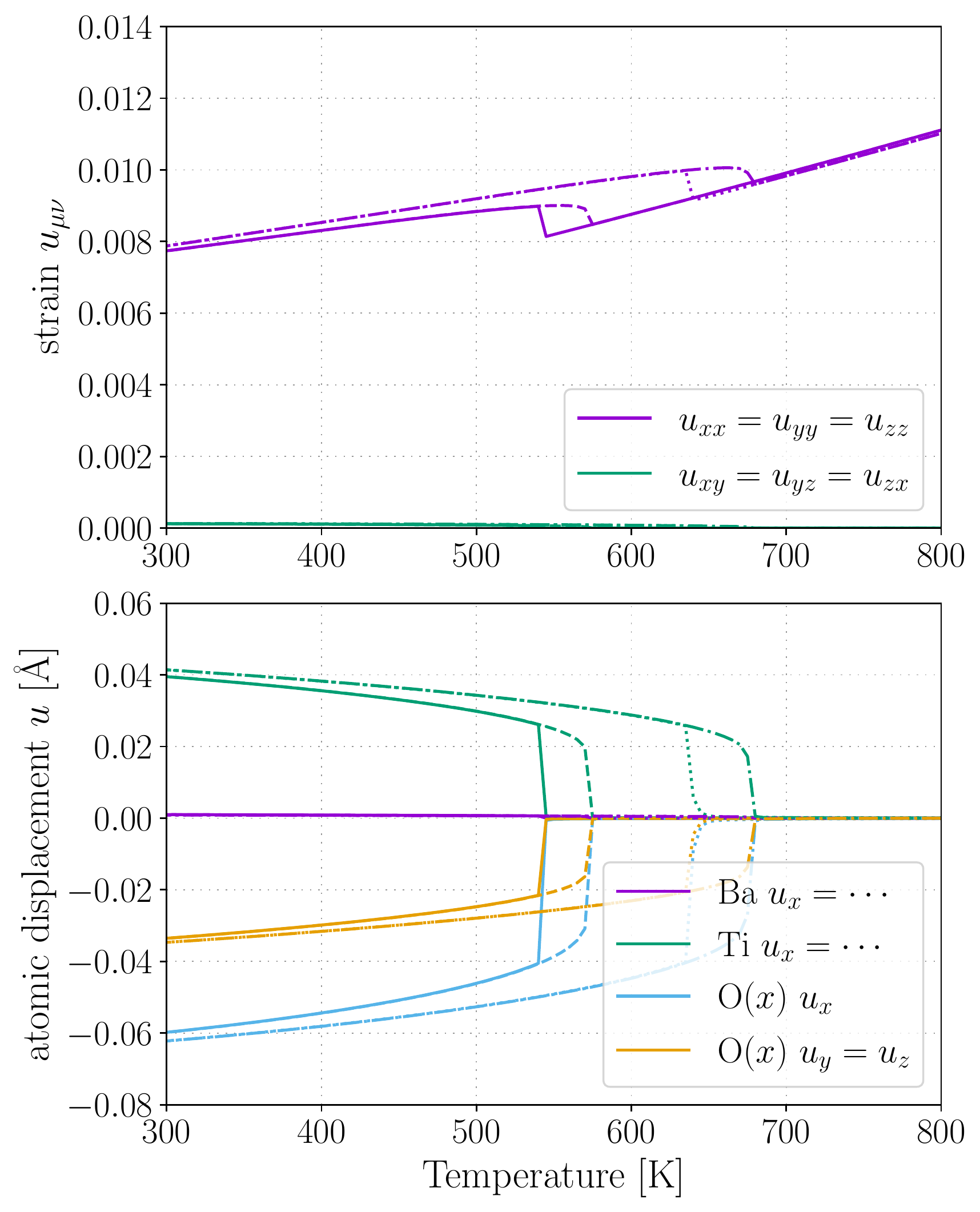}
\caption{
The comparison of the $T$-dependent crystal structure of BaTiO$_3$ in the virtual cubic-rhombohedral phase transition using IFCs calculated in different supercells. The solid line and the dashed lines represent the cooling and heating calculation with harmonic IFCs calculated in the 2$\times$2$\times$2 supercell, respectively. The dotted line and the chain line are the cooling and heating calculations with harmonic IFCs calculated in the 4$\times$4$\times$4 supercell.
The nuclear quantum effect is taken into account in the calculations.
}
\label{Fig_BTOCubicRhomboNdata300CompareSupercell}
\end{center}
\end{figure}

The comparison of the calculation results are shown in Figs.~\ref{Fig_BTOCubicTetraNdata300CompareSupercell}$\sim$\ref{Fig_BTOCubicRhomboNdata300CompareSupercell}. Although the transition temperatures are shifted when changing the supercell size, the $T$-dependence of the crystal structure agrees well except in the vicinity of the structural phase transitions. The differences in the transition temperatures are nearly 100 K, which is comparable to the error from the PBEsol functional. Note that the dependence on the DFT functionals will be much more prominent in general because the calculated transition temperatures are susceptible to the accuracy of the lattice constants.

\begin{figure}[h]
\vspace{0cm}
\begin{center}
\includegraphics[width=0.48\textwidth]{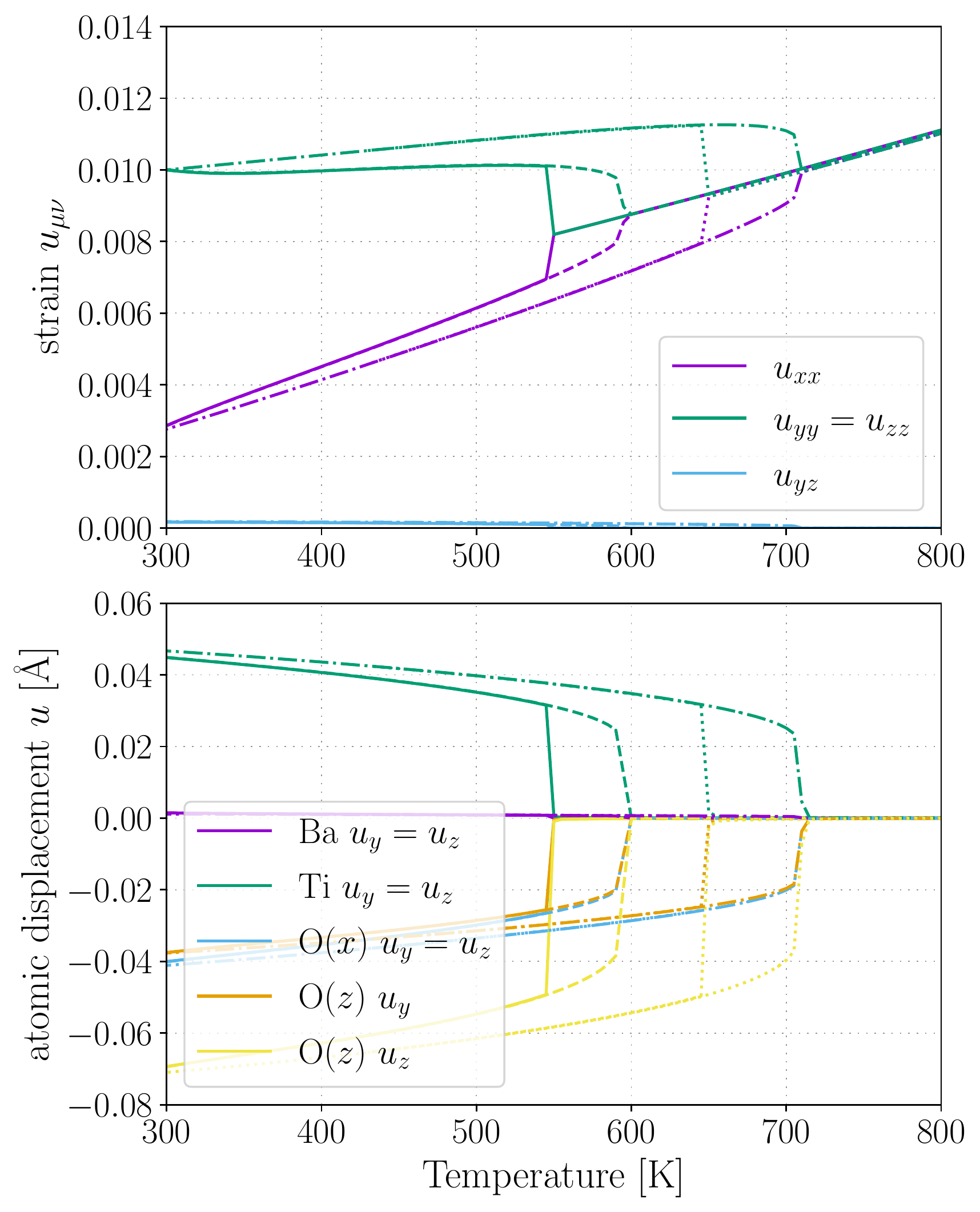}
\caption{
The comparison of the $T$-dependent crystal structure of BaTiO$_3$ in the virtual cubic-orthorhombic phase transition using IFCs calculated in different supercells. The solid line and the dashed lines represent the cooling and heating calculation with harmonic IFCs calculated in the 2$\times$2$\times$2 supercell, respectively. The dotted line and the chain line are the cooling and heating calculations with harmonic IFCs calculated in the 4$\times$4$\times$4 supercell.
The nuclear quantum effect is taken into account in the calculations.
}
\label{Fig_BTOCubicOrthoNdata300CompareSupercell}
\end{center}
\end{figure}

\section{Test on using the fixed Born effective charges in calculating the spontaneous polarization}
\label{Appendix_diffrence_of_Born_effcharge}
In Section \ref{subsec_str_opt_at_finiteT}, the spontaneous polarization is calculated with Eq. (\ref{Eq_spontaneous_polarization_Borneffcharge}). Here, the Born effective charges $Z^*_{\alpha, \mu\nu}$ are fixed to the values at the reference cubic structure. 
In this appendix, we check the change of the Born effective charges at different phases and estimate the error of the spontaneous polarization.

We calculate the crystal structure in the three low-temperature phases of BaTiO$_3$ by the structural optimization of DFT and calculate the Born effective charges at each phase. The optimum crystal structure in the tetragonal, orthorhombic, and rhombohedral phases are shown in Table~\ref{table_merged}. 
Compared to the crystal structures at finite temperatures, we can see that the deviations from the cubic structure are about twice as large.
The Born effective charges in the low-temperature phases are also shown in Tables~\ref{table_merged}. Compared to those in the reference cubic structure, the Born effective charges in the direction of polarization gets smaller by about 20 \% in the low-temperature phases. Thus, we estimate that the Born effective charges at the finite-temperature crystal structures are deviated by around $20/2= 10$ \% from those in the reference cubic structure. Since the spontaneous polarization is the integral of Born effective charges with respect to the structural changes, the error of the spontaneous polarization by using the fixed Born effective charge is around 5 \%, which does not significantly affect the calculation results.

\begin{table*}
\caption{Summary of the lattice vectors and atomic displacements of cubic, tetragonal, orthorhombic, and rhombohedral BaTiO$_3$ obtained by the structural optimization at zero-temperature based on DFT. The Born effective charges computed using the optimized structures are also shown in units of the elementary charge $e$. The fractional coordinates of each atom in the cubic phase are as follows: Ba:$(0,0,0)$, Ti:$(\frac{1}{2},\frac{1}{2},\frac{1}{2})$, O(1):$(0,\frac{1}{2},\frac{1}{2})$, O(2):$(\frac{1}{2},0,\frac{1}{2})$, O(3):$(\frac{1}{2},\frac{1}{2},0)$. }
\label{table_merged}
\begin{ruledtabular}
    \begin{tabular}{rccccc}
    & Cubic & Tetragonal & Orthorhombic & Rhombohedral \\ \hline
    Lattice vectors [\AA] & & & & & \\
    $a_1$ & (3.9855, 0.0000, 0.0000) & (3.9709, 0.0000, 0.0000) & (4.0170, 0.0062, 0.0000) & (4.0009, 0.0041, 0.0041) \\
    $a_2$ & (0.0000, 3.9855, 0.0000) & (0.0000, 3.9709, 0.0000) & (0.0062, 4.0170, 0.0000) & (0.0041, 4.0009, 0.0041)\\
    $a_3$ & (0.0000, 0.0000, 3.9855) & (0.0000, 0.0000, 4.0472) & (0.0000, 0.0000, 3.9658) & (0.0041, 0.0041, 4.0009) \\ \\
    Atomic displacements [\AA] & & & & &\\
    $u_{\text{Ba}}$ & (0, 0, 0) & (0, 0,  0.0382) & (0.0297, 0.0297, 0) & (0.0249, 0.0249, 0.0249) \\
    $u_{\text{Ti}}$ & (0, 0, 0) & (0, 0,  0.0993) & (0.0809, 0.0809, 0) & (0.0690, 0.0690, 0.0690) \\
    $u_{\text{O(1)}}$ & (0, 0, 0) & (0, 0, $-$0.0235) & ($-$0.0533, $-$0.0146, 0) & ($-$0.0435, $-$0.0151, $-$0.0151)\\ 
    $u_{\text{O(2)}}$ & (0, 0, 0) & (0, 0, $-$0.0235) & ($-$0.0146, $-$0.0533, 0) & ($-$0.0151, $-$0.0435, $-$0.0151) \\ 
    $u_{\text{O(3)}}$ & (0, 0, 0) & (0, 0, $-$0.0702) & ($-$0.0225, $-$0.0225, 0) & ($-$0.0151, $-$0.0151, $-$0.0435) \\ \\
    Born effective charges [$e$] & & & & \\
    \multirow{3}{*}{$Z^{*}_{\text{Ba}}$}  & (2.725, 0, 0) & (2.726, 0, 0) & (2.776, $-$0.016, 0) & (2.763, $-$0.011, $-$0.011)\\
                                          & (0, 2.725, 0) & (0, 2.726, 0) & ($-$0.016, 2.776, 0) & ($-$0.011, 2.763, $-$0.011) \\
                                          & (0, 0, 2.725) & (0, 0, 2.804) & (0, 0, 2.728) & ($-$0.011, $-$0.011, 2.763) \\
    \rule{0pt}{3ex}                                             
    \multirow{3}{*}{$Z^{*}_{\text{Ti}}$}  & (7.068, 0, 0) & (7.021, 0, 0) & (6.180, $-$0.319, 0) & (6.400, $-$0.244, $-$0.244) \\
                                          & (0, 7.068, 0) & (0, 7.021, 0) & ($-$0.319, 6.180, 0) & ($-$0.244, 6.400, $-$0.244) \\
                                          & (0, 0, 7.068) & (0, 0, 5.823) & (0, 0, 7.018) & ($-$0.244, $-$0.244, 6.400) \\
   \rule{0pt}{3ex}                                          
   \multirow{3}{*}{$Z^{*}_{\text{O(1)}}$} & ($-$5.576, 0, 0) & ($-$5.606, 0, 0) & ($-$4.954, 0.088, 0) & ($-$5.135, 0.072, 0.072) \\
                                          & (0, $-$2.109, 0) & (0, $-$2.098, 0) & (0.266, $-$1.991, 0) & (0.199, $-$2.014, $-$0.016)\\
                                          & (0, 0, $-$2.109) & (0, 0, $-$1.983) & (0, 0, $-$2.058) & (0.199, $-$0.016, $-$2.014) \\ 
    \rule{0pt}{3ex}                                          
   \multirow{3}{*}{$Z^{*}_{\text{O(2)}}$} & ($-$2.109, 0, 0) & ($-$2.098, 0, 0) & ($-$1.991, 0.266, 0) & ($-$2.014, 0.199, $-$0.016)\\
                                          & (0, $-$5.576, 0) & (0, $-$5.606, 0) & (0.088, $-$4.954, 0) & (0.072, $-$5.135, 0.072) \\
                                          & (0, 0, $-$2.109) & (0, 0, $-$1.983) & (0, 0, $-$2.058) & ($-$0.016, 0.199, $-$2.014)\\
   \rule{0pt}{3ex}                                          
   \multirow{3}{*}{$Z^{*}_{\text{O(3)}}$} & ($-$2.109, 0, 0) & ($-$2.043, 0, 0) & ($-$2.010, $-$0.020, 0) & ($-$2.014, $-$0.016, 0.199)\\
                                          & (0, $-$2.109, 0) & (0, $-$2.043, 0) & ($-$0.020, $-$2.010, 0) & ($-$0.016, $-$2.014, 0.199)\\
                                          & (0, 0, $-$5.576) & (0, 0, $-$4.661) & (0, 0, $-$5.629) & (0.072, 0.072, $-$5.135)
    \end{tabular}
\end{ruledtabular}
\end{table*}

\bibliography{apssamp}% Produces the bibliography via BibTeX.

\end{document}